\documentclass[amsmath,amssymb,showpacs,showkeywords,twocolumn]{revtex4}
\usepackage{graphicx}
\usepackage{times}
\usepackage{mathrsfs}
\usepackage{amsmath}
\usepackage{graphicx}
\usepackage{epsfig}
\usepackage{dcolumn}
\usepackage{bm}
\usepackage{multirow}
\usepackage[utf8x]{inputenc}
\begin{document}

\title{Spin Torque Oscillator and Magnetization Switching in Double-Barrier Rashba Zeeman Magnetic Tunnel Junction}
\author{Saumen Acharjee\footnote{saumenacharjee@dibru.ac.in},  Arindam Boruah\footnote{arindamboruah@dibru.ac.in}, Reeta Devi\footnote{reetadevi@dibru.ac.in} and Nimisha Dutta\footnote{nimishadutta@dibru.ac.in}}
\affiliation{Department of Physics, Dibrugarh University, Dibrugarh 786 004, 
Assam, India}

\begin{abstract}
In this work, we have studied the spin torque based magnetization oscillations and switching in presence of Rashba - Zeeman (RZ), Ruderman - Kittel - Kasuya - Yoside (RKKY) and Dzyaloshinskii - Moriya (DM) interactions in a double-barrier RZ$|$Heavy Metal (HM)$|$RZ magnetic tunnel junction (MTJ). The system has stable magnetization oscillations and can work as an oscillator or a switcher for a significant difference in the strength of RKKY and DM interaction under suitable spin transfer torque (STT). For the proposed system with same order of RKKY and DM interaction, a nonlinear characteristic of the magnetization oscillation is observed. However, this nonlinearity of oscillations can be reduced by an external magnetic field or considering a material with suitable RZ interaction. In addition to this, our study reveals the magnetization switching can be tuned by using suitable STT. A dependence of switching time on layer thickness is also observed. Also, the switching speed increases with the thickness for systems having either same order of RKKY and DM interaction or dominated by RKKY interaction. An opposite characteristic is seen when DM interaction dominates over RKKY interaction.
\end{abstract}

\pacs{72.25.Dc, 72.25.-b, 75.78.-n, 75.75.−c, 85.75.-d}
\maketitle

\section{Introduction}
Spin torque oscillators (STOs) are nano-sized electronic devices based on magnetic tunnel junctions (MTJ) that have recently received considerable attention due to their wide range of applications \cite{slonczewski1,slonczewski2,berger,locatelli, grollier, torrejon,kudo,tsunegi,romera,acharjee1,taniguchi,johansen}.
STOs are essentially magnetoresistive stacks, where the polarized spin current generates the spin torque \cite{slonczewski1,slonczewski2,berger}. Thus, it leads to self-sustaining magnetization oscillations in the free layer. This characteristic of STOs can be utilized to use them as microwave generators, field sensors, phased array antennae etc \cite{locatelli, grollier, torrejon,kudo,tsunegi,romera,acharjee1,taniguchi,johansen}. Another essential phenomenon witnessed in an MTJ is magnetization switching, which is the backbone of current non-volatile magnetic memories \cite{katine, kiselev, kubota, krivorotov1, krivorotov2}. 

Usually, a conventional MTJ consists of a tunnel barrier sandwiched between two ferromagnetic layers, namely, pinned and free layers. However, such arrangements have non-adequate thermal stability below $40$ nm \cite{li}. This can be achieved by using double interface MTJs, typically an arrangement of a heavy metal (HM) between two ferromagnetic layers \cite{choi,sato,li}. The FM$|$HM$|$FM arrangement is vital in enhancing the spin-orbit coupling (SOC) and generating Ruderman - Kittel - Kasuya - Yoside (RKKY) interaction. Due to the presence of RKKY interaction, the magnetizations of the two layers are ferromagnetically coupled \cite{parkin, li} . So they can behave like identical layers. Moreover, the strong SOC of HM can also even induce  Dzyaloshinskii - Moriya (DM) interaction, which is an antisymmetric exchange coupling, \cite{dzyaloshinsky, moriya}. Recent studies have shown that RKKY interactions may suppress the detrimental effects of DM interactions in STT switching \cite{li}. Thus it is noteworthy to understand the interplay of SOC, RKKY and DM interaction in STT magnetization dynamics.

Recently discovered Rashba - Zeeman (RZ) effect in  Ag$_2$Te / Cr$_2$O$_3$ composite can provide some novel features that are not found in pure Rashba or Zeeman systems \cite{tao1}. These materials have some unique characteristics and also have the ability to trigger insulator to conductor via an exchange field \cite{tao1} and can also be used as spin filters \cite{xiao,wojcik,acharjee4}. In general, the Rashba SOC (RSOC) is an antisymmetric SOC responsible for splitting of energy sub-bands \cite{rashba,liu11,zhang111, chico,ganguly1,ganguly2, fouladi1,zhang121,liu111,lashell,acharjee,acharjee2,acharjee3,cavigilia,ishizaka}. The impact of RSOC on STOs is worth mentioning, which shows the utilization of RSOC in magnetization switching and also its contribution to self-oscillations in STOs. Moreover, the RSOC of the multilayer systems can substantially increase the size of the STO phase \cite{johansen, duan}. So, it is necessary to understand the role of RSOC and its interplay with RKKY and DMI in STOs and magnetization switching. Though STOs and magnetization switching had been studied in double-barrier MTJs earlier, the introduction of RZ material as a free layer can significantly change auto oscillation and magnetization switching conditions. Moreover, STOs and magnetization switching were not studied in the same frame earlier. Thus in this work, we consider a double-barrier Rashba Zeeman Magnetic Tunnelling Junction (RZ-MTJ) where we consider an HM sandwiched between two RZs.

The organization of this paper are as follows: we present a minimal theory to study magnetization dynamics and switching in double interface RZ-MTJ in section II. In Section III, we present the results of our work by considering the effect of RKKY, DMI, RSOC and other important parameters like external magnetic field. We have analysed the effect of the parameter on magnetization dynamics and switching too in this section. Finally, a brief summary of our work is presented in Section IV.

\section{Minimal Theory}
\begin{figure}[hbt]
\centerline
\centerline{
\includegraphics[scale=0.68]{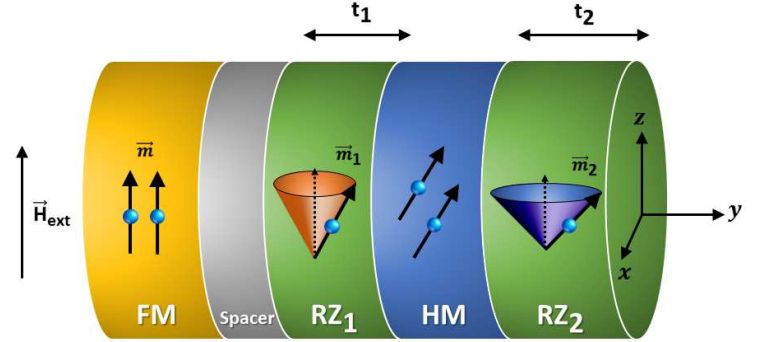}
\vspace{-0.1cm}
}
\caption{Schematic illustration of the proposed double-barrier MTJ consisting of Rashba-Zeeman (RZ)$|$Heavy metal (HM)$|$Rashba-Zeeman (RZ) hybrid as a composite free layer sandwiched between Ferromagnetic (FM) reference and control layers. The reference and control layers act as polarisers whose magnetizations can be controlled independently of the free layer. As a result an STT is induced in the RZ$|$HM$|$RZ composite free layer.  $\mathbf{m}_1$ and $\mathbf{m}_2$ represent the magnetizations while $t_1$ and $t_2$ are the thickness of the RZ$_1$ and RZ$_2$ layers respectively. The magnetization of FM layer is denoted by $\mathbf{m}$.
}
\label{fig1}
\end{figure}

The schematic representation of a double-barrier RZ based MTJ is shown in Fig. \ref{fig1}. In general a double-barrier MTJ is composed of three magnetic layers viz. reference layer, free/storage layer and control layer.  The reference and control layers act as polarisers whose magnetizations can be controlled independently of the free layer \cite{coelho}. We have considered an FM reference layer and an FM control layer with a RZ$|$HM$|$RZ composite free layer for our analysis. A Spin Transfer Torque (STT) is induced in the free layer because of the FM polarisers \cite{li, coelho} as shown in Fig. \ref{fig1}. We choose the easy axis anisotropy along the y-direction, while an external magnetic field is considered along the z-direction in our analysis. The time evolution of the magnetization in RZ layers can be studied by using two coupled Landau - Lifshitz - Gilbert - Slonczewski (LLGS) equations.
\begin{equation}
\label{eq1}
\partial_t\mathbf{m}_j = - \gamma\mathbf{m}_j\times (\mathbf{H}_{\text{eff}} + \mathbf{H}^{\text{R}}_j)+\alpha_j\mathbf{m}_j\times\partial_t\mathbf{m}_j+\mathbf{\mathcal{T}}_j^\text{STT}
\end{equation}
where, the indices $j = 1,2$ corresponds to first and second RZ layers respectively. The parameter $\gamma$ is the gyromagnetic ratio and $\alpha_j$ is the Gilbert damping parameter of the RZ layers while $\mathbf{H}_{\text{eff}}$ is the effective field of the system can be obtained using the effective Hamiltonian $\mathcal{H}_{\text{eff}}$ of the system
\begin{multline}
\label{eq2}
\mathcal{H}_\text{eff} = \text{K}_\text{exc}(\nabla \mathbf{m}_j)^2-\mathbf{D}_{12}.(\mathbf{m}_1\times\mathbf{m}_2) + \mathcal{J}_{12}(1-\mathbf{m}_1.\mathbf{m}_2)\\ - \text{H}_\text{ext} ( \hat{z}.\mathbf{m}_j) + h_0 \mathbf{\sigma}_j.\hat{m}_j
\end{multline}

where, $\text{K}_\text{exc}$ incorporate the exchange coupling between the FM and RZ layers, $\mathbf{D}_{12}$ is the DMI vector, $\mathcal{J}_{12} = \frac{\sigma_{exc}}{\Delta_{12}}$ is the ratio of bilinear exchange coefficient between two surfaces with discretion cell dimension \cite{li}. Here, $\text{H}_\text{ext}$ is the external magnetic field strength and the last term of Eq. (\ref{eq2}) gives the contribution of Zeeman energy with $h_0$ being the strength of Zeeman energy. The Rashba field $\mathbf{H}^{\text{R}}_j$ in Eq. (\ref{eq1}) is given by \cite{johansen}
\begin{equation}
\label{eq3}
\mathbf{H}^{\text{R}}_j = -\frac{1}{1+\beta^2}\frac{\alpha_\text{R}m_e\mathcal{P}_jj_0}{eM_0\hbar\mu_0}(\hat{y}\times\hat{z})
\end{equation}
where, $\alpha_\text{R}$ is the RSOC strength with $\mathcal{P}_j$ being the polarization of the current $j_0$ through the RZ-layers. $M_0$ is the saturation magnetization and $\beta$ is the adiabatic damping parameter. It is to be noted that the anisotropy of the system can be incorporated by considering an easy-axis anisotropy field $\mathbf{H}_\text{an} = \frac{1}{2}\frac{\text{K}_\text{an}m_y}{M_0}\hat{y}$ in Eq. (\ref{eq1}) \cite{acharjee1}. 
\begin{figure*}[hbt]
\centerline
\centerline{
\includegraphics[scale = 0.3]{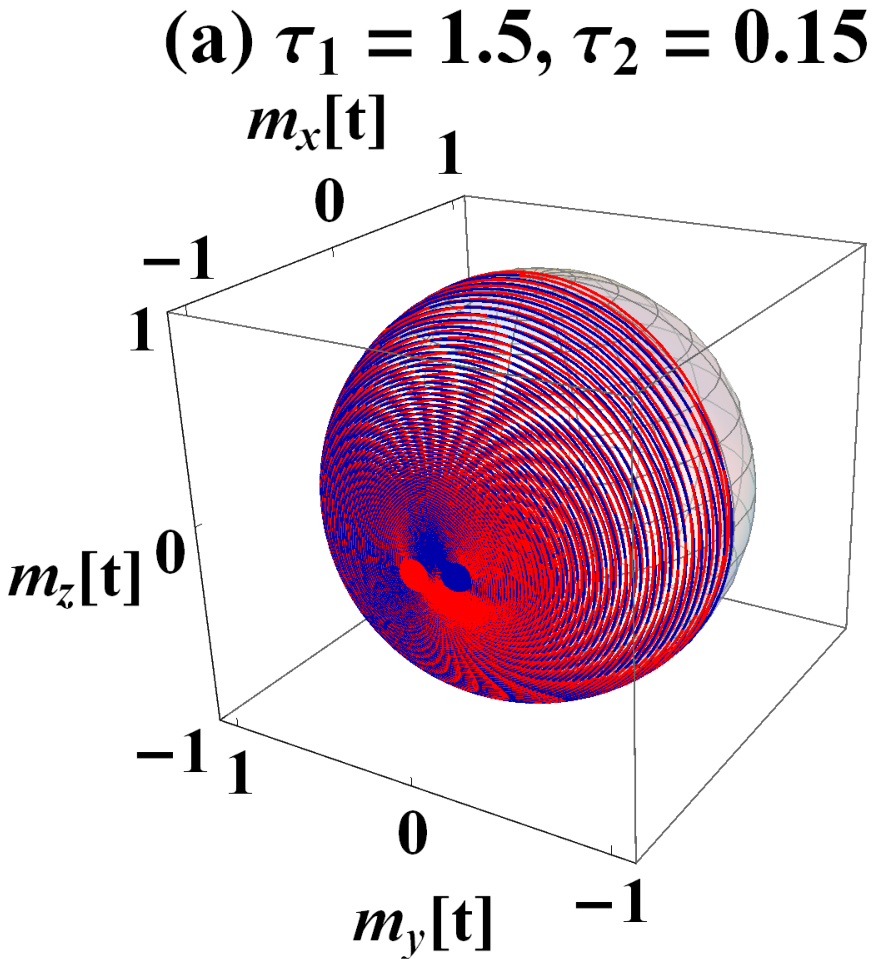}
\hspace{-0.09cm}
\vspace{0.02cm}
\includegraphics[scale = 0.3]{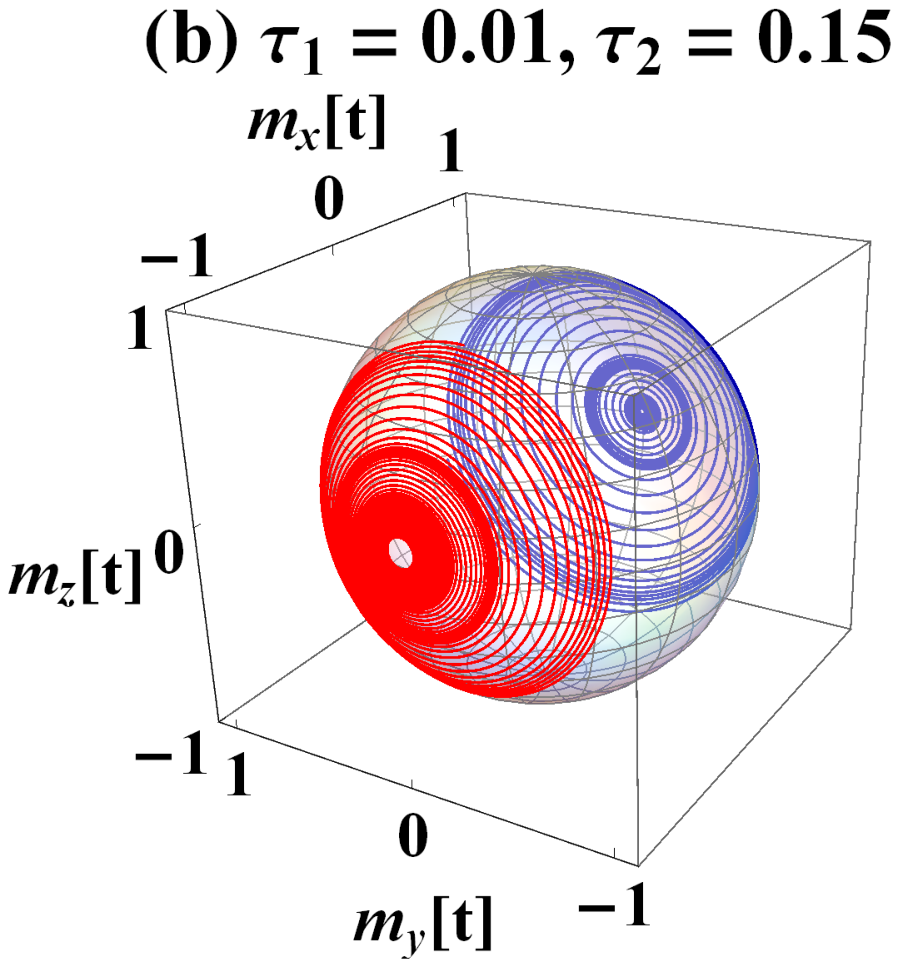}
\hspace{-0.09cm}
\vspace{0.02cm}
\includegraphics[scale = 0.3]{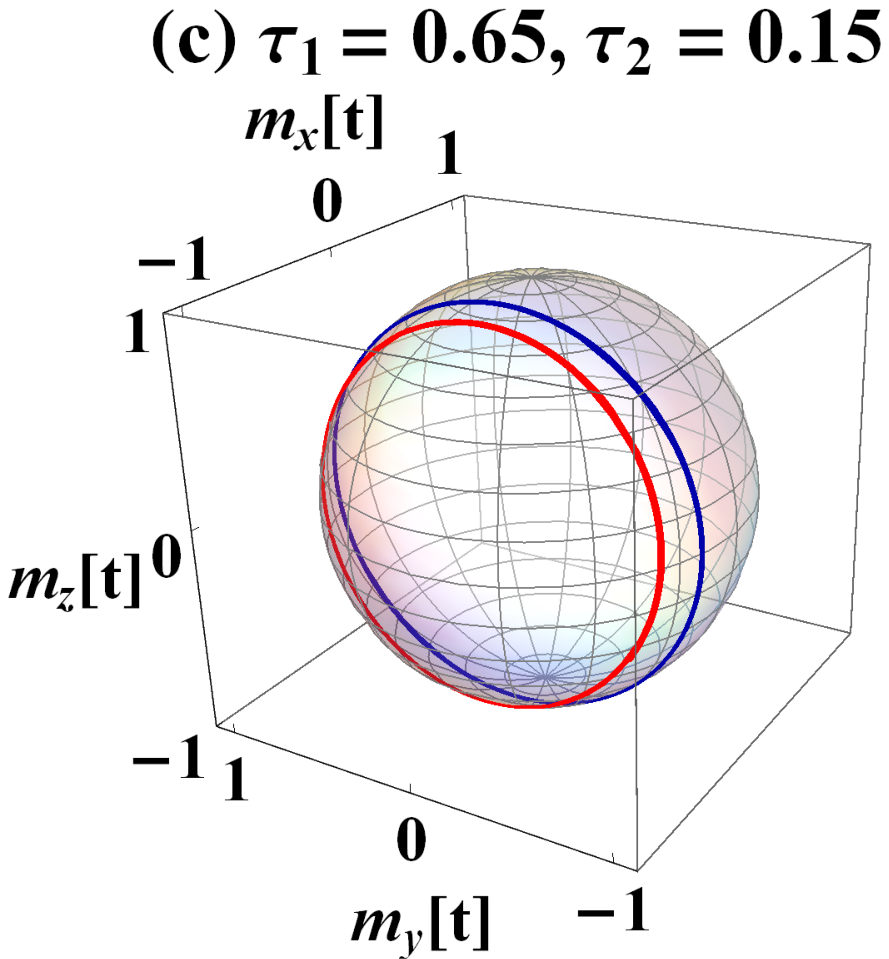}
\hspace{-0.09cm}
\vspace{0.02cm}
\includegraphics[scale = 0.3]{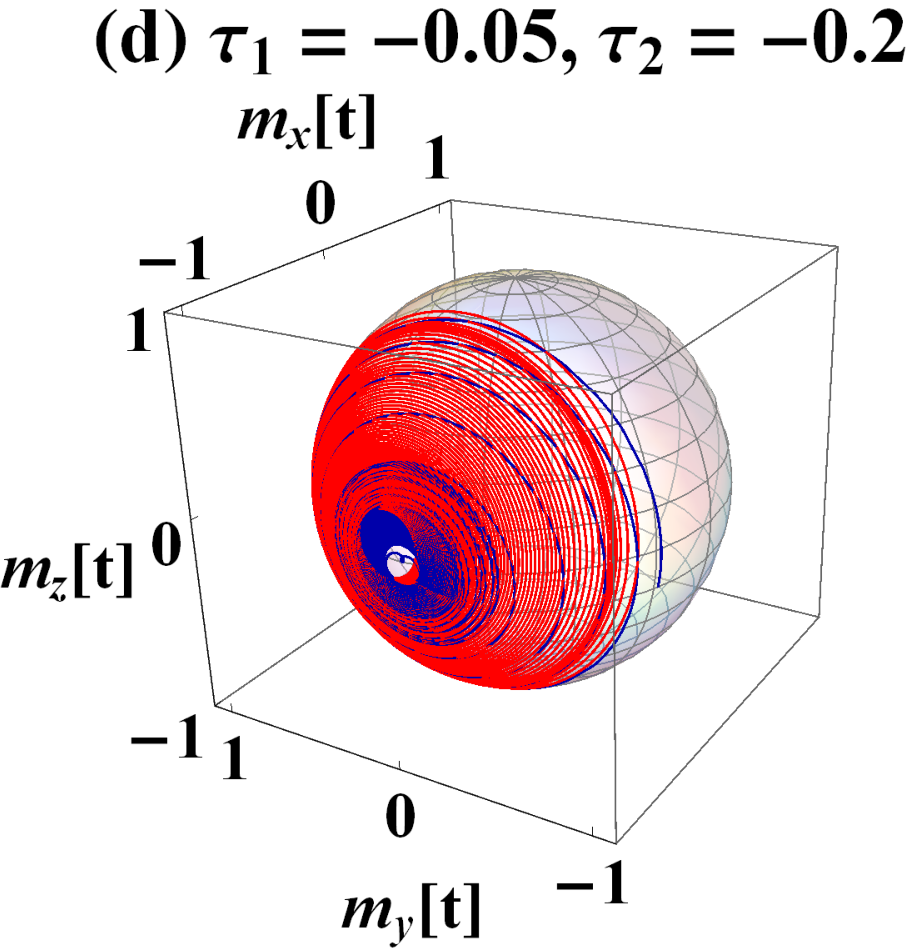}
\hspace{-0.09cm}
\vspace{0.02cm}
\includegraphics[scale = 0.3]{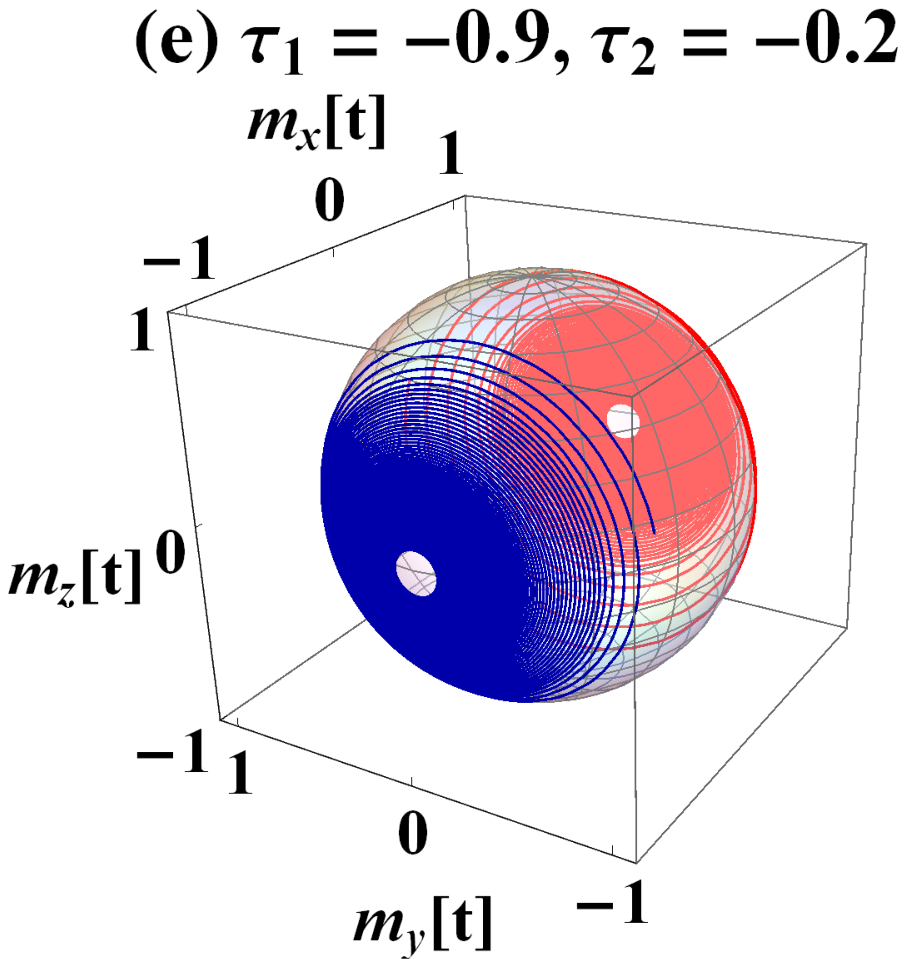}
\hspace{-0.09cm}
\vspace{0.02cm}
\includegraphics[scale = 0.3]{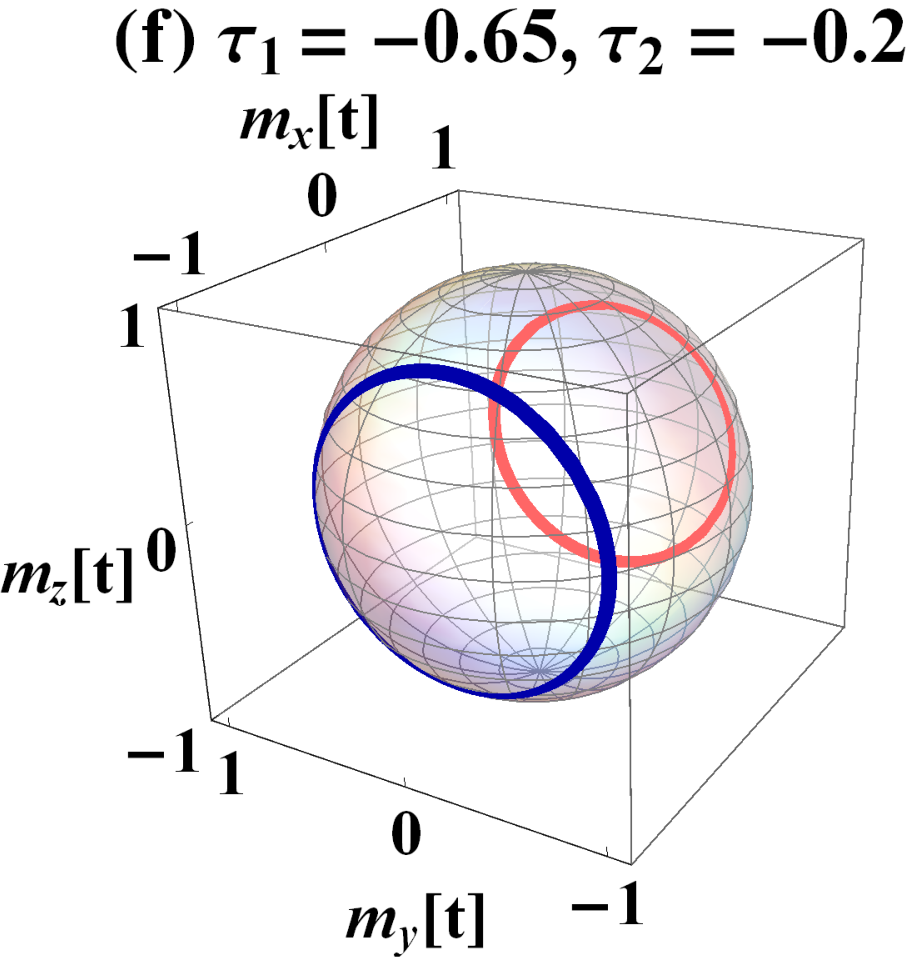}
\hspace{-0.09cm}
\vspace{0.02cm}
\includegraphics[scale = 0.275]{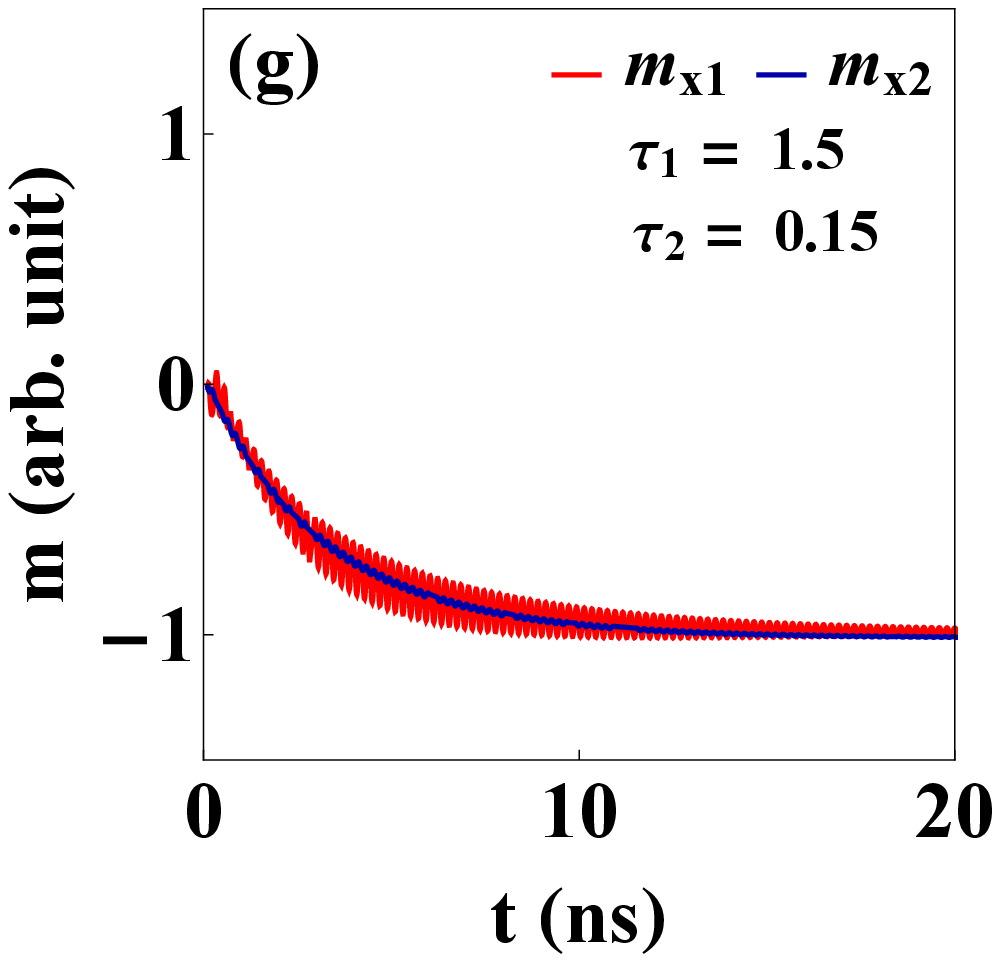}
\hspace{-0.09cm}
\includegraphics[scale = 0.275]{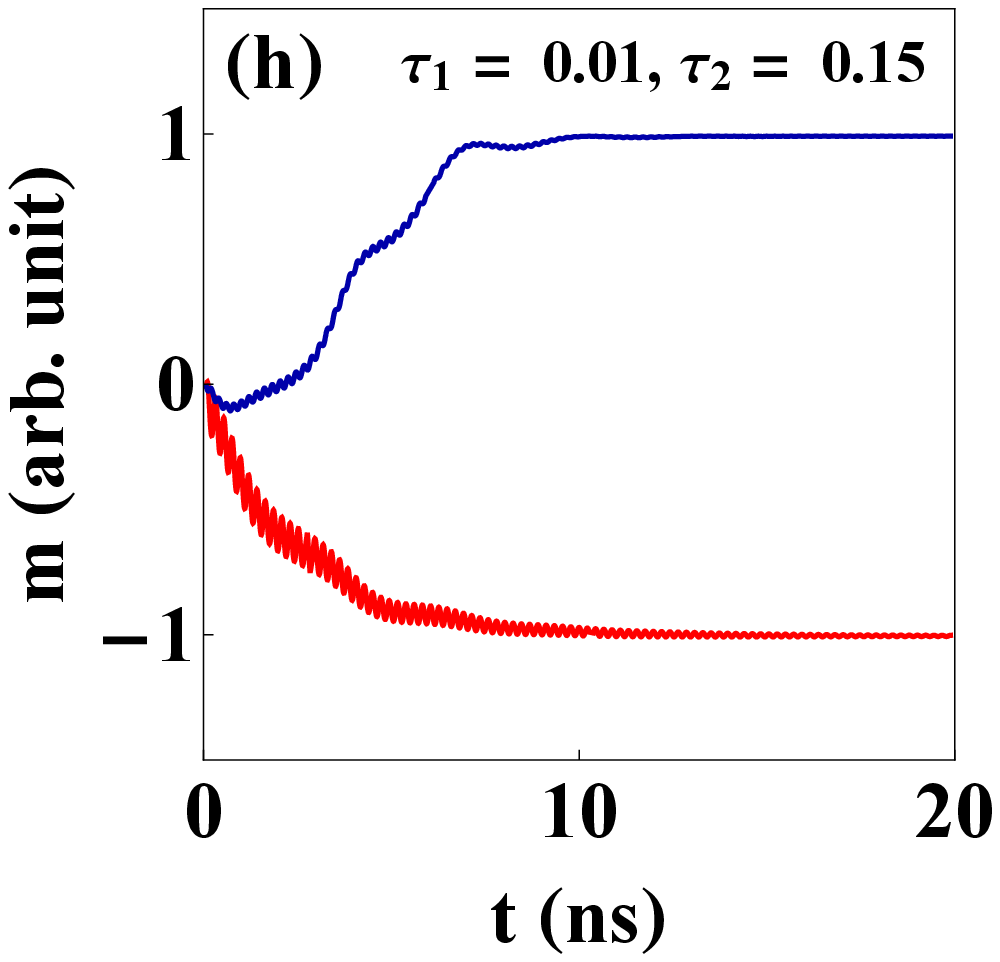}
\hspace{-0.09cm}
\includegraphics[scale = 0.275]{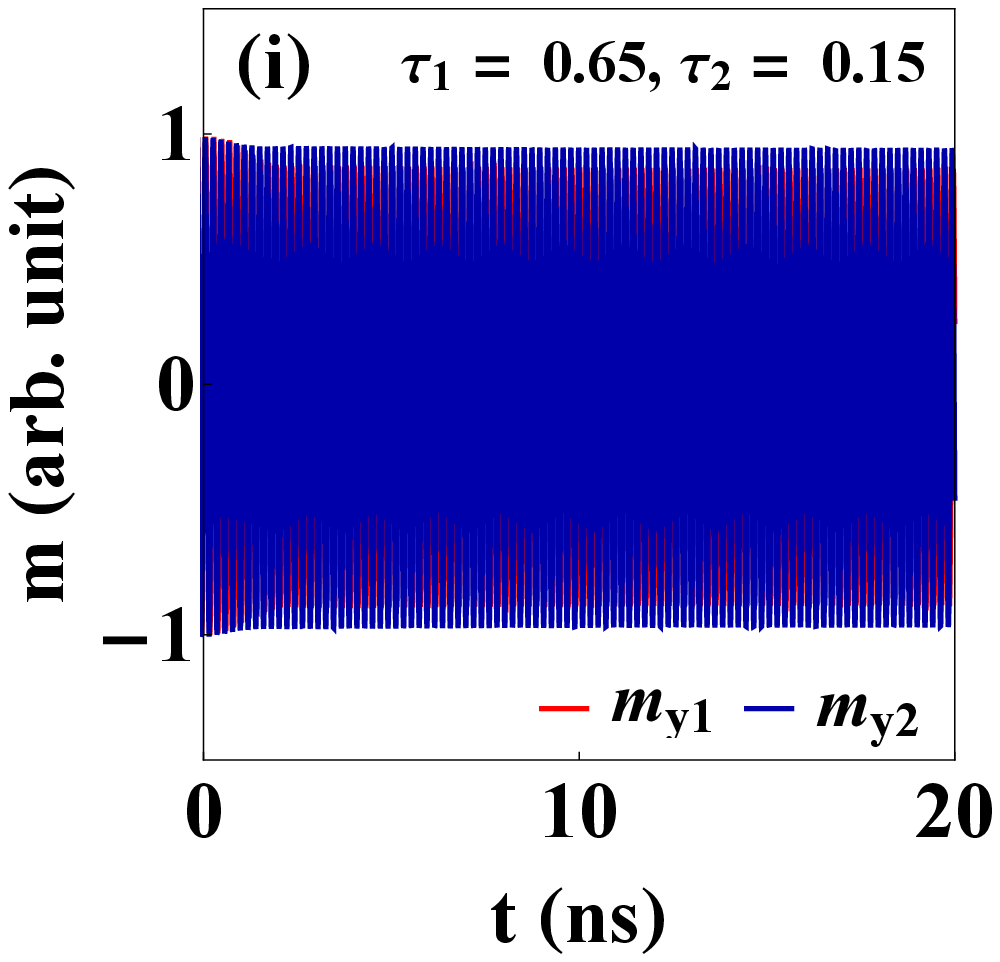}
\hspace{-0.09cm}
\includegraphics[scale = 0.275]{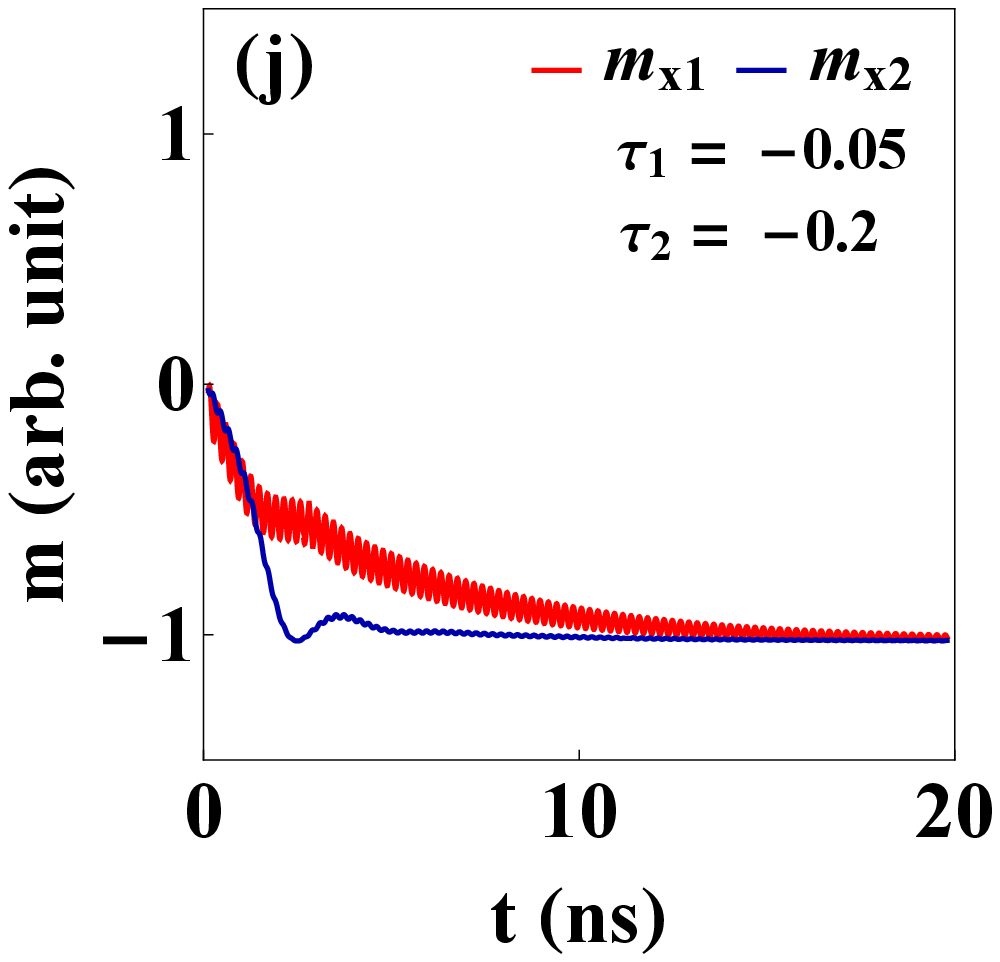}
\hspace{-0.09cm}
\includegraphics[scale = 0.275]{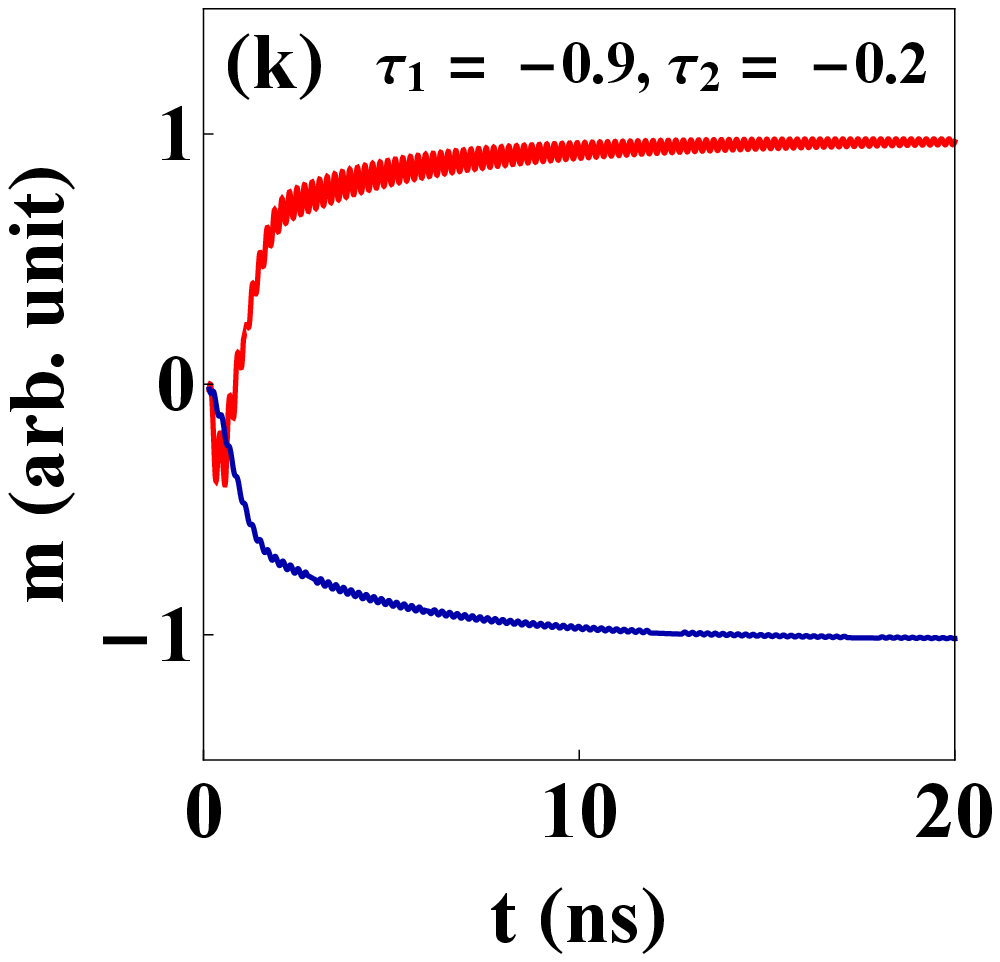}
\hspace{-0.09cm}
\includegraphics[scale = 0.275]{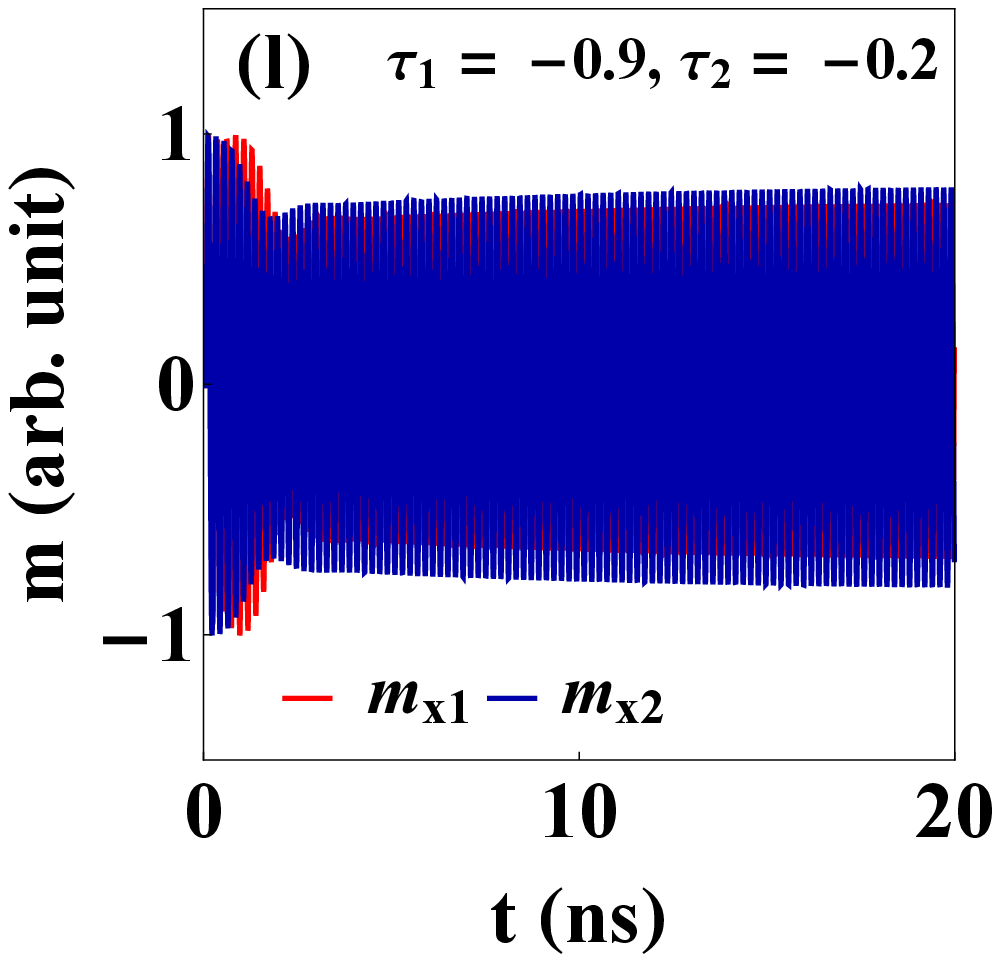}
\hspace{0.06cm}
\includegraphics[scale = 0.275]{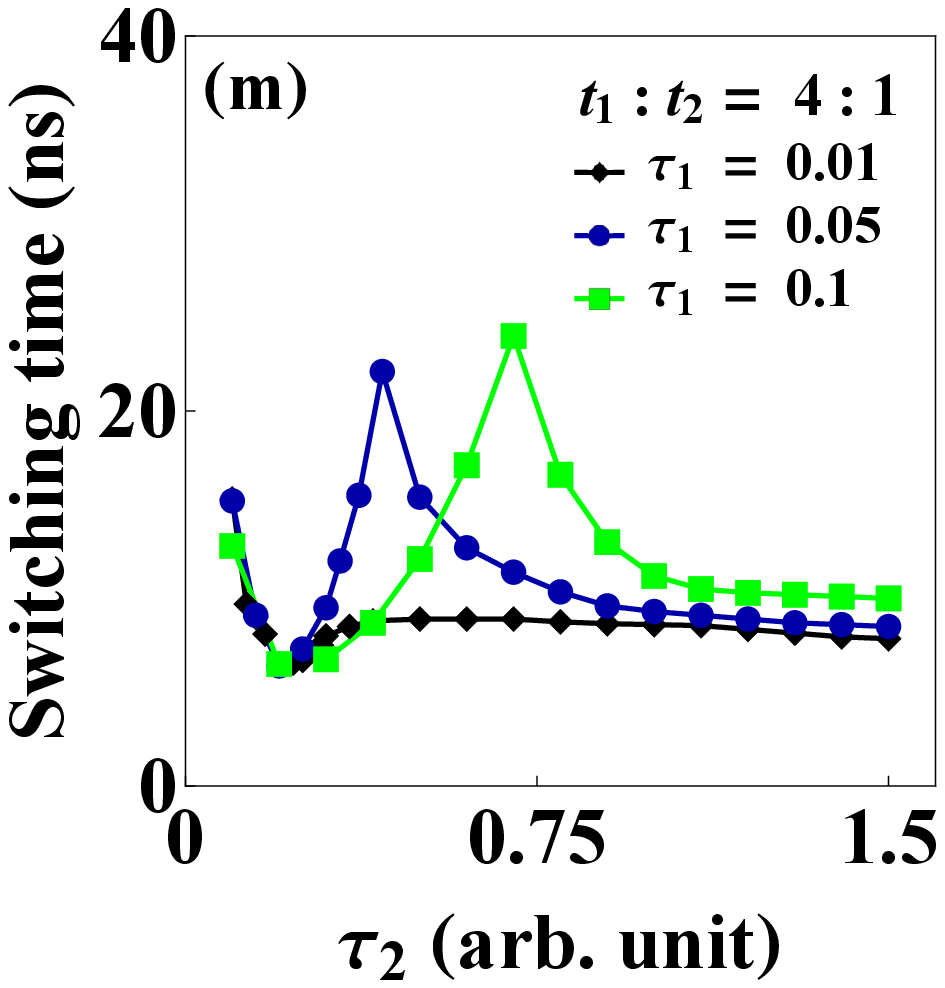}
\hspace{0.06cm}
\includegraphics[scale = 0.275]{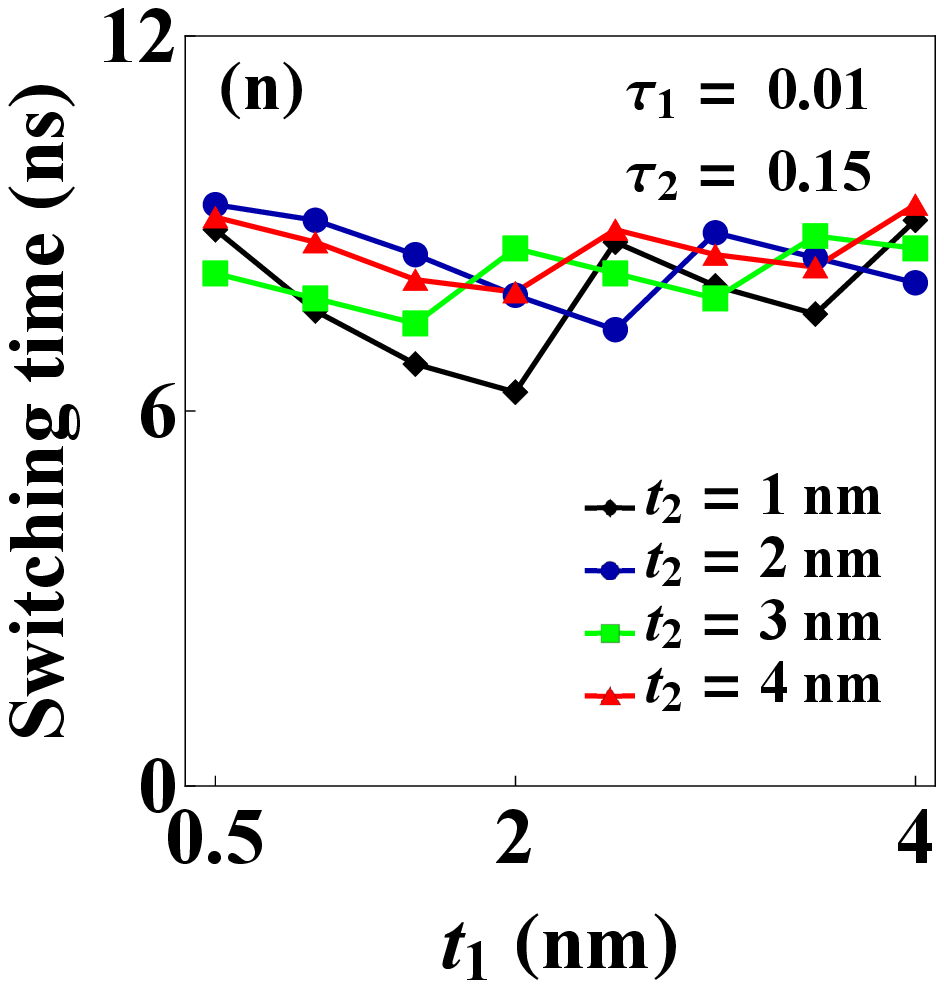}
\hspace{0.06cm}
\includegraphics[scale = 0.275]{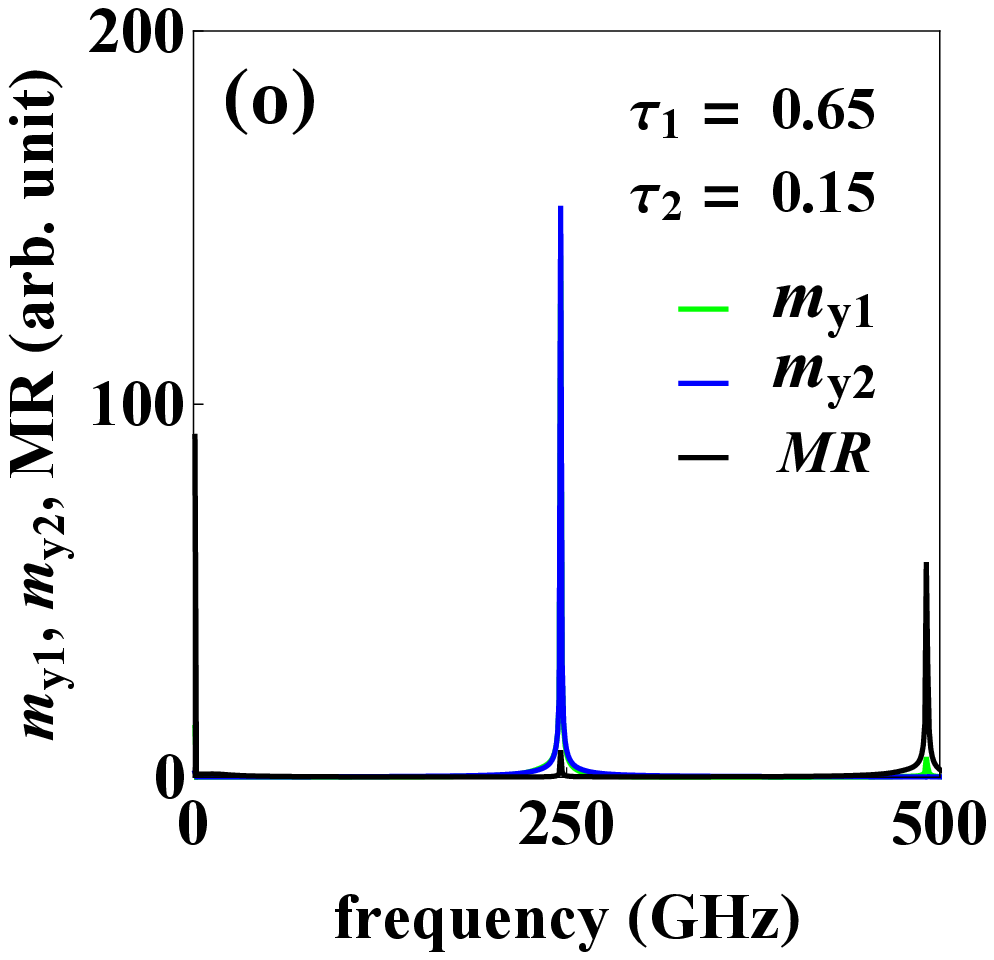}
\hspace{0.06cm}
\includegraphics[scale = 0.275]{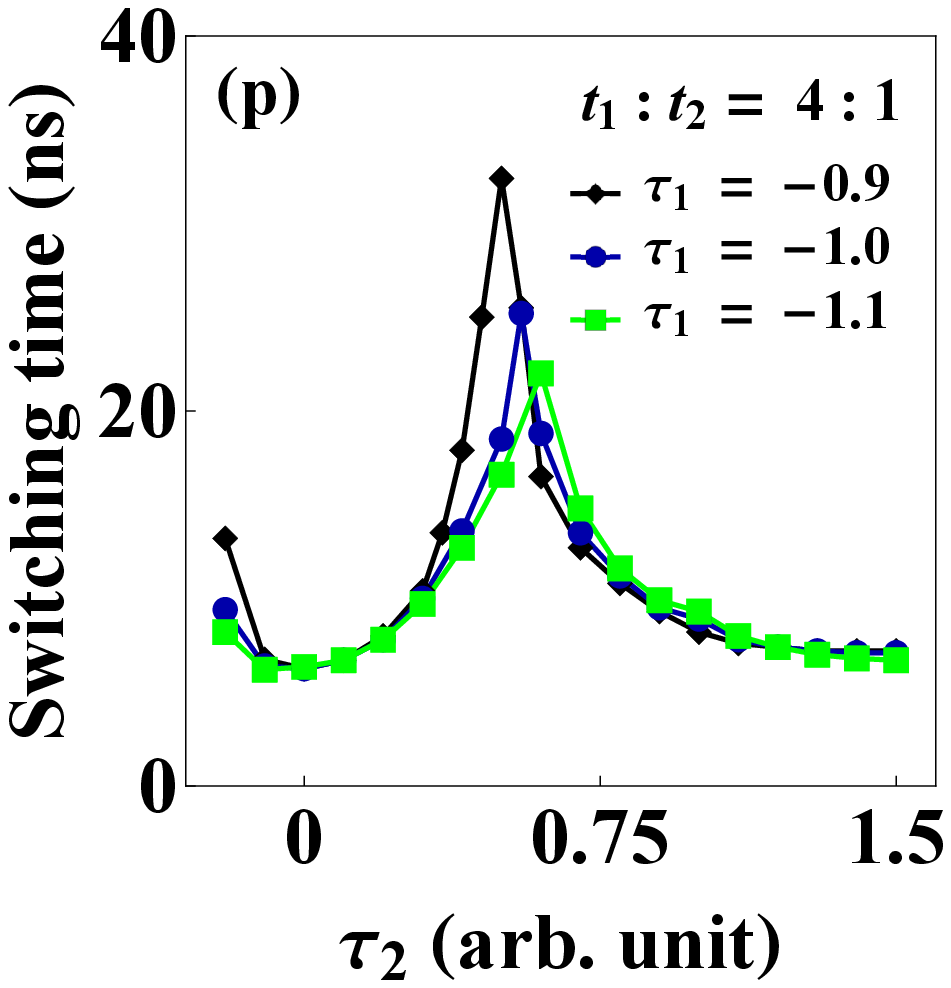}
\hspace{0.06cm}
\includegraphics[scale = 0.275]{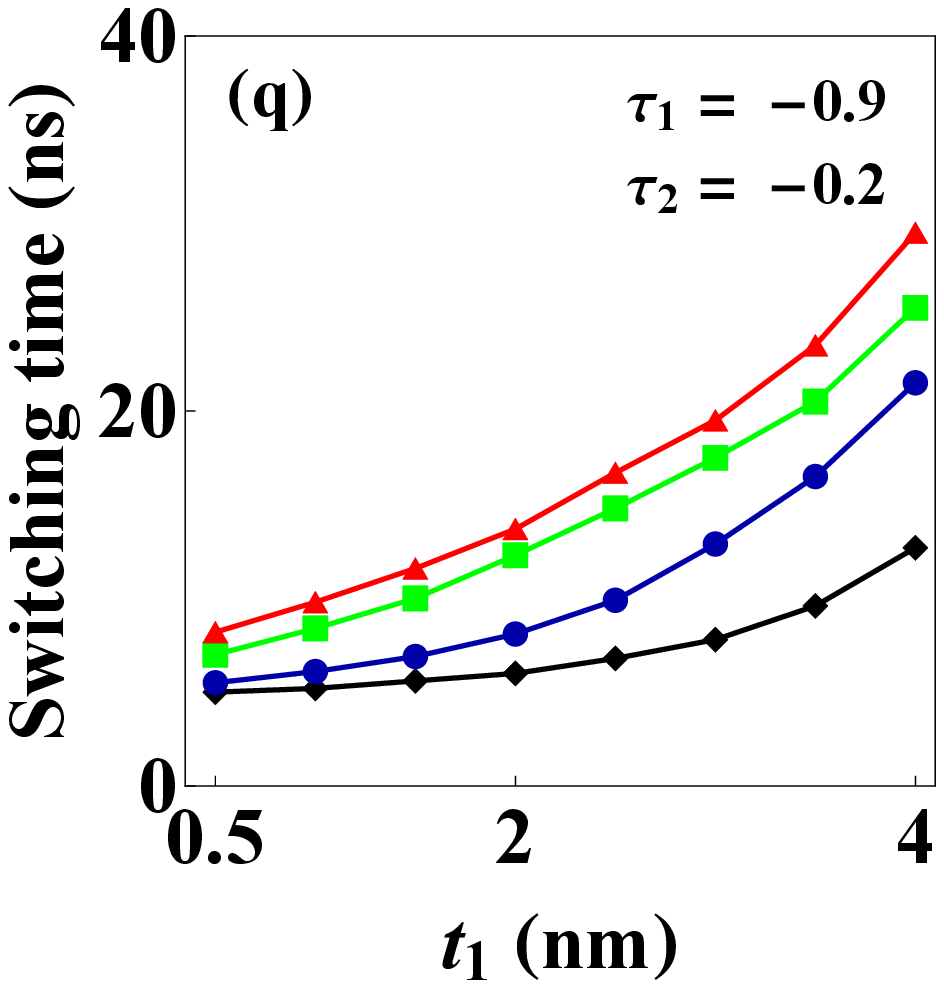}
\hspace{0.1cm}
\includegraphics[scale = 0.275]{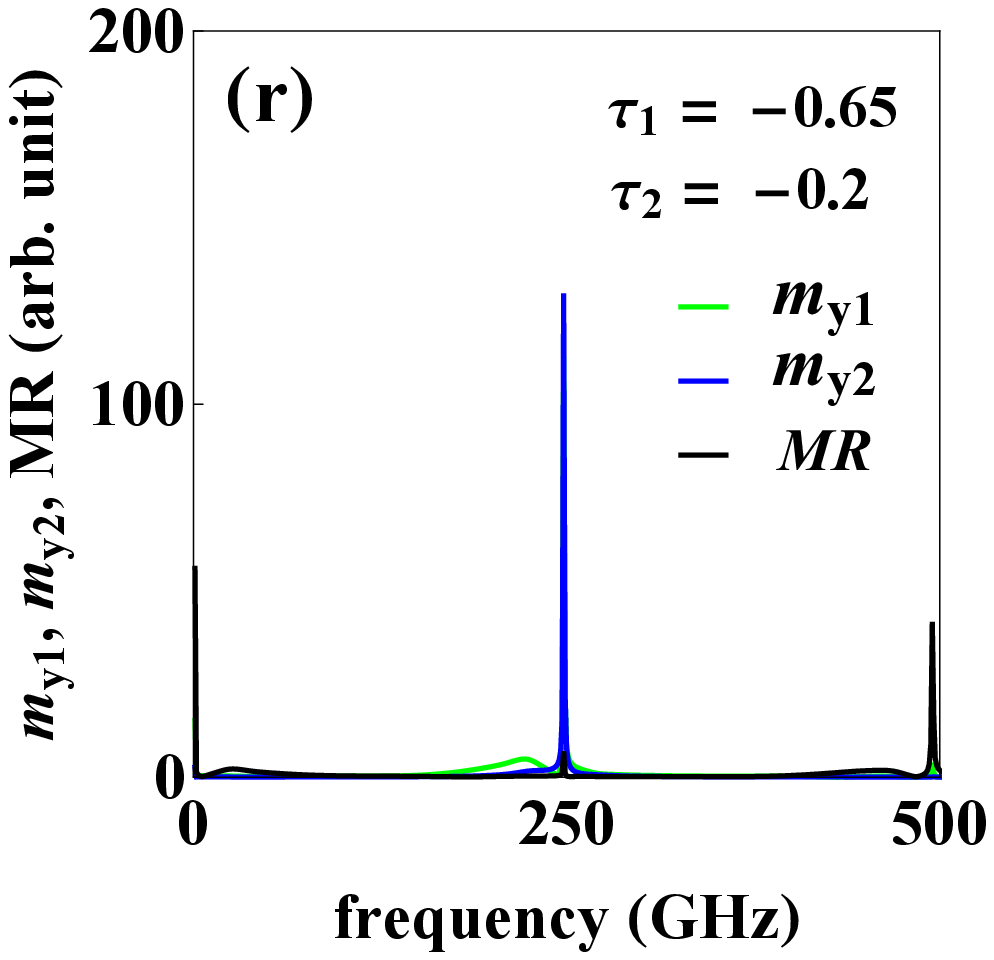}
}
\caption{(a) - (f) Oscillation trajectories of $\textbf{m}_1$ (blue) and $\textbf{m}_2$ (red) for different values of $\tau_1$ and $\tau_2$ considering K$_\text{R}$ = $0.1$K$_\text{DM}$. Plots (g) - (l) represent the time evolution of the magnetization component $\text{m}_{x1}$ (red) and $\text{m}_{x2}$ (blue). The variation of switching time with $(\tau_1, \tau_2)$ and $(t_1, t_2)$ are shown in plots (m), (p) and (n), (r) respectively. The Fourier transform of magnetization for (o) $(\tau_1, \tau_2) = (0.65, 0.15)$ and (r) $(\tau_1, \tau_2) = (-0.65, -0.2)$ respectively.
}
\label{fig2}
\end{figure*}

The last term of Eq. (\ref{eq1}) corresponds to the spin-transfer torques $\mathbf{\mathcal{T}}_j^\text{STT}$ acting on the magnetization $\mathbf{m}_j$ and can be given by \cite{johansen}
\begin{equation}
\label{eq4}
\mathbf{\mathcal{T}}_1^\text{STT} = -\tau_1\left[\mathbf{m}_1\times(\mathcal{P}_0\mathbf{m}-\mathcal{P}_1\mathbf{m}_2)\times \mathbf{m}_1\right]
\end{equation}
\begin{equation}
\label{eq5}
\mathbf{\mathcal{T}}_2^\text{STT} = -\tau_2\mathcal{P}_1(\mathbf{m}_2 \times \mathbf{m}_1
\times \mathbf{m}_2)
\end{equation}
where, $\tau_i = \frac{\gamma\hbar j_{1}}{2eM_0\mu_0 t_i}$ with the index $i = 1, 2$ corresponds to first and second RZ layers respectively and measured in $Jm^5/A^4s$. Here, $t_i$ are the thickness while $\mathcal{P}_0$, $\mathcal{P}_1$ are the polarization of the current current in
the respective RZ layers. Here, $\mathbf{m}$ is the magnetization of the fixed FM layer of our system. Using Eqs. (\ref{eq2})-(\ref{eq4}) in Eq. (\ref{eq1}), we obtained six non linear coupled first order differential equations:
\begin{multline}
\label{eq6a}
m_{\text{x1}}'[t] = \frac{\gamma}{1+\alpha ^2}  \{\alpha \mathcal{H}_{\text{eff1}} (m_{\text{y1}}^2+m_{\text{z1}}^2)+\mathcal{H}_{\text{eff2}} (m_{\text{z1}}\\-\alpha  m_{\text{x1}} m_{\text{y1}})-\mathcal{H}_{\text{eff3}} (\alpha  m_{\text{x1}} m_{\text{z1}}+m_{\text{y1}})\}+\mathcal{P} \tau_1[(\alpha  (m_{\text{x1}}^2+m_{\text{y1}}^2\\+m_{\text{z1}}^2) \{m_{\text{y1}} (m_{\text{z2}}
-1)-m_{\text{y2}} m_{\text{z1}}\}+m_{\text{x1}} \{-m_{\text{y1}} m_{\text{y2}}\\-m_{\text{z1}} (m_{\text{z2}}-1)\}+m_{\text{x2}} (m_{\text{y1}}^2+m_{\text{z1}}^2)]
\end{multline}
\begin{multline}
\label{eq7}
m_{\text{y1}}'[t] = \frac{\gamma}{1+\alpha^2} \{-\mathcal{H}_{\text{eff1}} (\alpha  m_{\text{x1}} m_{\text{y1}}+m_{\text{z1}})+\alpha  \mathcal{H}_{\text{eff2}} (m_{\text{x1}}^2\\+m_{\text{z1}}^2)+\mathcal{H}_{\text{eff3}} (m_{\text{x1}}-\alpha  m_{\text{y1}} m_{\text{z1}})\}+\mathcal{P} \tau_1[(-\alpha  (m_{\text{x1}}^2+m_{\text{y1}}^2\\+m_{\text{z1}}^2) \{m_{\text{x1}} (m_{\text{z2}}-1)-m_{\text{x2}} m_{\text{z1}}\}-m_{\text{x1}} m_{\text{x2}} m_{\text{y1}}+m_{\text{x1}}^2 m_{\text{y2}}\\
-m_{\text{y1}} m_{\text{z1}} m_{\text{z2}}+m_{\text{y1}} m_{\text{z1}}+m_{\text{y2}} m_{\text{z1}}^2)]
\end{multline}
\begin{multline}
\label{eq8}
m_{\text{z1}}'[t] = \frac{\gamma}{1+\alpha ^2}  \{\mathcal{H}_{\text{eff1}} (m_{\text{y1}}-\alpha  m_{\text{x1}} m_{\text{z1}})-\mathcal{H}_{\text{eff2}} (m_{\text{x1}}\\
+\alpha  m_{\text{y1}} m_{\text{z1}})+\alpha \mathcal{H}_{\text{eff3}} (m_{\text{x1}}^2+m_{\text{y1}}^2)\}+\mathcal{P} \tau_1[\alpha  (m_{\text{x1}}^2+m_{\text{y1}}^2\\+m_{\text{z1}}^2) (m_{\text{x1}} m_{\text{y2}}-m_{\text{x2}} m_{\text{y1}})-m_{\text{x1}} m_{\text{x2}} m_{\text{z1}}+m_{\text{x1}}^2 m_{\text{z2}}
\\-m_{\text{x1}}^2-m_{\text{y1}} m_{\text{y2}} m_{\text{z1}}+m_{\text{y1}}^2 m_{\text{z2}}-m_{\text{y1}}^2)]
\end{multline}
\begin{multline}
\label{eq9}
m_{\text{x2}}'[t] = \frac{\gamma }{1+\alpha^2} \{\alpha \mathcal{H}_{\text{eff4}} (m_{\text{y2}}^2+m_{\text{z2}}^2)+\mathcal{H}_{\text{eff5}} (m_{\text{z2}}\\-\alpha  m_{\text{x2}} m_{\text{y2}})-\mathcal{H}_{\text{eff6}} (\alpha  m_{\text{x2}} m_{\text{z2}}+m_{\text{y2}})\}
 +\mathcal{P}\tau _2[-(m_{\text{y2}}^2
 \\+m_{\text{z2}}^2) (m_{\text{x1}}-\alpha  m_{\text{y1}} m_{\text{z2}}+\alpha  m_{\text{y2}} m_{\text{z1}})+\alpha  m_{\text{x2}}^2 (m_{\text{y1}} m_{\text{z2}}\\-m_{\text{y2}} m_{\text{z1}})+m_{\text{x2}} (m_{\text{y1}} m_{\text{y2}}+m_{\text{z1}} m_{\text{z2}})]
\end{multline}
\begin{multline}
\label{eq10}
m_{\text{y2}}'[t] =  \frac{\gamma}{1+\alpha ^2} \{-\mathcal{H}_{\text{eff4}} (\alpha  m_{\text{x2}} m_{\text{y2}}+m_{\text{z2}})+\alpha  \mathcal{H}_{\text{eff5}} (m_{\text{x2}}^2\\+m_{\text{z2}}^2)+\mathcal{H}_{\text{eff6}} (m_{\text{x2}}-\alpha  m_{\text{y2}} m_{\text{z2}})\}
 +\mathcal{P} \tau _2[\alpha  (m_{\text{x2}}^2+m_{\text{y2}}^2
 \\+m_{\text{z2}}^2) (m_{\text{x2}} m_{\text{z1}}-m_{\text{x1}} m_{\text{z2}})+m_{\text{x1}} m_{\text{x2}} m_{\text{y2}}-m_{\text{x2}}^2 m_{\text{y1}}\\-m_{\text{y1}} m_{\text{z2}}^2+m_{\text{y2}} m_{\text{z1}} m_{\text{z2}})]
\end{multline}
\begin{multline}
\label{eq11}
m_{\text{z2}}'[t] =  \frac{\gamma }{1+\alpha^2} \{\mathcal{H}_{\text{eff4}} (m_{\text{y2}}-\alpha  m_{\text{x2}} m_{\text{z2}})-\mathcal{H}_{\text{eff5}} (m_{\text{x2}}
\\+\alpha  m_{\text{y2}} m_{\text{z2}})+\alpha  \mathcal{H}_{\text{eff6}} (m_{\text{x2}}^2+m_{\text{y2}}^2)\})+\mathcal{P}\tau _2[-\alpha  (m_{\text{x2}}^2+m_{\text{y2}}^2\\+m_{\text{z2}}^2) (m_{\text{x2}} m_{\text{y1}}-m_{\text{x1}} m_{\text{y2}})+m_{\text{x1}} m_{\text{x2}} m_{\text{z2}}-m_{\text{x2}}^2 m_{\text{z1}}\\
+m_{\text{y1}} m_{\text{y2}} m_{\text{z2}}-m_{\text{y2}}^2 m_{\text{z1}})]
\end{multline}
where, 
\begin{align}
\mathcal{H}_{\text{eff1}} &= -D_1 t_1 K_{\text{DM}}+h_0-K_{\text{R}} m_{\text{x2}}-\alpha_{\text{R}},\nonumber\\
\mathcal{H}_{\text{eff2}} &= -D_1 t_1 K_{\text{DM}}+h_0-K_{\text{R}} m_{\text{y2}}\nonumber\\
\mathcal{H}_{\text{eff3}} &= K_{\text{an}} m_{\text{z1}}-\text{H}_{\text{extz}} m_{\text{z1}}-D_1 t_1 K_{\text{DM}}+h_0-K_{\text{R}} m_{\text{z2}}\nonumber\\
\mathcal{H}_{\text{eff4}} &= -D_2 t_2 K_{\text{DM}}+h_0-K_{\text{R}} m_{\text{x1}}-\alpha_{\text{R}}\nonumber\\
\mathcal{H}_{\text{eff5}} &= -D_2 t_2 K_{\text{DM}}+h_0-K_{\text{R}} m_{\text{y1}}\nonumber\\
\mathcal{H}_{\text{eff6}} &= K_{\text{an}} m_{\text{z2}}-\text{H}_{\text{extz}} m_{\text{z2}}-D_2 t_2 K_{\text{DM}}+h_0-K_{\text{R}} m_{\text{z1}}\nonumber
\end{align}

\begin{figure*}[hbt]
\centerline
\centerline{
\includegraphics[scale = 0.3]{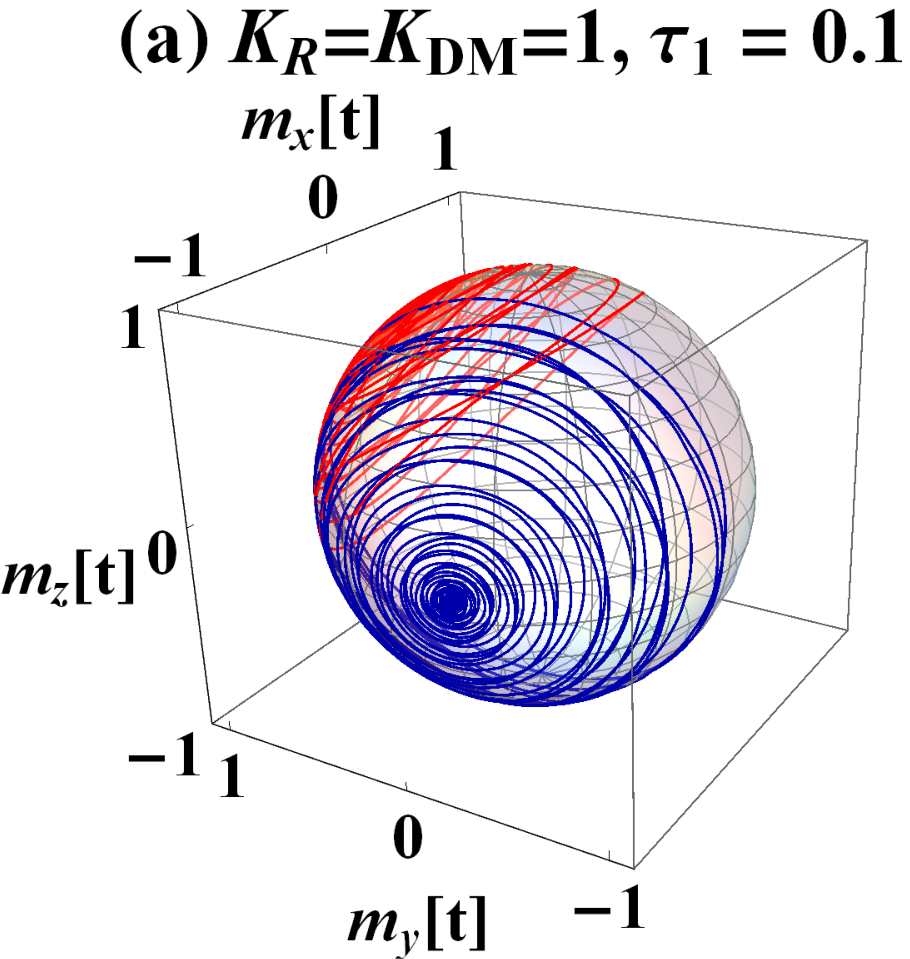}
\hspace{-0.09cm}
\vspace{0.02cm}
\includegraphics[scale = 0.3]{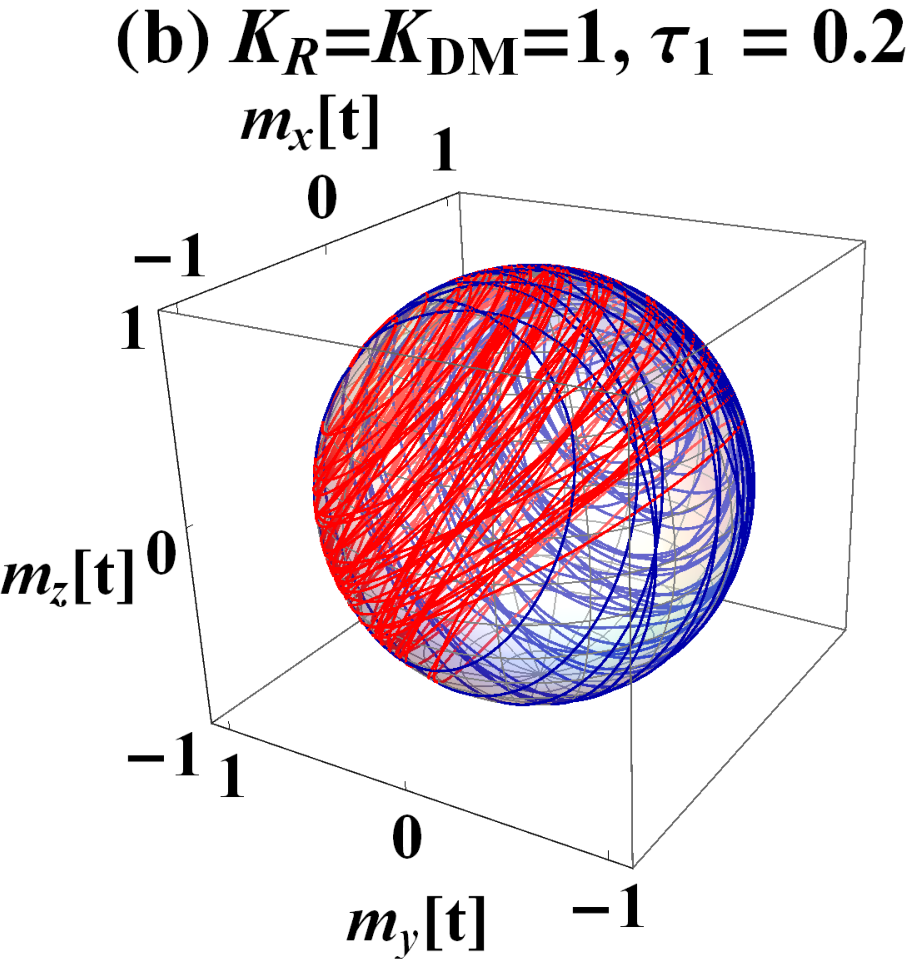}
\hspace{-0.09cm}
\vspace{0.02cm}
\includegraphics[scale = 0.3]{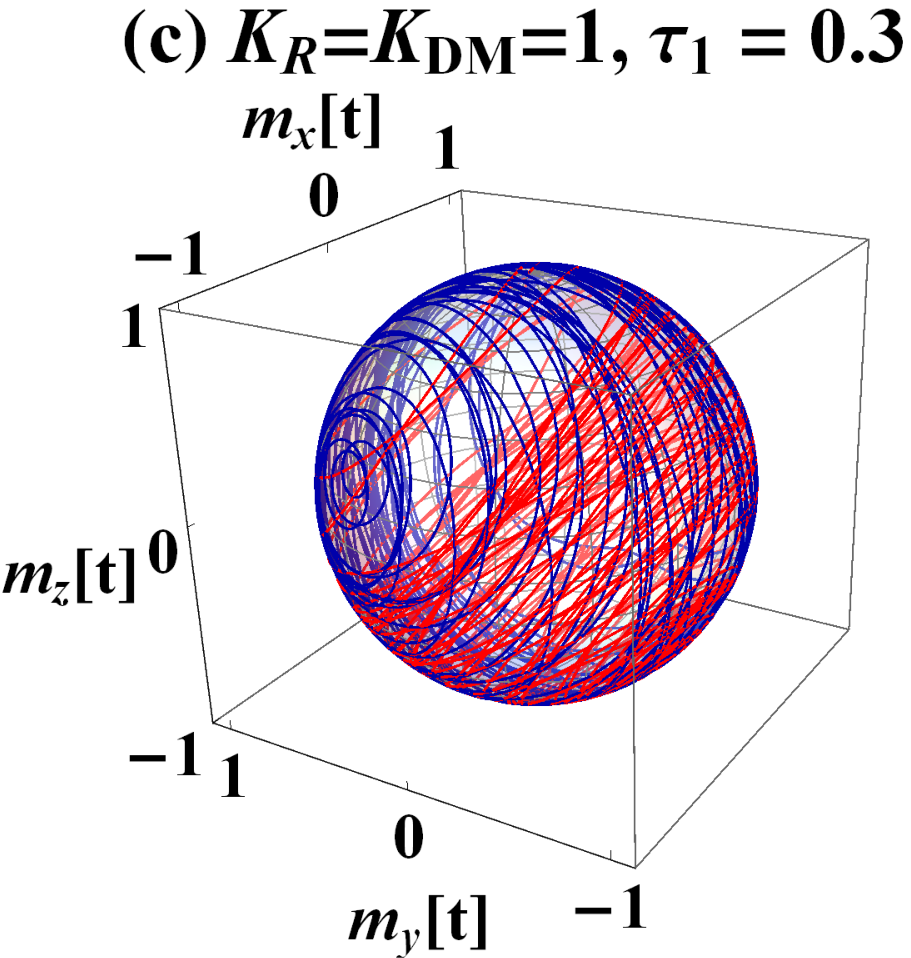}
\hspace{-0.09cm}
\vspace{0.02cm}
\includegraphics[scale = 0.3]{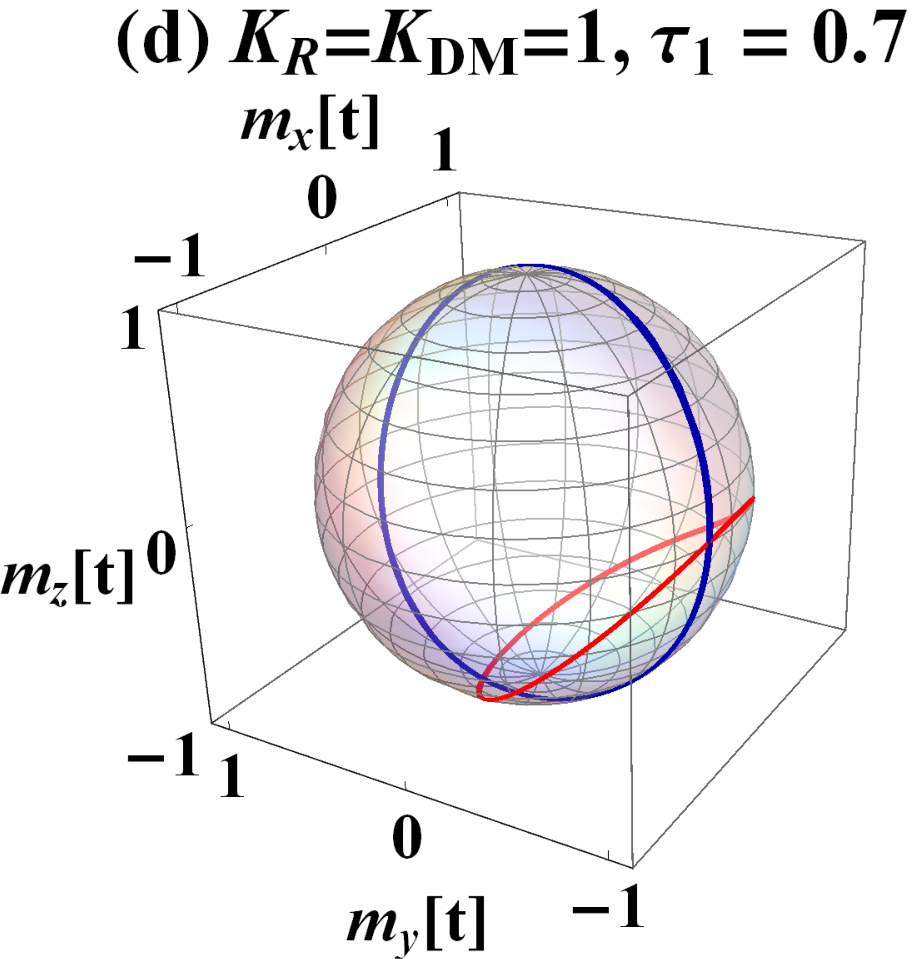}
\hspace{-0.09cm}
\vspace{0.02cm}
\includegraphics[scale = 0.3]{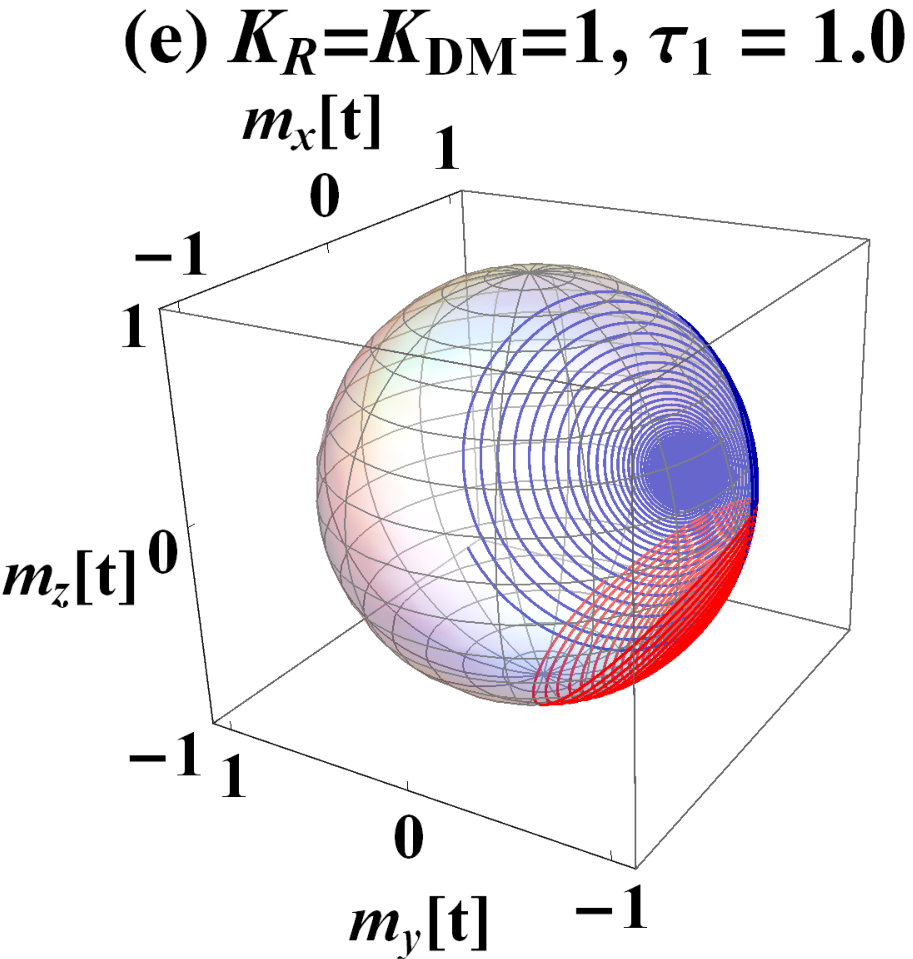}
\hspace{-0.09cm}
\vspace{0.02cm}
\includegraphics[scale = 0.3]{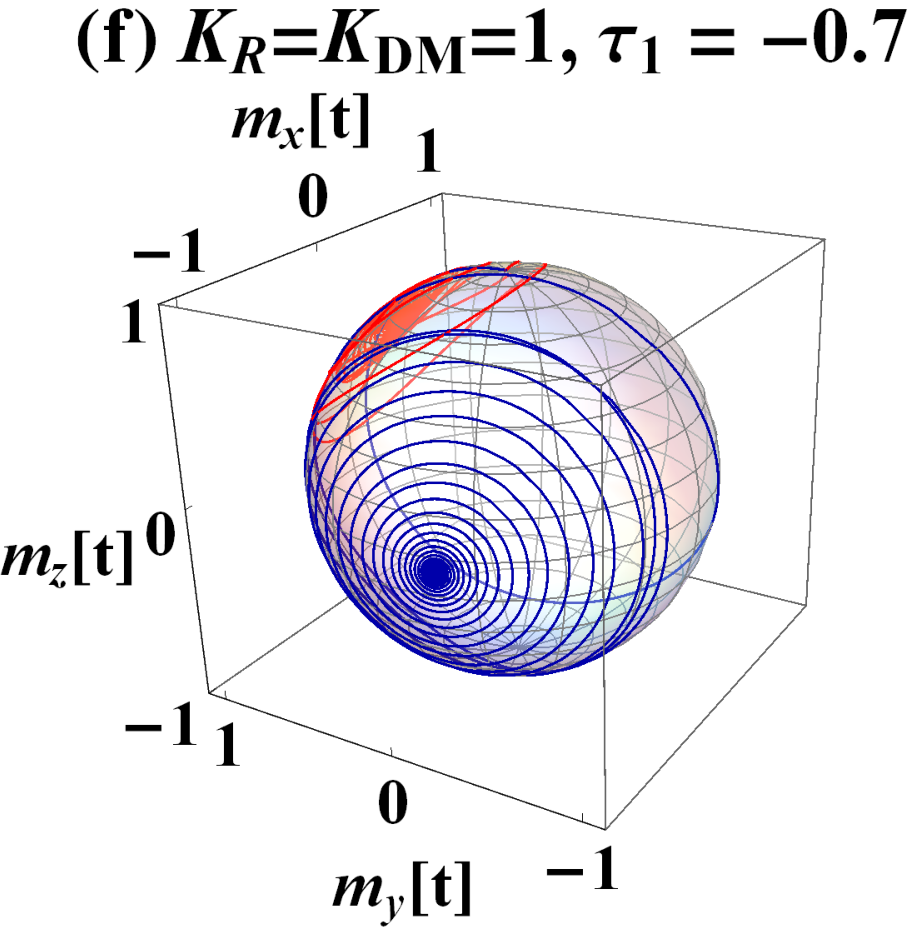}
\hspace{-0.09cm}
\vspace{0.02cm}
\includegraphics[scale = 0.275]{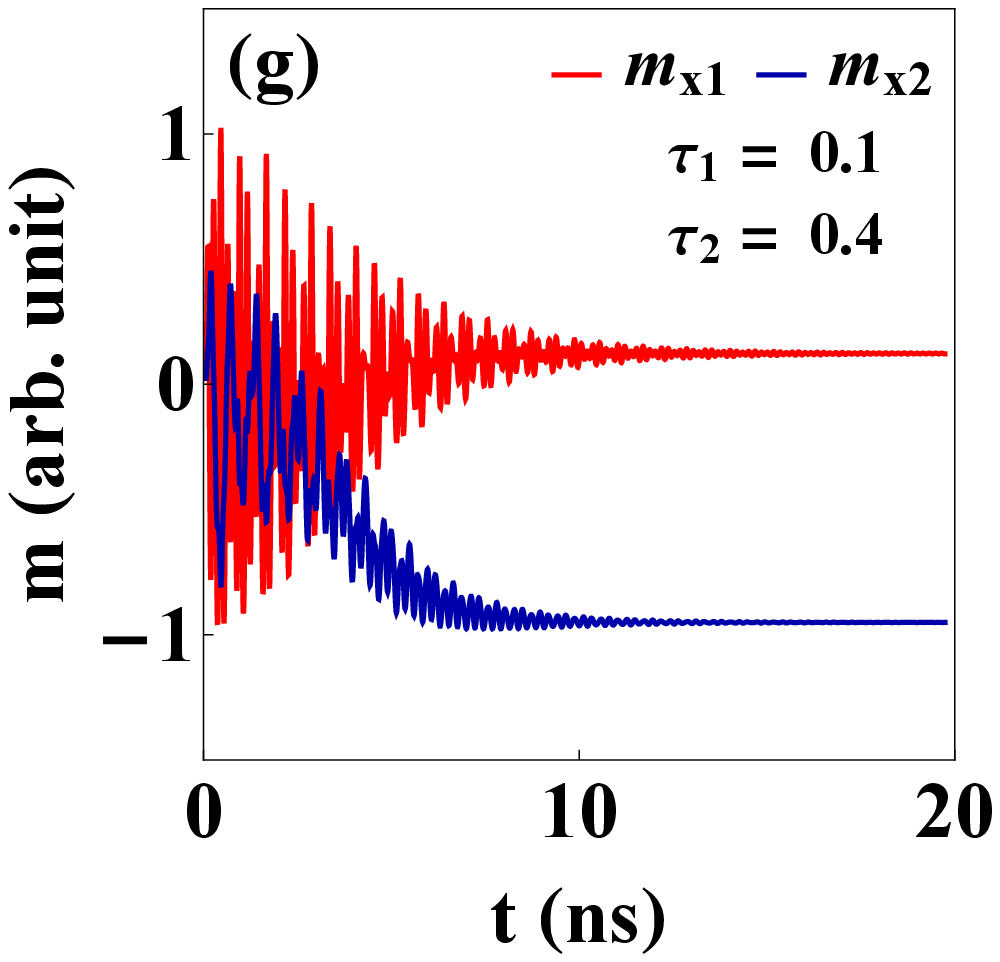}
\hspace{-0.09cm}
\includegraphics[scale = 0.275]{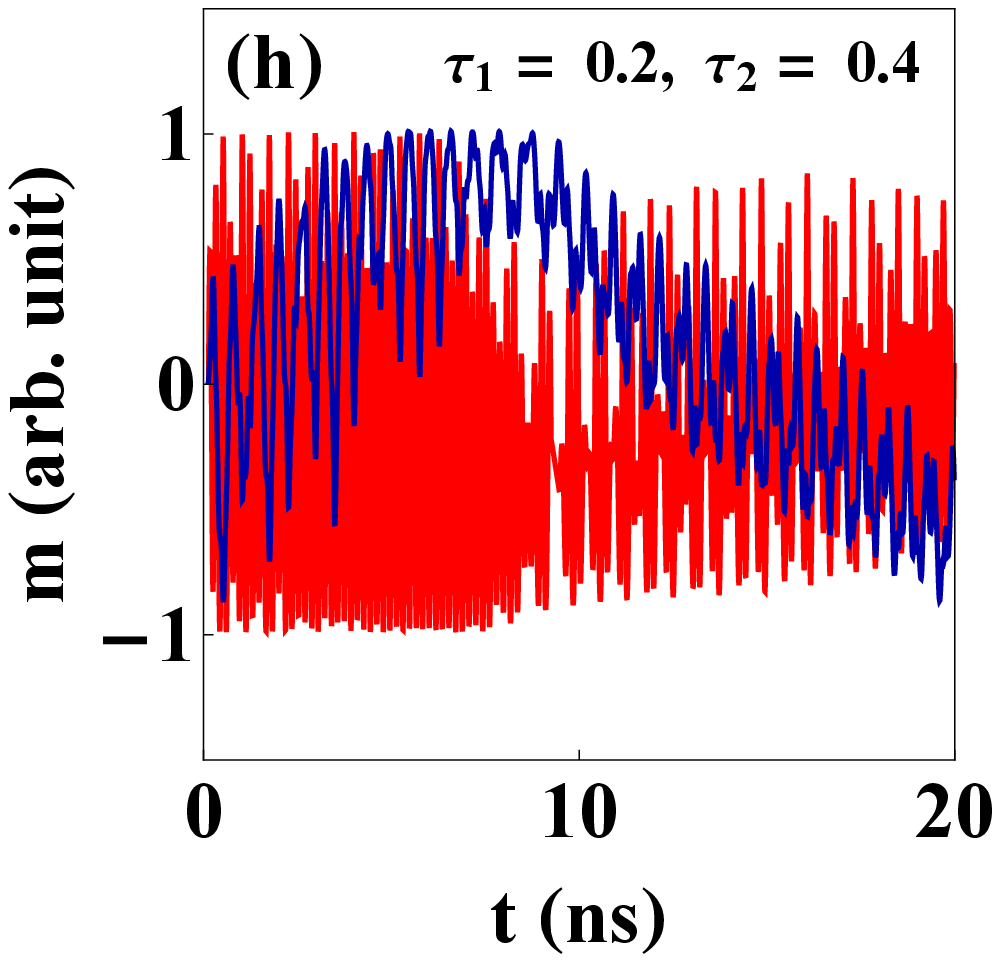}
\hspace{-0.09cm}
\includegraphics[scale = 0.275]{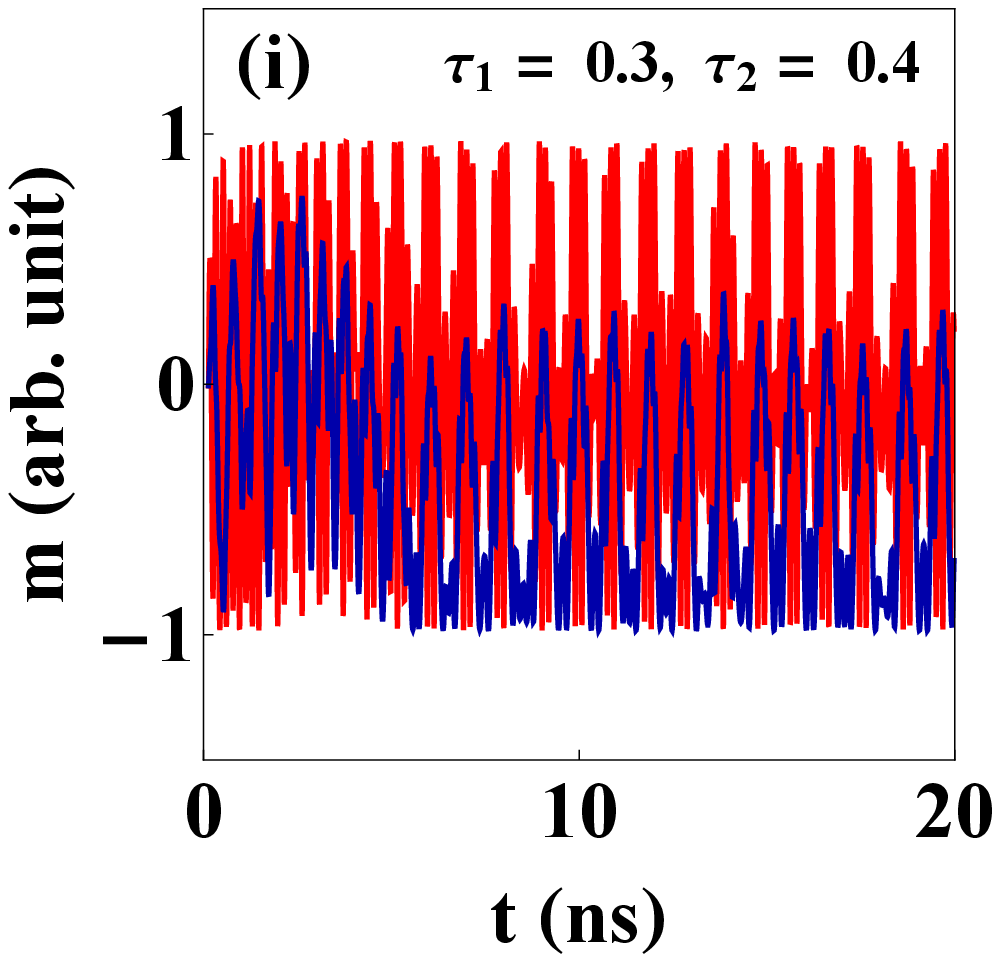}
\hspace{-0.09cm}
\includegraphics[scale = 0.275]{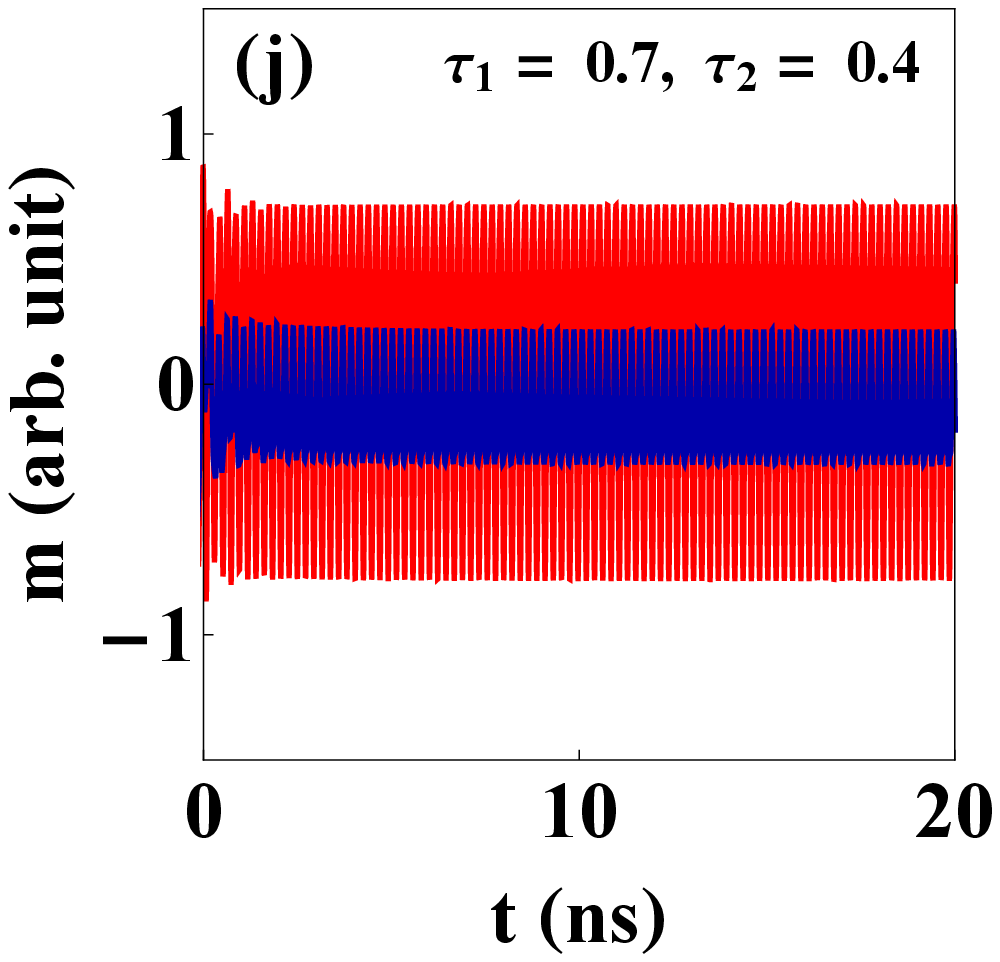}
\hspace{-0.09cm}
\includegraphics[scale = 0.275]{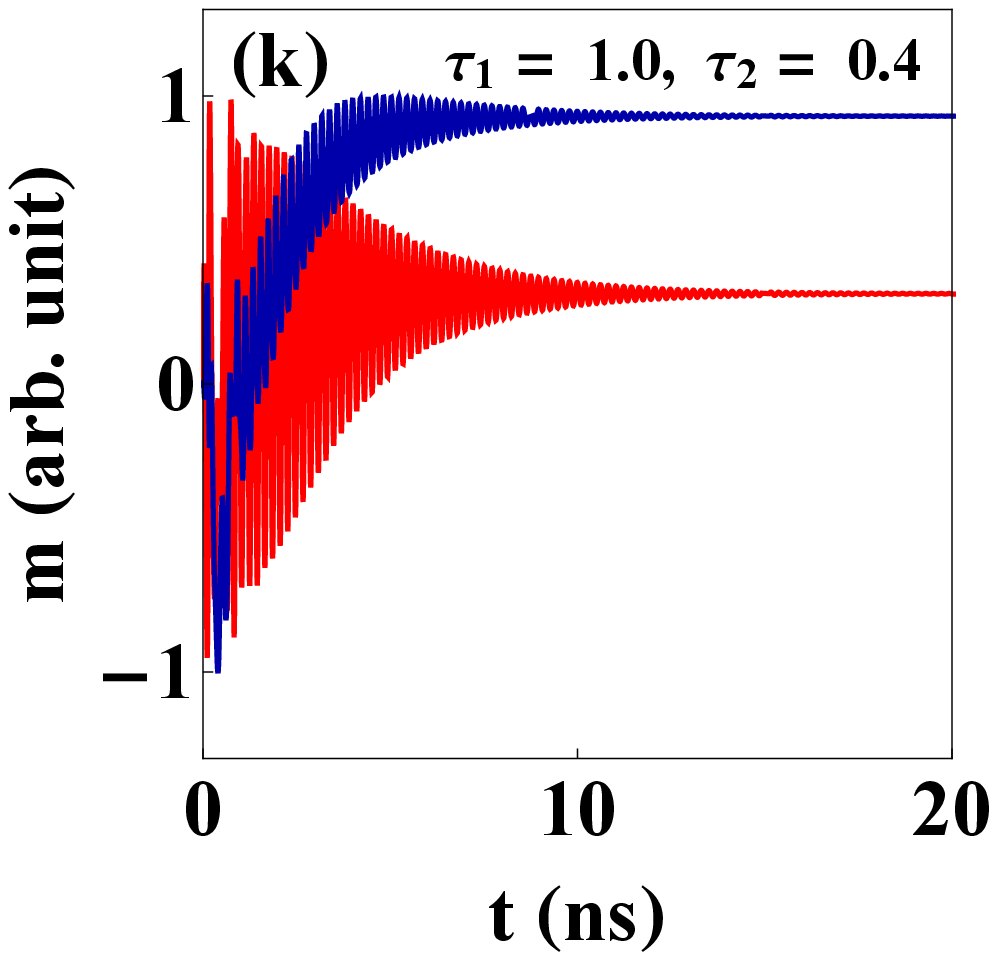}
\hspace{-0.09cm}
\includegraphics[scale = 0.275]{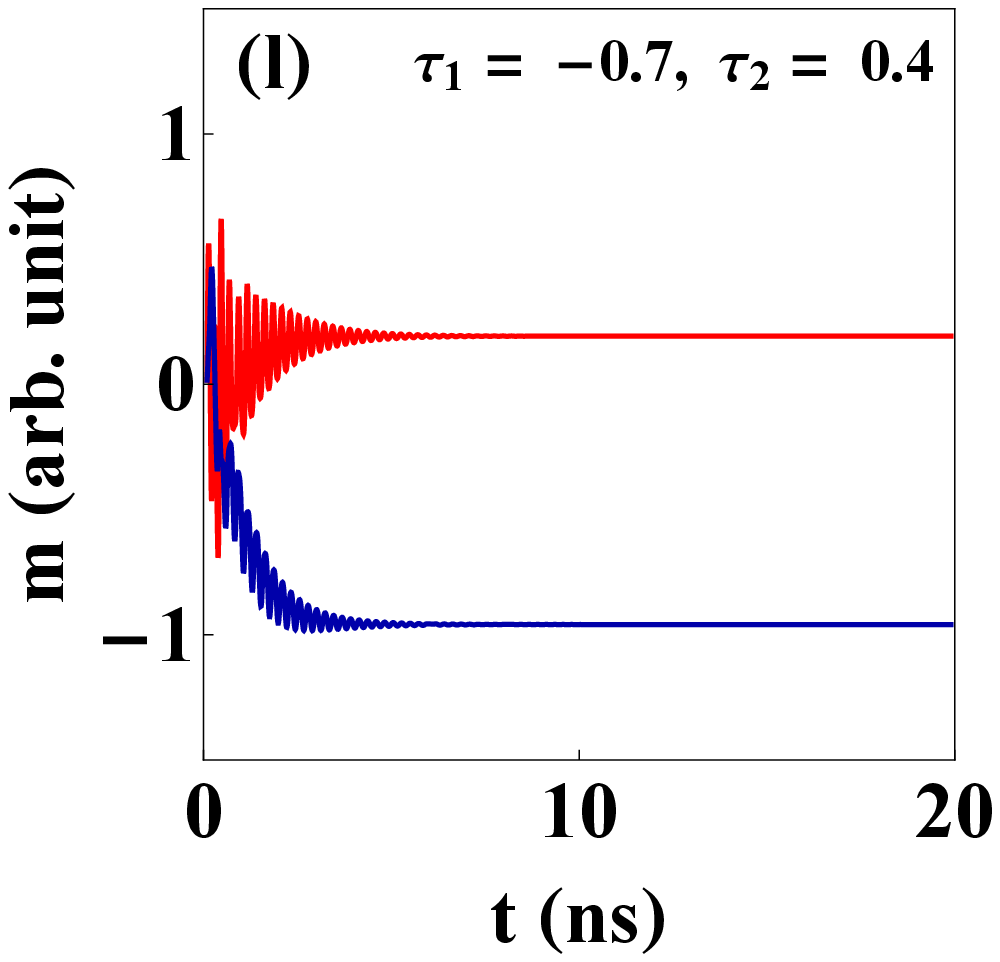}
\hspace{-0.09cm}
\includegraphics[scale = 0.273]{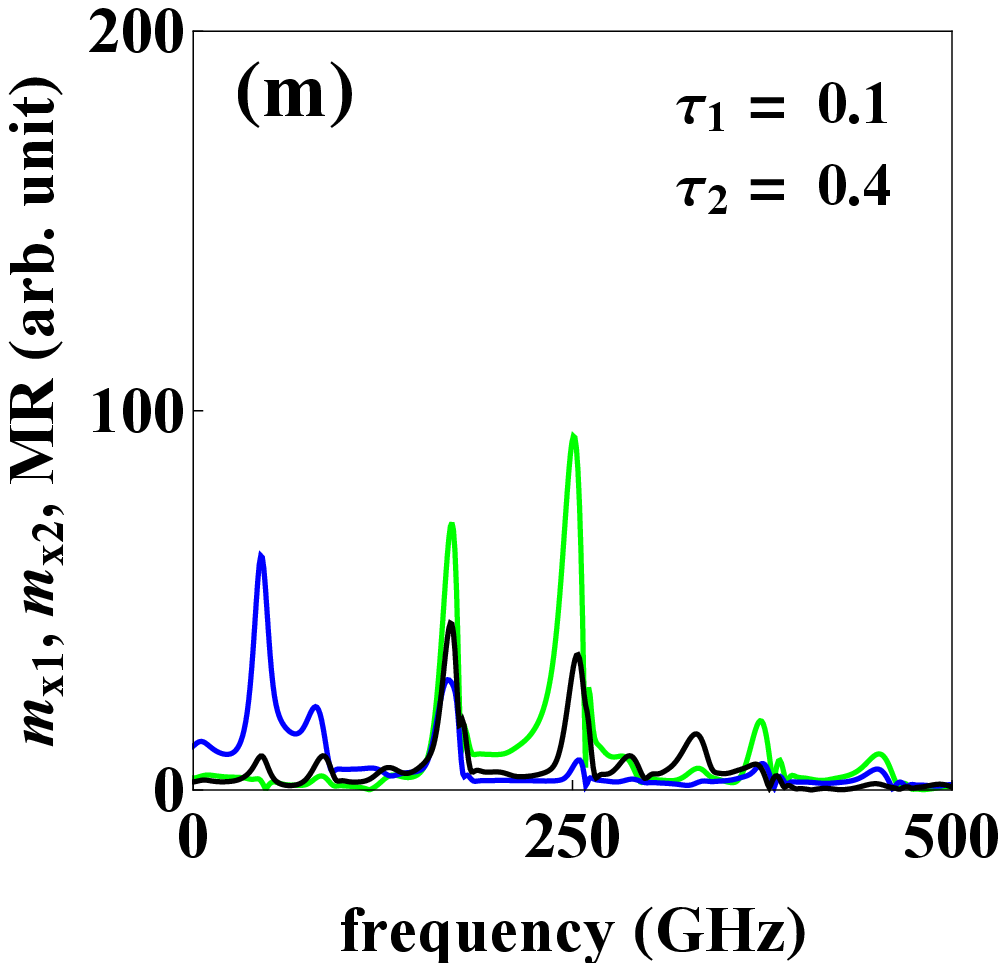}
\includegraphics[scale = 0.273]{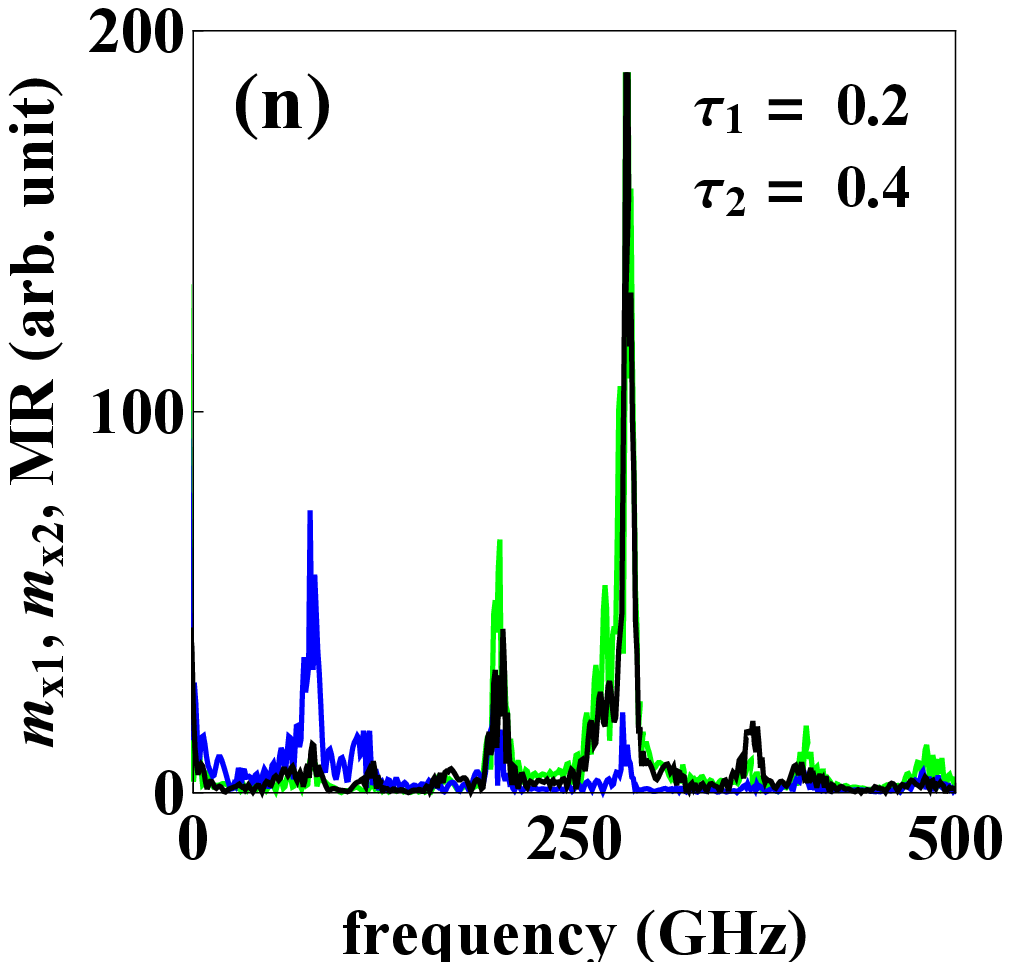}
\includegraphics[scale = 0.273]{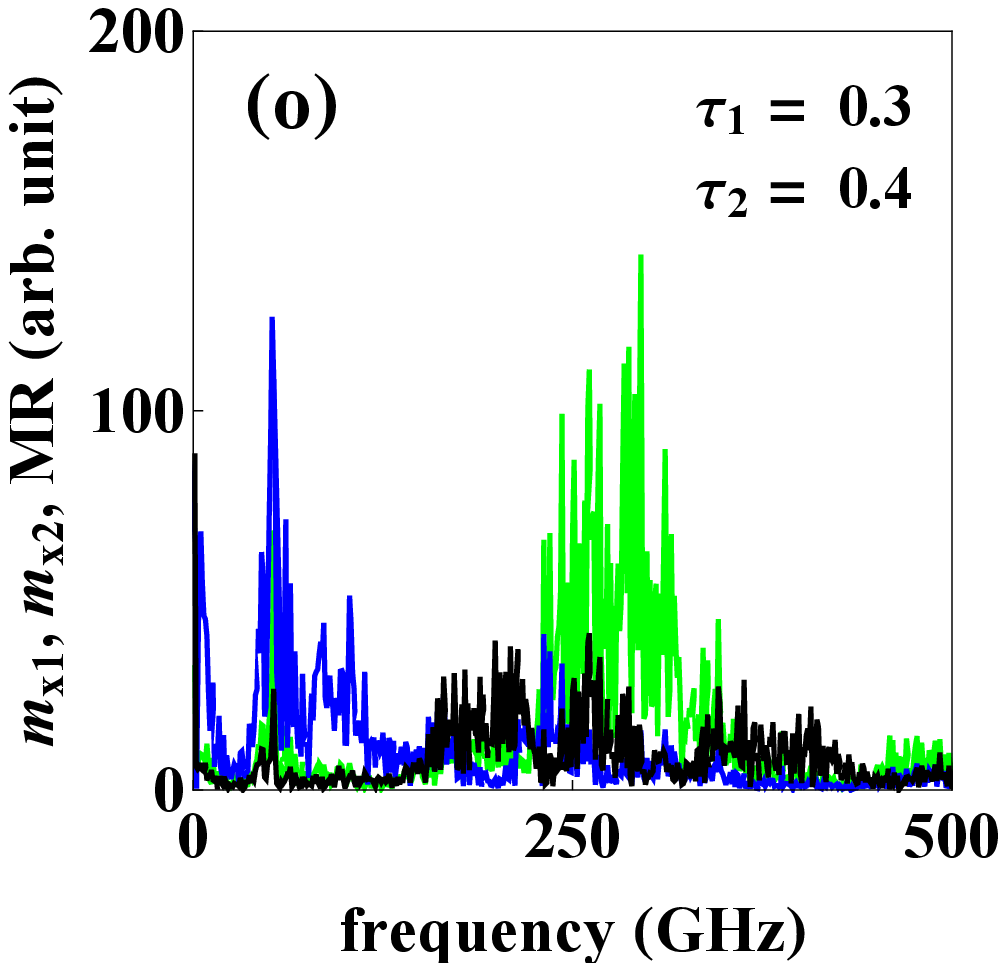}
\includegraphics[scale = 0.273]{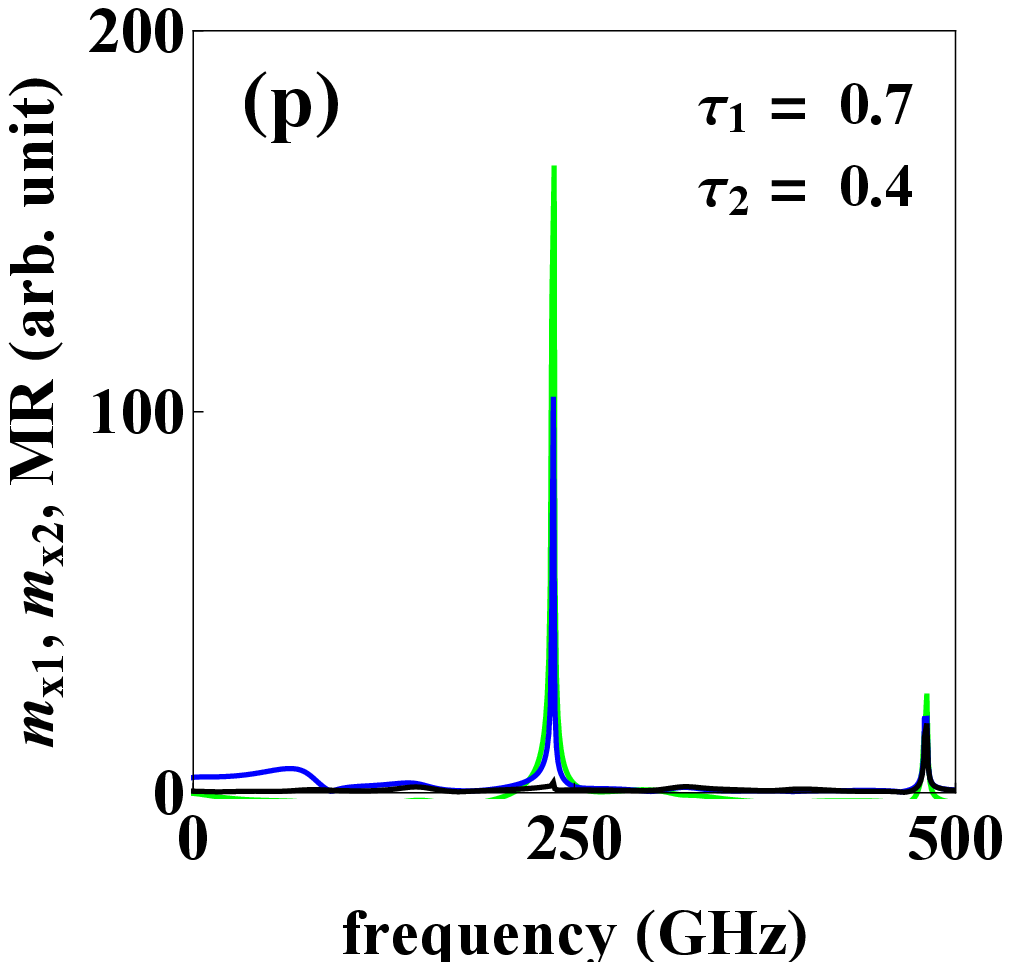}
\includegraphics[scale = 0.273]{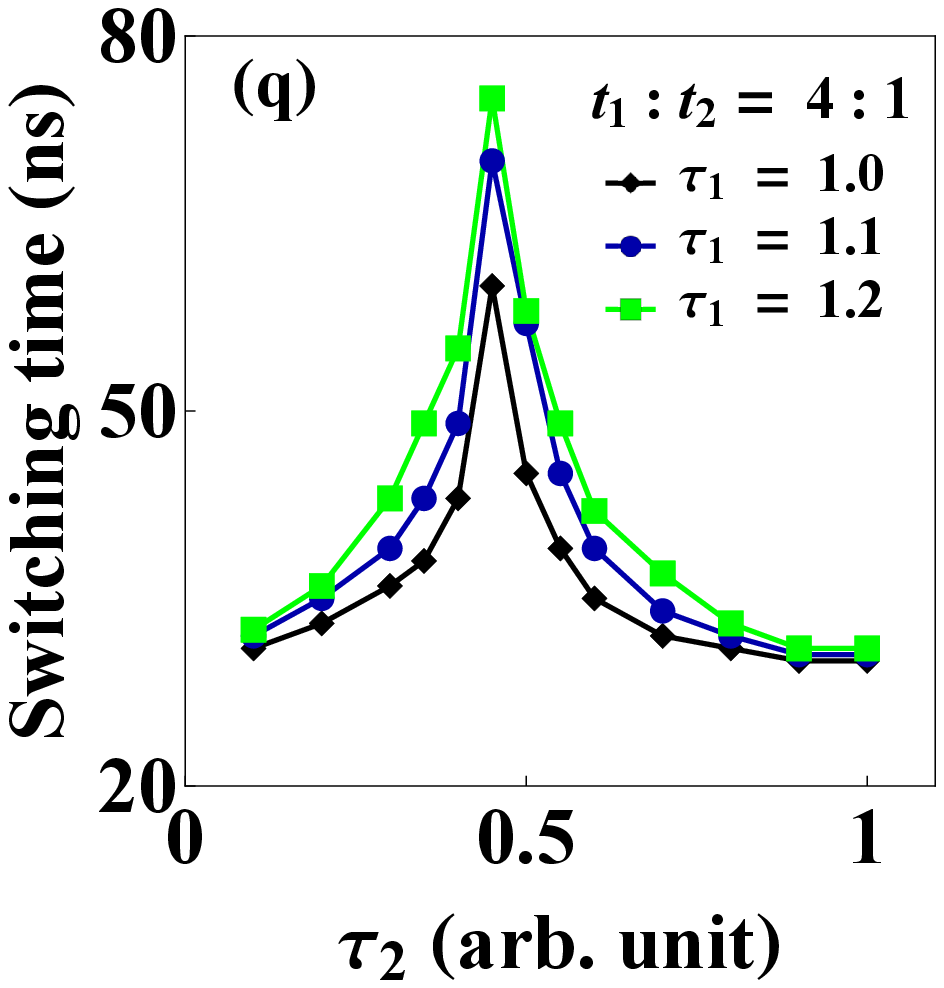}
\includegraphics[scale = 0.273]{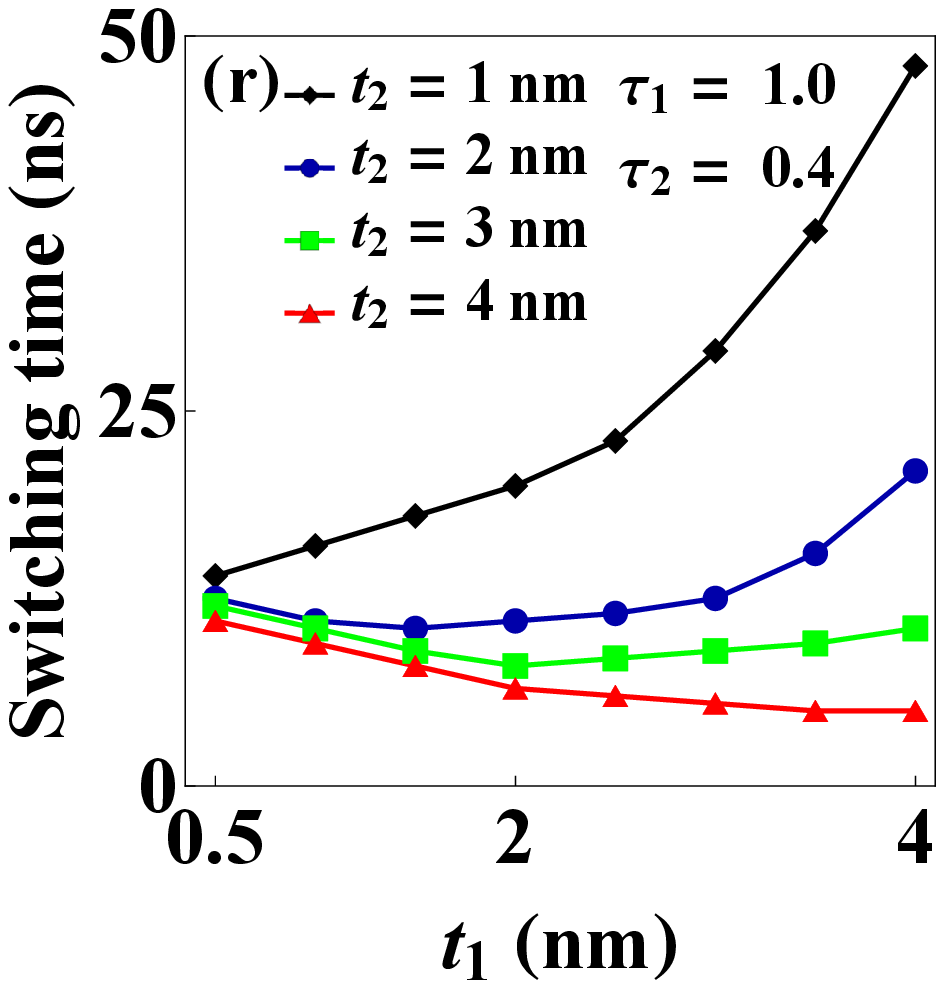}
}
\caption{(a) - (f) Oscillation trajectories of $\textbf{m}_1$ (blue) and $\textbf{m}_2$ (red) for different values of $\tau_1$ considering $\tau_2 = 0.4$ and K$_\text{R}$ = K$_\text{DM}$ = $1$. Plots (g) - (l) represent the time evolution of the magnetization component $\text{m}_{x1}$ (red) and $\text{m}_{x2}$ (blue). Plots (m) - (p) are the Fourier transform of the magnetization corresponding to the plots (g) - (j). The variation of switching time with $(\tau_1, \tau_2)$ and $(t_1, t_2)$ are shown in plots (q) and (r).}
\label{fig3}
\end{figure*}

\section{Results and Analysis}
We investigate the magnetization dynamics and switching behaviour quantitatively by solving Eqs. (\ref{eq6a})-(\ref{eq11}) numerically. We set the layer thickness, $t_1:t_2 = 4:1$ and their respective polarizations as $\mathcal{P}_0 = \mathcal{P}_1 = 0.5$ for all our analysis \cite{taniguchi}. This makes an asynchronism between $m_{x1}$ and $m_{x2}$ even in presence of RKKY. Moreover, other characteristics of magnetization dynamics like chaotic oscillations, magnetization switching etc can also be investigated under such condition. We also have explored different ratios of layer thickness and their impact on magnetization switching in our analysis. The RZ strength are set as $\alpha_\text{R} = 3 \times 10^{-10}$ eV.m and $h_0 = 0.7$ for all our analysis in Figs. \ref{fig2} - \ref{fig4}. Furthermore, we set the DM coefficients $(D_1, D_2)$ $= (-0.43, 0.43)$ are measured in $mJ/m^2$ for all our analysis \cite{li}. The opposite sign of $D_1$ and $D_2$ are considered due to the opposite chirality of the induced DM interactions in RZ$_1|$HM and HM$|$RZ$_2$ interfaces.

\subsection{Interplay of RKKY with DM interaction}
\subsubsection{For systems with RKKY $<$ DM interaction}
The magnetization oscillation and switching has been studied in Fig. \ref{fig2} for K$_\text{R}$ = $0.1$K$_\text{DM}$ for different choices of $(\tau_1, \tau_2)$. We found that the system display magnetization switching in both positive and negative spin torque conditions, which are characterized by the positive and negative sign of $\tau_1$ and $\tau_2$. Though the precession of the magnetic moments $m_{x1}$ and $m_{x2}$ are found to be identical for $(\tau_1, \tau_2)$ = $(1.5, 0.15)$, but magnetization switching of $m_{x2}$ is observed around $\tau_1 = 0.01$ as seen from Figs. \ref{fig2}(b) and (h). Moreover, the system is found to have stable magnetization oscillation in $m_{x1}$ and $m_{x2}$ for $\tau_1 = 0.65$, indicated by the Figs. \ref{fig2}(c) and (i). The sustained oscillations can be confirmed by a sharp peak around $250$ GHz in the Fourier transform spectra of $m_{x1}$ and $m_{x1}$. Thus for $\tau_1 = 0.65$ and $\tau_2 = 0.15$, the moments display a stable oscillation. It is to be noted that auto oscillation and the synchronization of the magnetizations can be measured by calculating GMR or TMR of the MTJ , where the resistance of the multilayer is related to magnetization through $\mathbf{m}_{1}$ and $\mathbf{m}_2$ as \cite{taniguchi}
\begin{equation}
\label{eq6}
\mathcal{R} = \frac{\mathcal{R}_\text{AP} + \mathcal{R}_\text{P}}{2} - \frac{\mathcal{R}_\text{AP} - \mathcal{R}_\text{P}}{2}\mathbf{m}_{1}.\mathbf{m}_{2}
\end{equation}

where, $\mathcal{R}_\text{P}$ ($\mathcal{R}_\text{AP}$) is the resistance of the system corresponding to parallel (antiparallel) orientations of magnetization and the vector product $\mathbf{m}_{1}.\mathbf{m}_{2}$ can be termed as MR of the system.
The synchronization of the magnetization $\mathbf{m}_{1}$ and $\mathbf{m}_2$ can be understood via two sharp MR peaks around $0$ and $500$ GHz as shown in Fig. \ref{fig2}(o). A similar characteristic of MR is also observed in Fig. \ref{fig2}(r) for negative biasing condition. The switching of $m_{x1}$ for STT,  $(\tau_1, \tau_2) = (-0.9, -0.2)$, is due to the opposite biasing of RZ$_1$ layer. However, the system can also behave like an oscillator even in negative biasing conditions for $\tau_1 = -0.65$ as seen from Figs. \ref{fig2}(f) and (l). 

The variation of switching speed with $\tau_2$ for both positive and negative values of $\tau_1$ is studied in Figs. \ref{fig2}(m) and (p) respectively. The switching time (ST) can be numerically obtained by calculating the time corresponding to $97.5\%$ of the saturation magnetization. It should be noted that for $\tau_1 = 0.01$, the ST gradually decreases for low values of $\tau_2$, but it remains constant in moderate and high $\tau_2$ regions. Moreover, the switching speed decreases for $\tau_1 = 0.05$ and $0.1$. In this condition, the maximum ST is found at $\tau_2 = 0.5$ and $0.7$, respectively. An identical but opposite characteristic of magnetization switching is also observed for negative $\tau_1$ values, as seen from Fig. \ref{fig2}(p). However, in this case, a delayed switching of $m_{x1}$ is observed for lower values of $\tau_1$, which may be due to the opposite current biasing of the RZ$_1$ layer. Although the magnetization switching can occur for different $(\tau_1, \tau_2)$ combinations as seen from Figs. \ref{fig2}(m) and \ref{fig2}(p), yet the interplay of STT with layer thickness and switching can be understood by exploring the variation of ST with layer thickness. Figs. \ref{fig2}(n) and \ref{fig2}(q) display the switching speed for ($\tau_1$, $\tau_2$) = $(0.01, 0.15)$ and ($\tau_1$, $\tau_2$) = $(-0.9, -0.2)$ respectively. For ($\tau_1$, $\tau_2$) = $(0.01, 0.15)$, the ST is found to be nearly same for all $(t_1, t_2)$. Nevertheless, the maximum switching speed is found for the thickness $t_1 = 2$nm and $t_2 = 1$nm. The ST gradually increases with the increase in $t_1$ and $t_2$ for ($\tau_1$, $\tau_2$) = $(-0.9, -0.2)$ as seen from Fig. \ref{fig2}(q). Moreover, the switching speed slowed down with the increase in the thickness of the RZ$_2$ layer, as seen from Fig. \ref{fig2}(q). 
\begin{figure*}[hbt]
\centerline
\centerline{
\includegraphics[scale = 0.3]{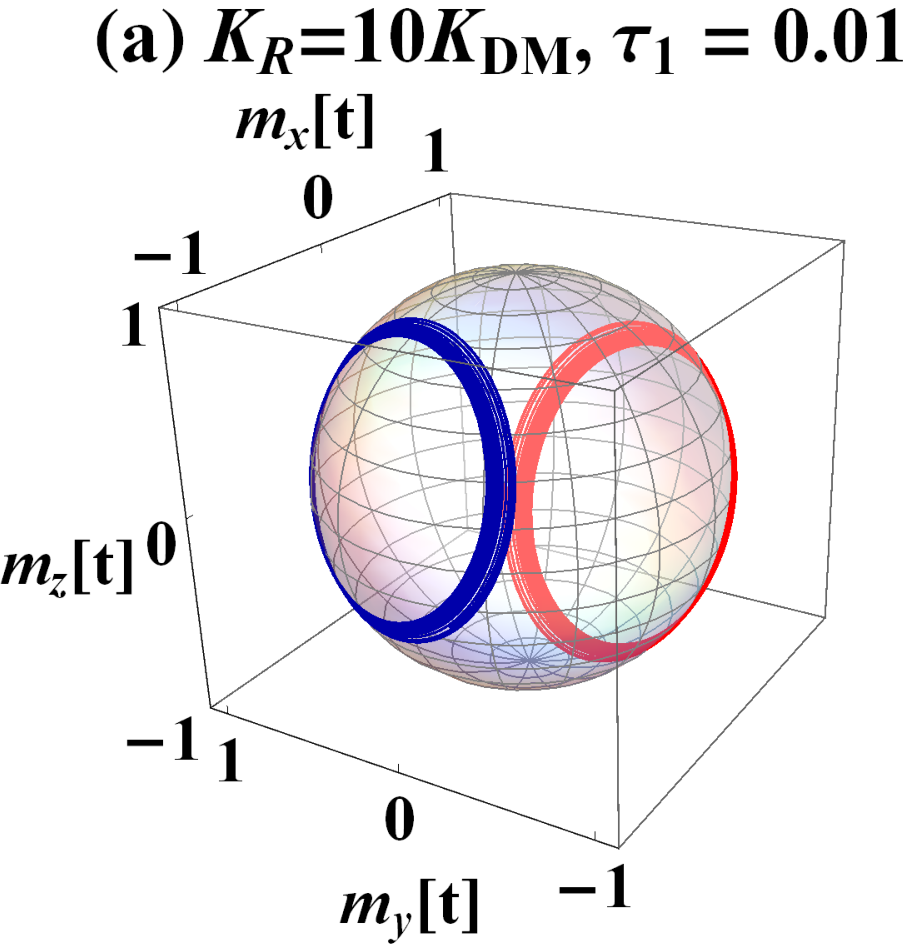}
\hspace{-0.09cm}
\vspace{0.02cm}
\includegraphics[scale = 0.3]{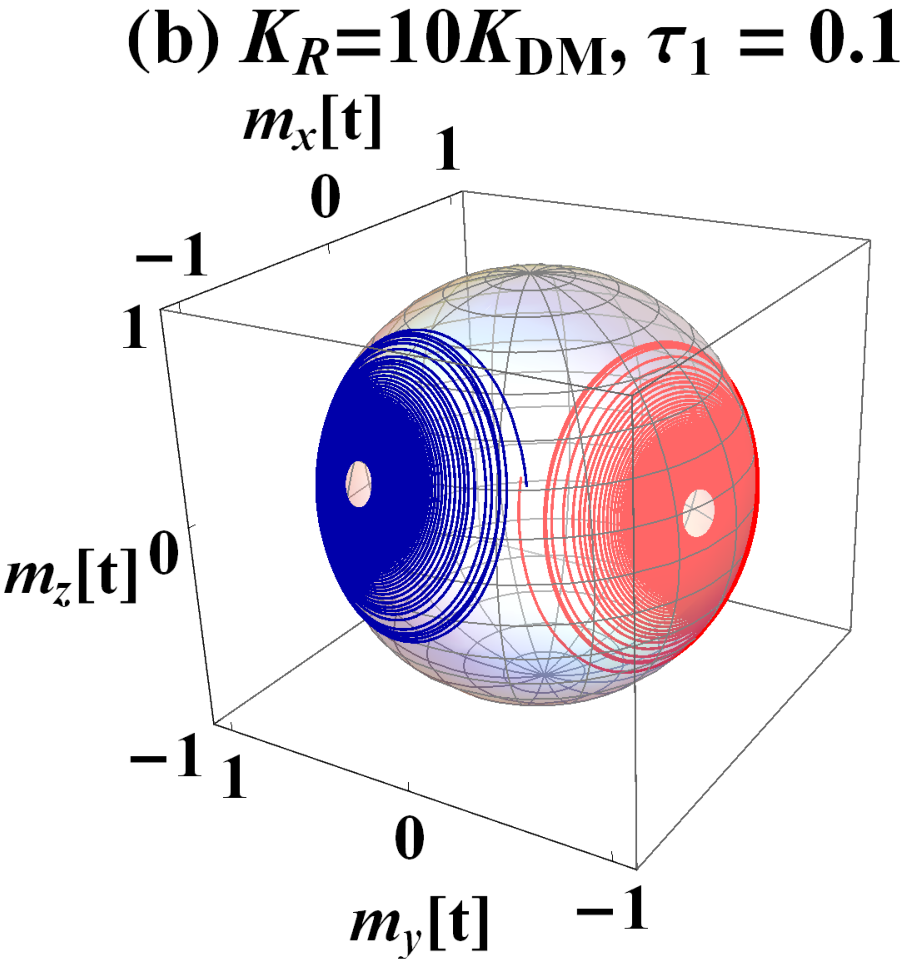}
\hspace{-0.09cm}
\vspace{0.02cm}
\includegraphics[scale = 0.3]{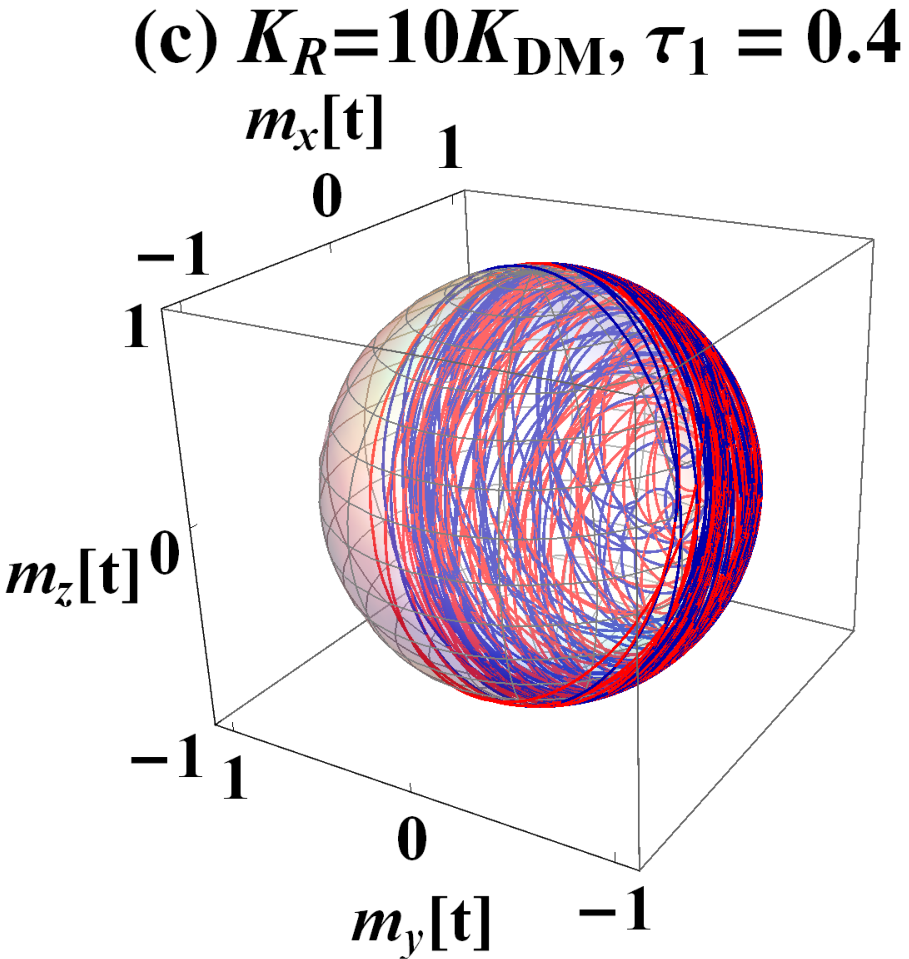}
\hspace{-0.09cm}
\vspace{0.02cm}
\includegraphics[scale = 0.3]{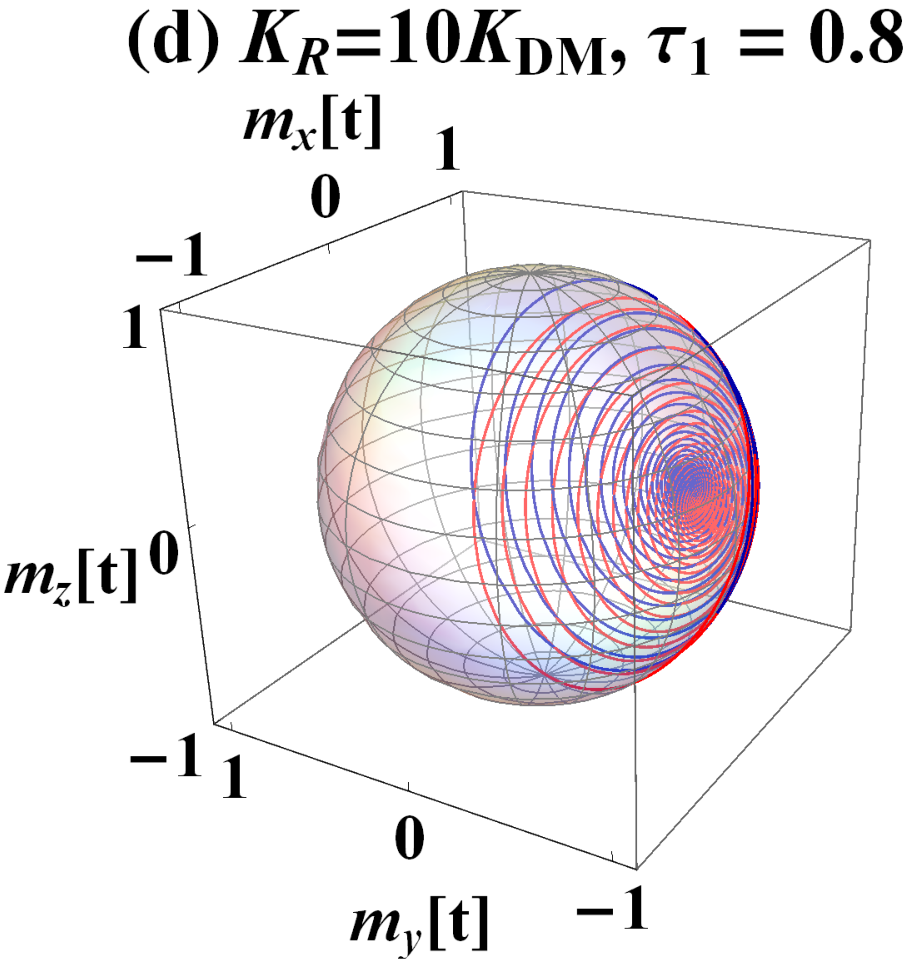}
\hspace{-0.09cm}
\vspace{0.02cm}
\includegraphics[scale = 0.3]{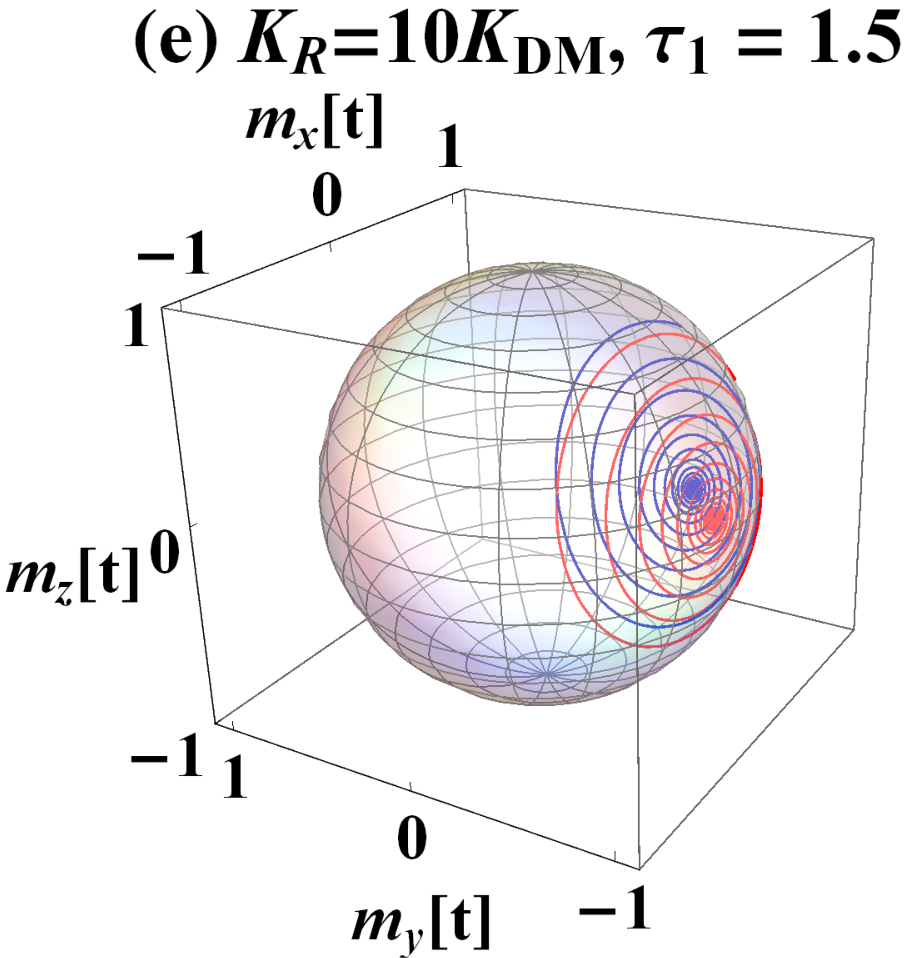}
\hspace{-0.09cm}
\vspace{0.02cm}
\includegraphics[scale = 0.3]{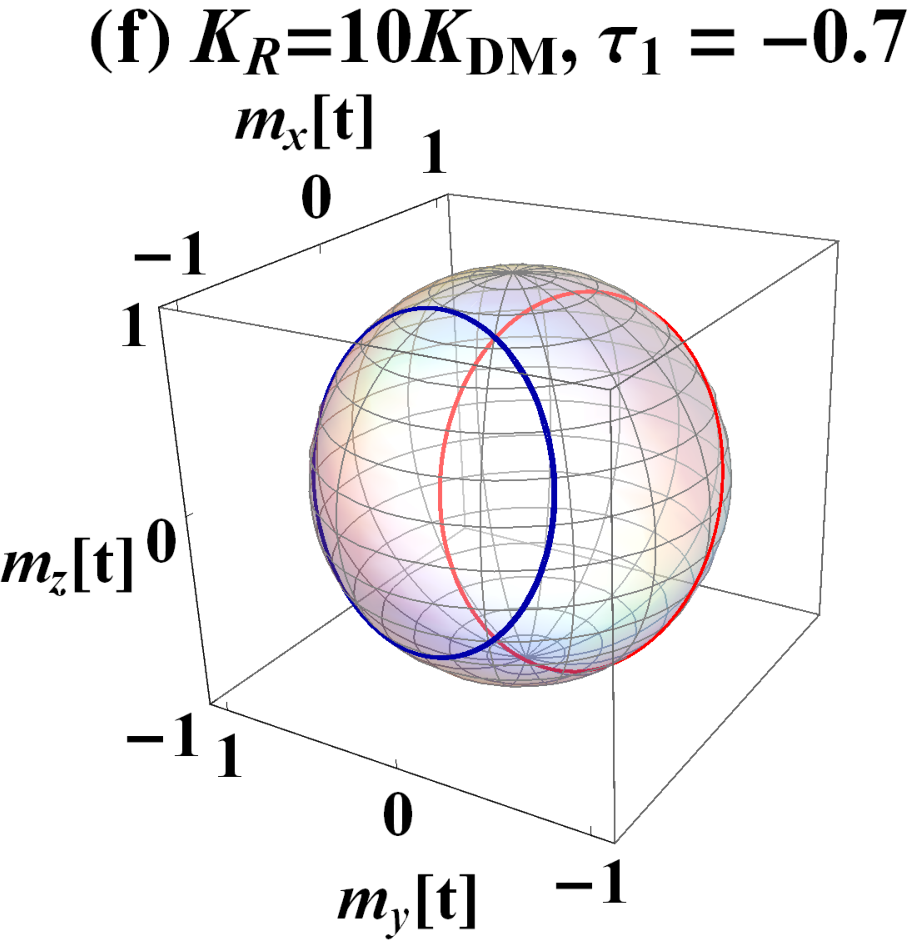}
\hspace{-0.09cm}
\vspace{0.02cm}
\includegraphics[scale = 0.275]{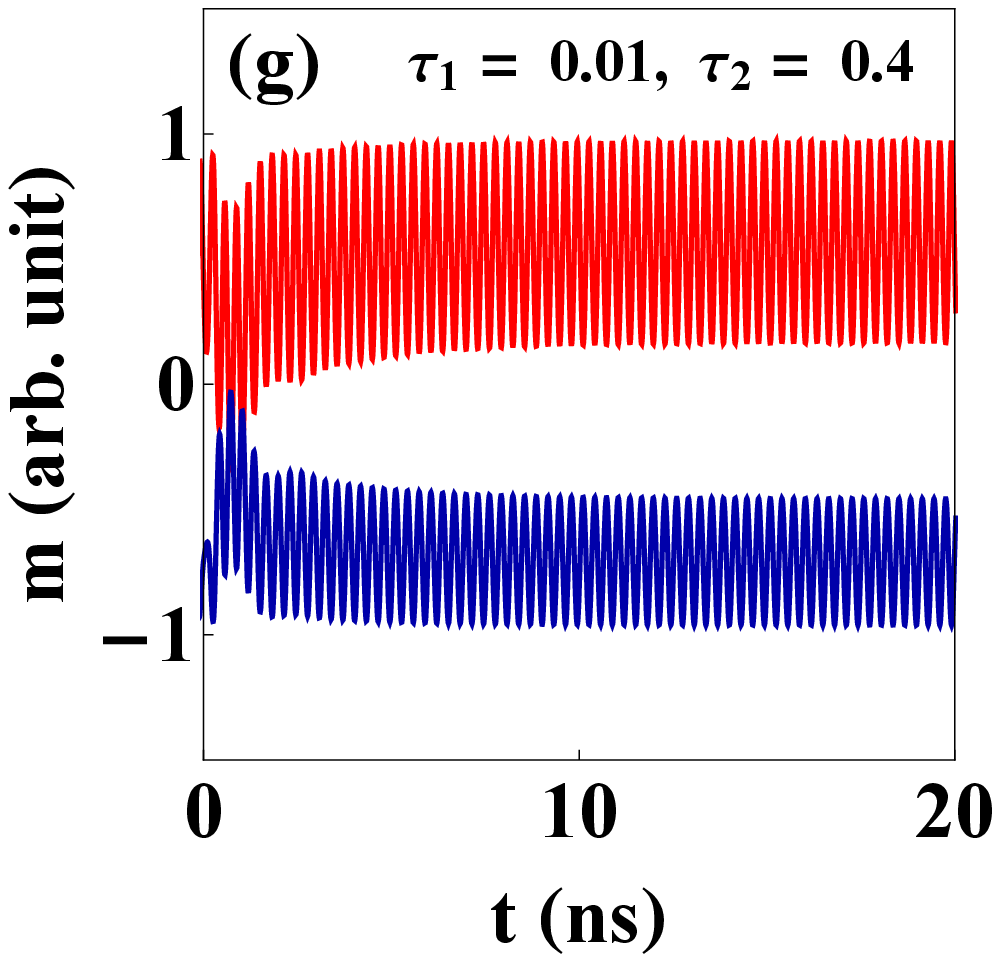}
\hspace{-0.09cm}
\includegraphics[scale = 0.275]{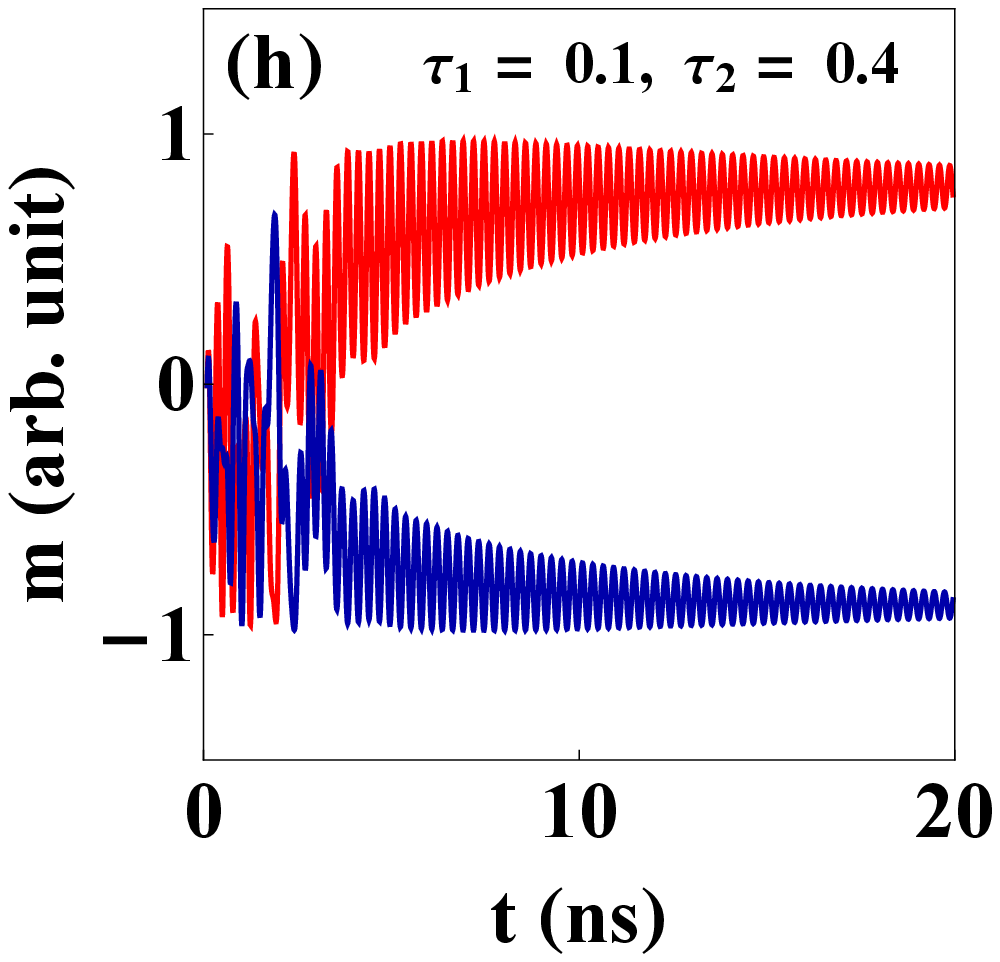}
\hspace{-0.09cm}
\includegraphics[scale = 0.275]{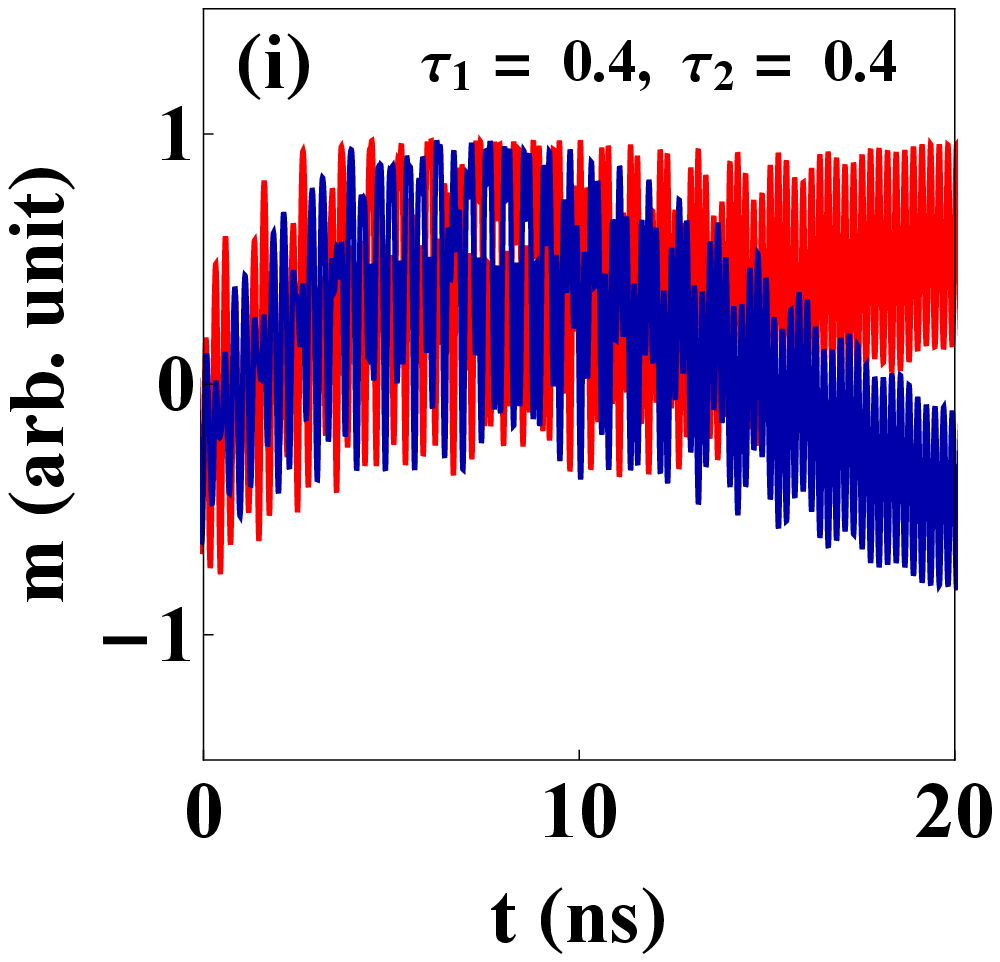}
\hspace{-0.09cm}
\includegraphics[scale = 0.275]{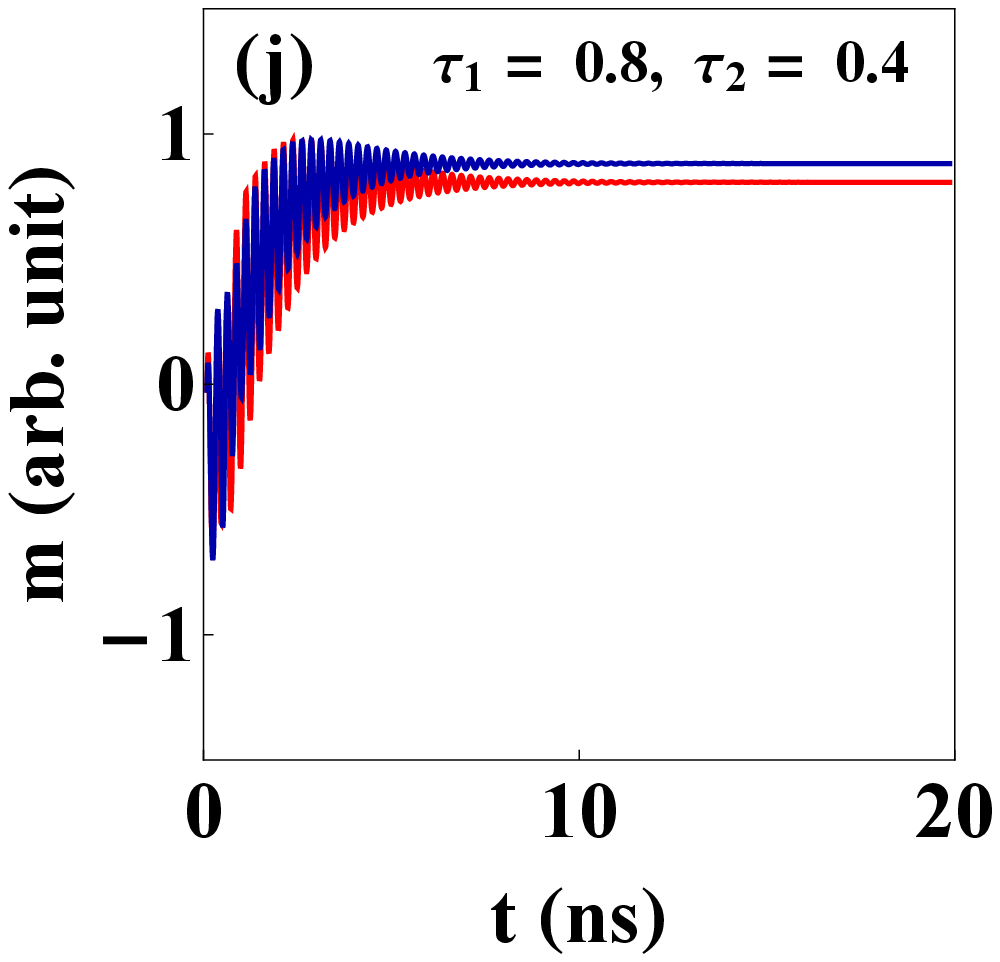}
\hspace{-0.09cm}
\includegraphics[scale = 0.275]{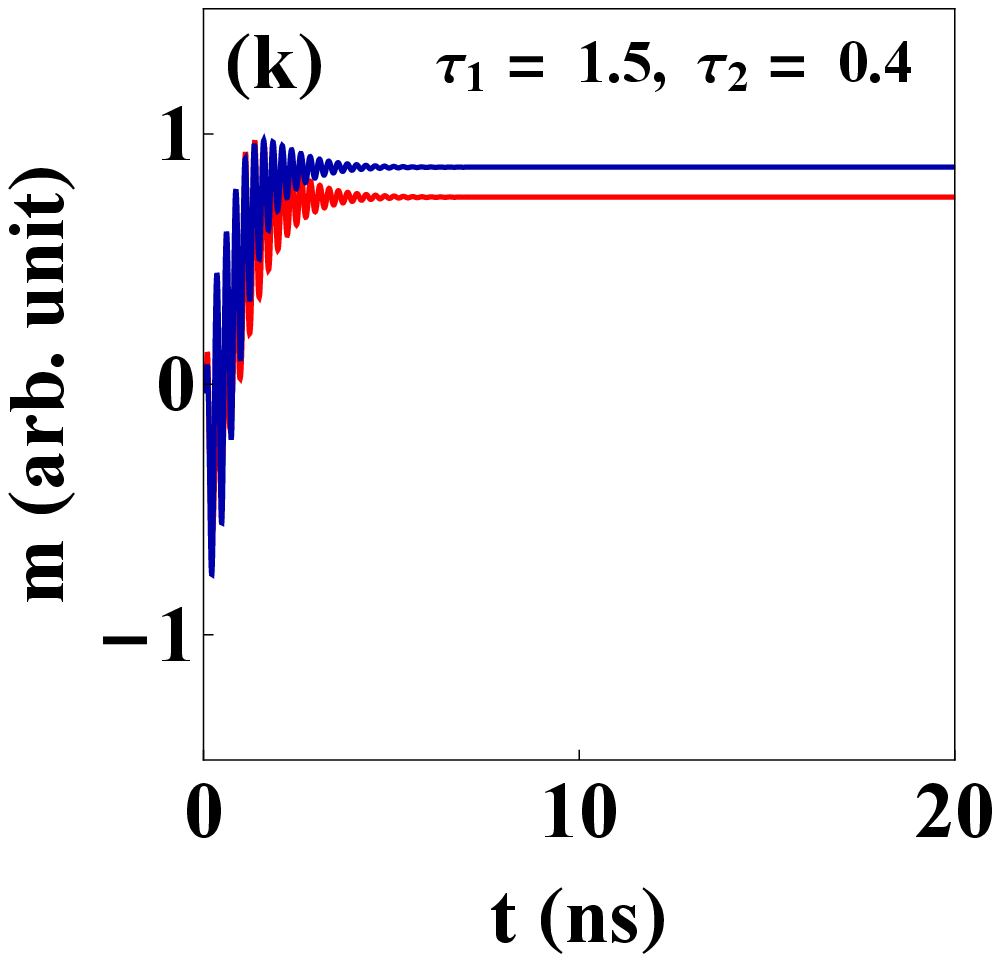}
\hspace{-0.09cm}
\includegraphics[scale = 0.275]{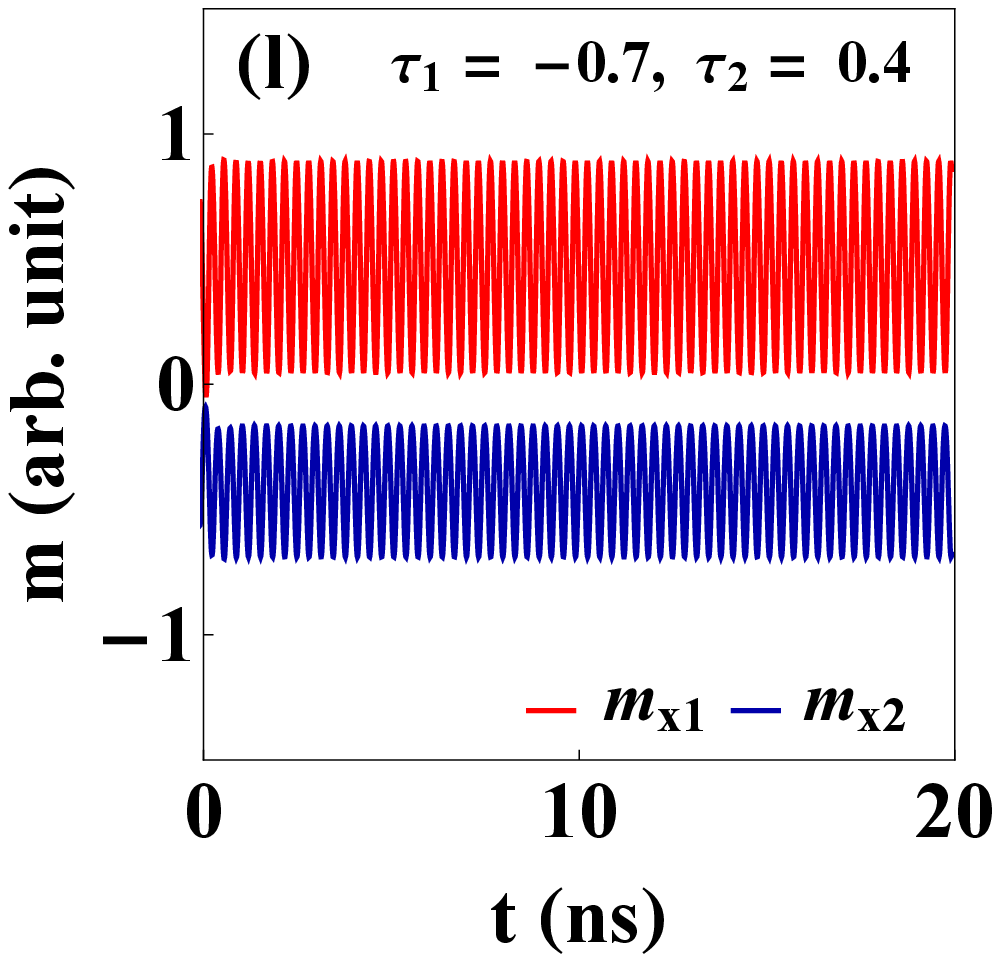}
\hspace{-0.09cm}
\includegraphics[scale = 0.273]{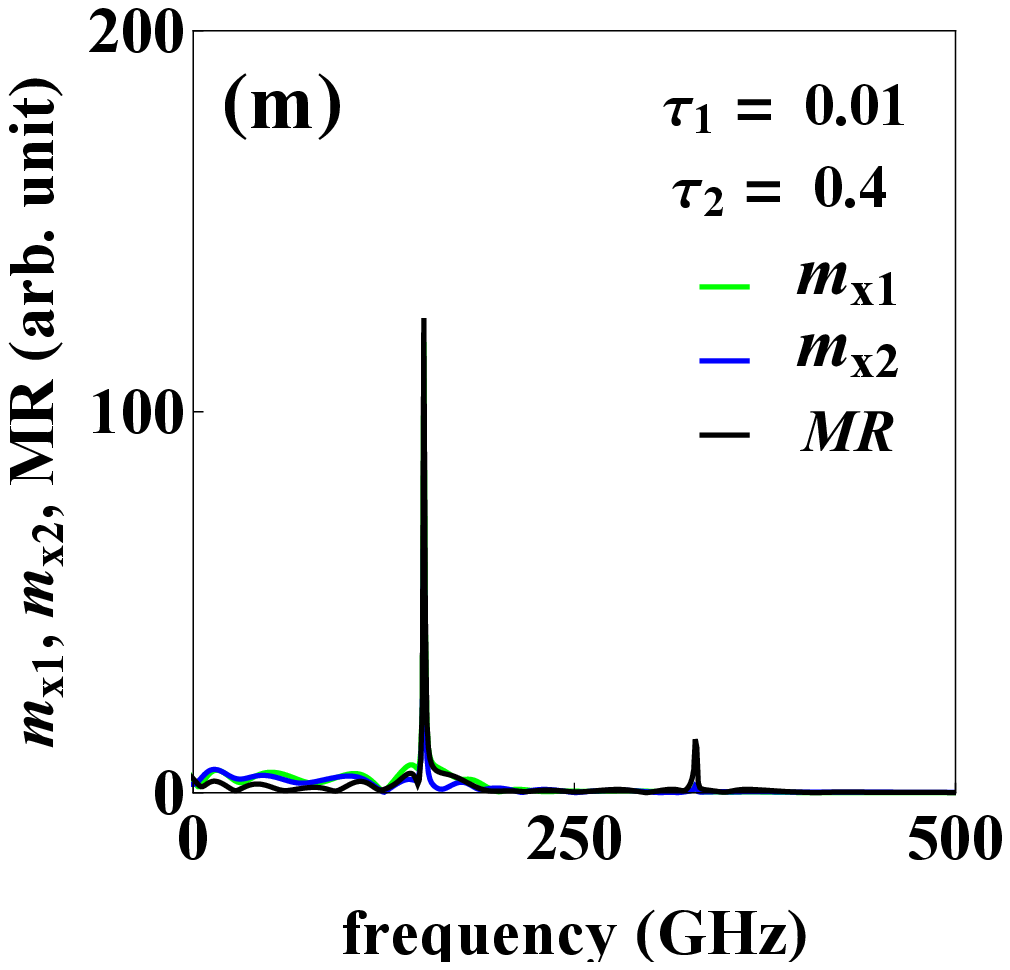}
\includegraphics[scale = 0.273]{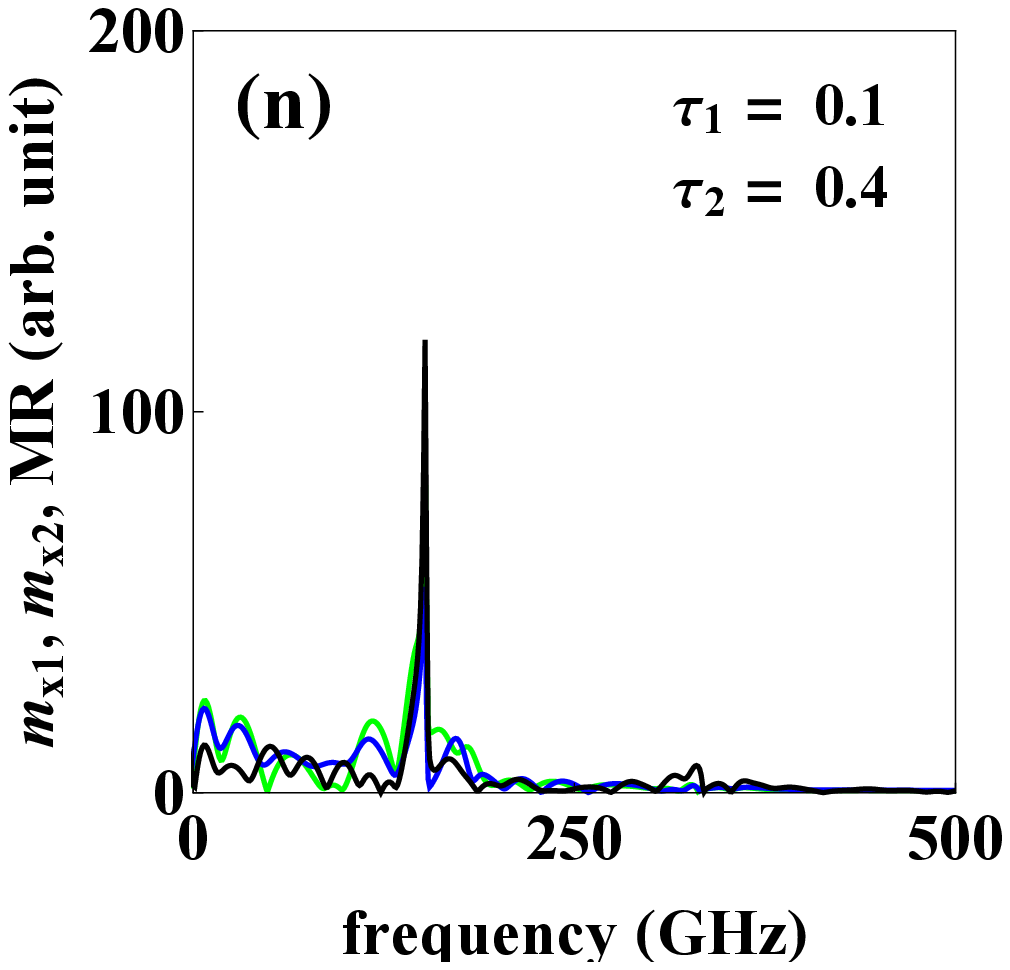}
\includegraphics[scale = 0.273]{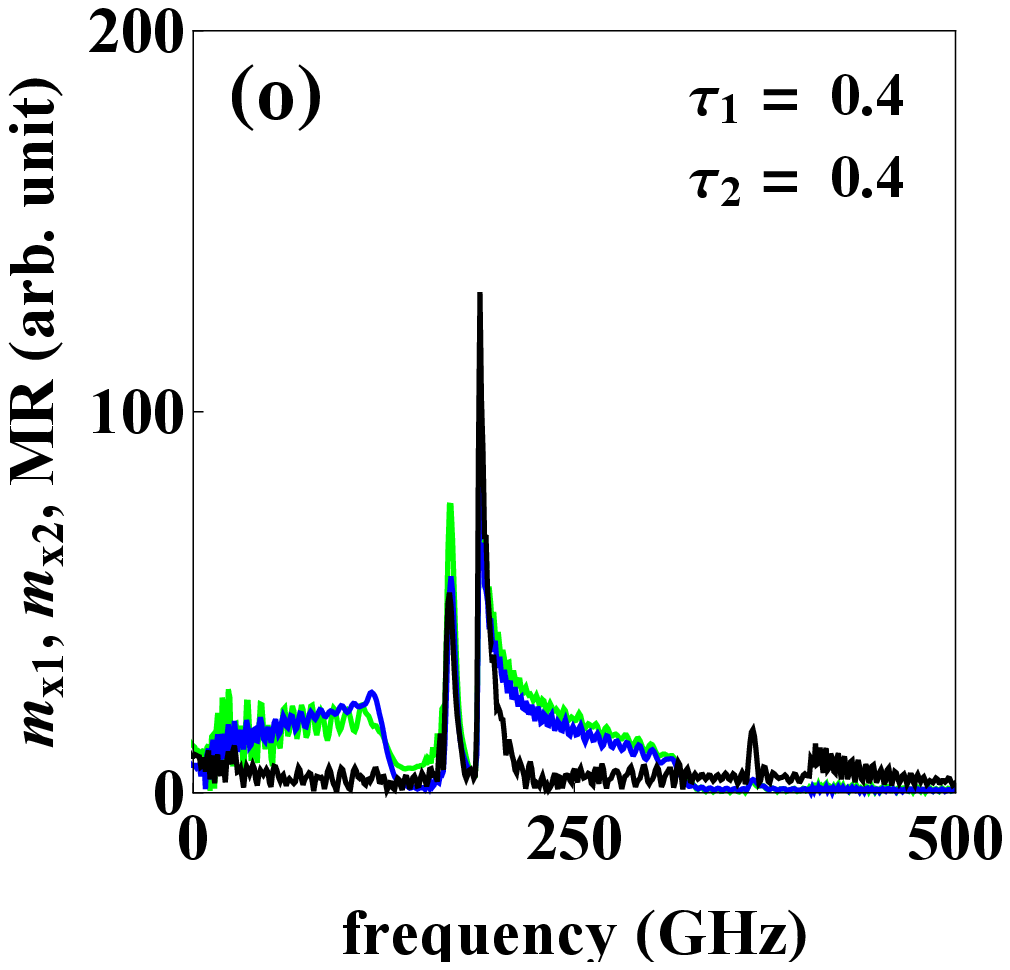}
\includegraphics[scale = 0.273]{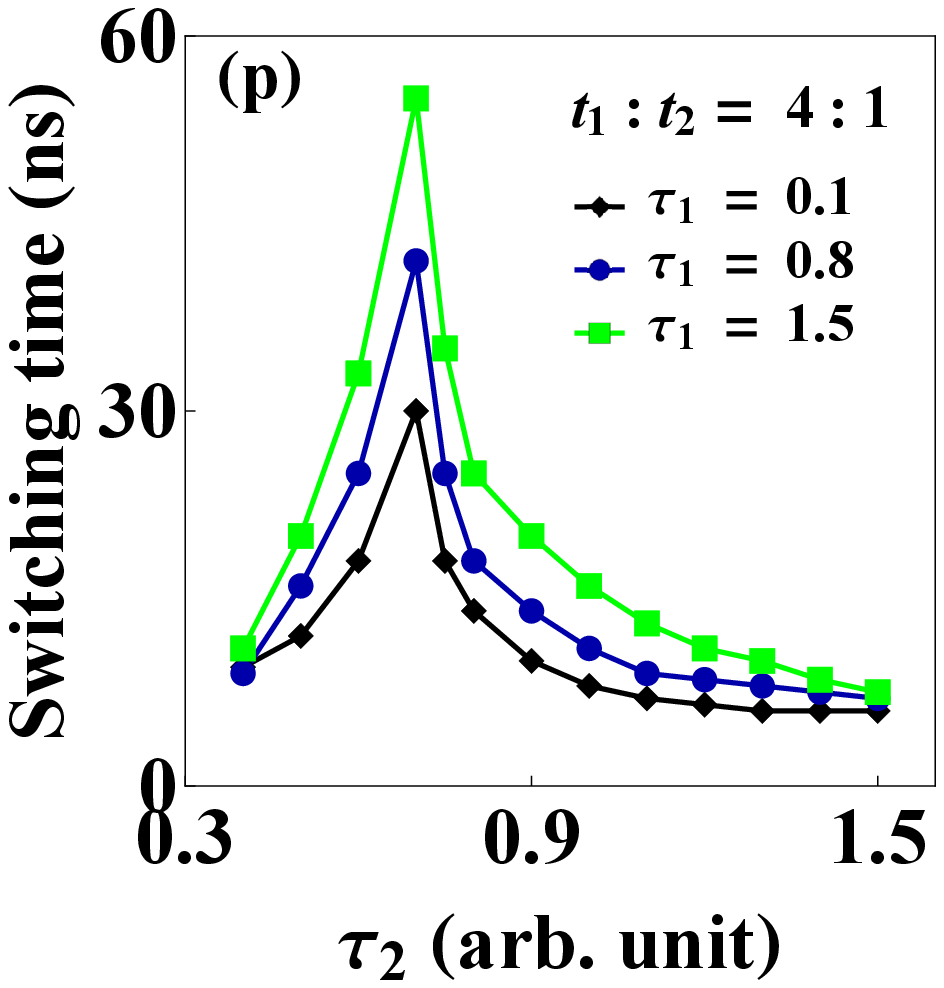}
\includegraphics[scale = 0.273]{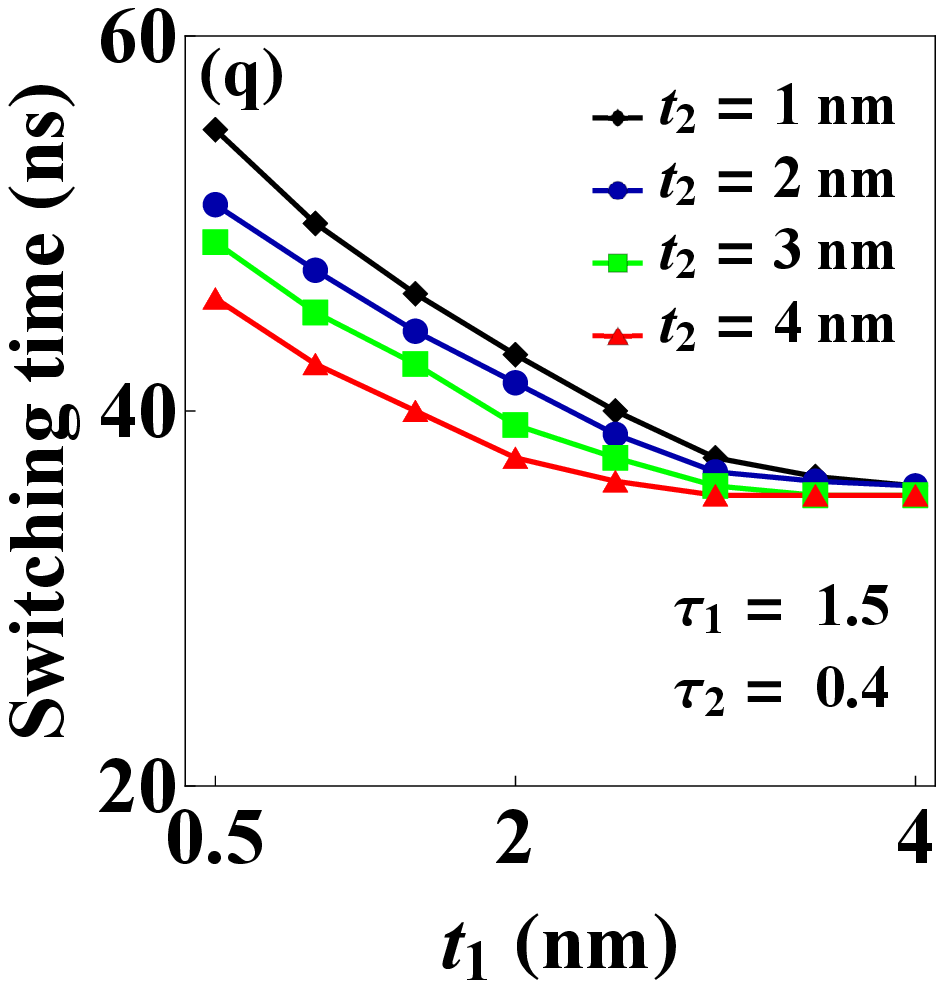}
\includegraphics[scale = 0.273]{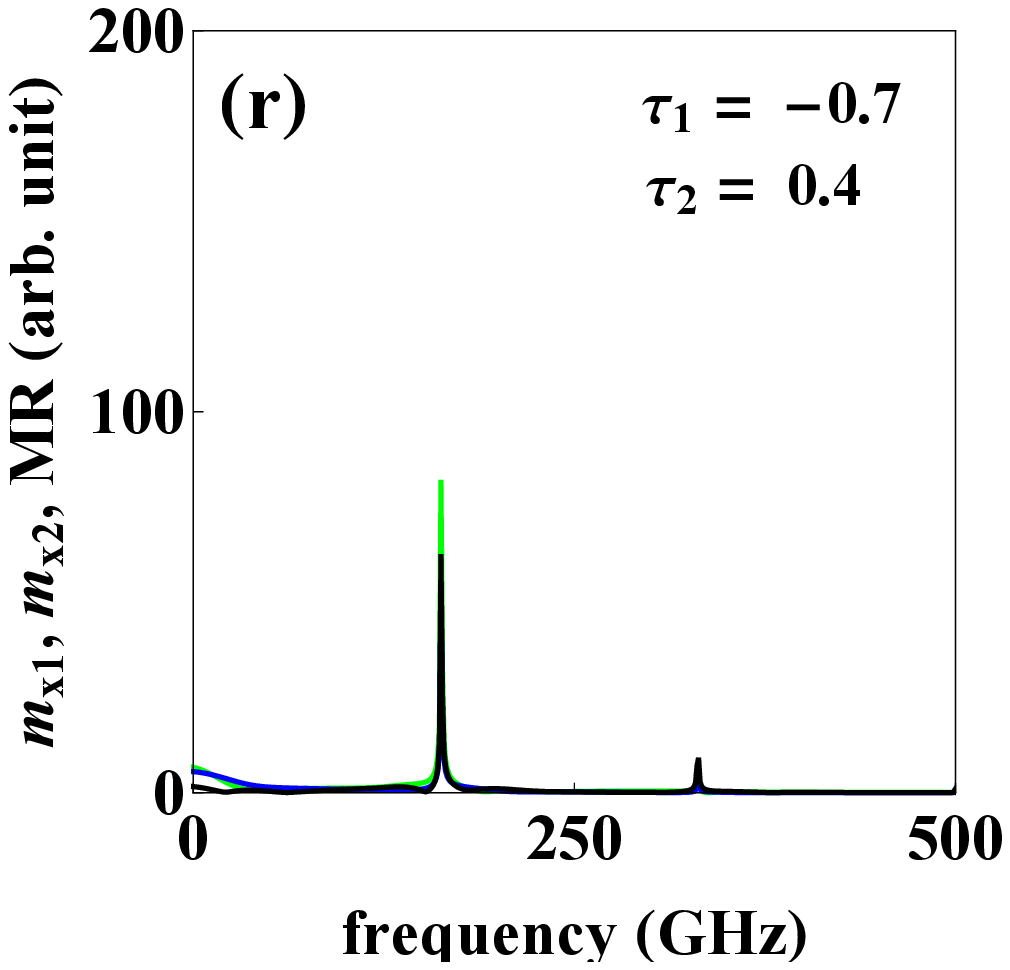}
}
\caption{(a) - (f) Oscillation trajectories of $\textbf{m}_1$ (blue) and $\textbf{m}_2$ (red), (g) - (l) the time evolution of the magnetization component $\text{m}_{x1}$ (red) and $\text{m}_{x2}$ (blue) for different values of $\tau_1$ considering $\tau_2 = 0.4$ and K$_\text{R} = 10$K$_\text{DM}$. Plots (m) - (o) and (r) are the Fourier transform of the magnetization component, while plots (p) and (q) depicts the variation of switching time with $\tau_i$ and $t_i$ respectively.}
\label{fig4}
\end{figure*}

\subsubsection{For systems with RKKY $\sim$ DM interaction}
The magnetization oscillations of the system are found to be very sensitive to $\tau_1$ As the RKKY interaction is of the order of DMI, as seen from Figs. \ref{fig3}(a) and (g). For $\tau_1 = 0.1$, the synchronization of $\mathbf{m}_1$ and $\mathbf{m}_2$ cease to exist. In this condition, $\mathbf{m}_1$ precesses about the z-axis while $\mathbf{m}_2$ about the x-axis. However, the oscillations are associated with multiple frequencies, as seen in Fig. \ref{fig3}(m). With a little increase in $\tau_1$ to $0.2$, the system executes non-linear oscillations as seen from Figs. \ref{fig3}(b) and (h). This nonlinearity is further enhanced as $\tau_1$ increases to $0.3$ as indicated by Figs. \ref{fig3}(c) and (i). These non-linear oscillations can be characterized by the multiple frequencies in the Fourier spectra in \ref{fig3}(n) and (o). This may be due to the same order of RKKY and DM interaction which destabilizes the oscillations. However, sustained magnetization oscillations at about $250$ GHz can be achieved with a further increase in $\tau_1$  to $0.7$ as suggested by Figs. \ref{fig3}(d) and (j). The switching characteristics can be studied by choosing $\tau1 = 1.0$ and $-0.7$ as observed from Figs. \ref{fig3}(e), (f), (k) and (l). The ST vs $\tau_2$ for different $\tau_1$ has been studied in Fig. 3(q). It is seen that the rapid switching can be obtained by considering low values of $\tau_1$. It should be noted that the ST is identical with Fig. \ref{fig2}, but rapid switching is obtained for low values of $\tau_1$. Moreover, rapid switching is obtained for $t_1 = t_2 = 4$ nm. This indicates that by controlling the bias, current magnetization oscillations and switching can be achieved even when RKKY interaction is of the order DM interaction.

\subsubsection{For systems with RKKY $>$ DM interaction}
In Fig. \ref{fig4}, we study the magnetization dynamics for K$_\text{R} = 10$K$_\text{DM}$ with different choices of $\tau_1$ considering $\tau_2 = 0.4$. The system executes sustained oscillations about the frequency $150$ GHz for $\tau_1 = 0.01$ and $-0.7$ as seen from Figs. \ref{fig4}(a), (m), (f), and (r). However, for $\tau_1 = 0.01$, the system stabilizes after a short time, as indicated by Fig. \ref{fig4}(g). The switching characteristics are observed for $\tau_1 = 0.1$ and $0.8$. However, it is to be noted that in this case, the magnetization reversal of $\mathbf{m}_2$ is observed as suggested by Figs. \ref{fig4}(b), (d), (h) and (j). With increase in $\tau_1$ to $1.5$, a rapid switching of $\mathbf{m}_2$ is seen in Fig. \ref{fig4}(e) and (k). For moderate $\tau_1$, the system executes non-linear oscillations as seen from Fig. \ref{fig4}(c), (i) and (o). The dependence of ST with $\tau_1$ and $\tau_2$ is identical to that of Fig. \ref{fig3}. However, a gradual decrease in ST is observed with the rise in $t_1$ and $t_2$ as indicated by Fig. \ref{fig4}(p) and (q). Thus, the non-linear characteristics of the magnetization oscillations significantly decreases for K$_\text{R} > $ K$_\text{DM}$ as compared to K$_\text{R} \sim$ K$_\text{DM}$. Moreover, the frequency of oscillations also decreases in this scenario.   

We observe that for RKKY $<$ DM interaction, the system behaves like a switcher for $0.01<\tau_1<0.6$, oscillator for $0.6<\tau_1<0.7$, while it remains unswitched for $0.7<\tau_1<1$ in case of $\tau_2>0$. When $\tau_2<0$, the system remains unswitched for $-0.6<\tau_1<0$ while behaves as an oscillator for $-0.65<\tau_1<-0.6$ and as a switcher for $-1<\tau_1<-0.65$ as seen from Fig. \ref{fig2}. For the systems with RKKY $\sim$ DM interaction, it shows unswitched characteristics for $-0.7<\tau_1<0.2$, behaves as an oscillator for $0.6<\tau_1<0.7$ and as a switcher for $0.7<\tau_1<1$ considering $\tau_2 = 0.4$. Moreover, under this condition, the system display chaotic oscillations for $0.2<\tau_1<0.6$ as seen from Fig. \ref{fig3}. Likewise, for the systems with RKKY $>$ DM interaction, oscillatory characteristics are seen for $-0.7<\tau_1<0.1$, chaotic oscillations are found for $0.1<\tau_1<0.4$ and the system remain unswitched for $0.4<\tau_1<1.5$ considering $\tau_2 = 0.4$ as seen from Fig. \ref{fig4}.

\begin{figure}[hbt]
\centerline
\centerline{
\includegraphics[scale = 0.30]{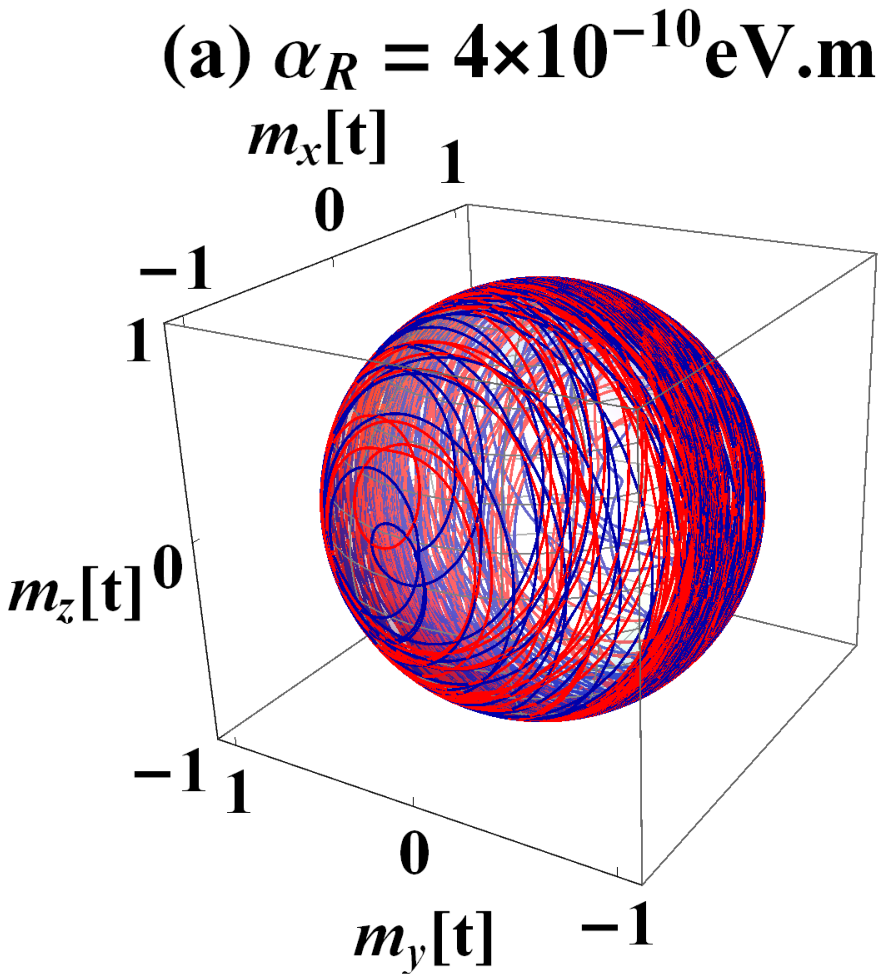}
\hspace{-0.09cm}
\vspace{0.15cm}
\includegraphics[scale = 0.30]{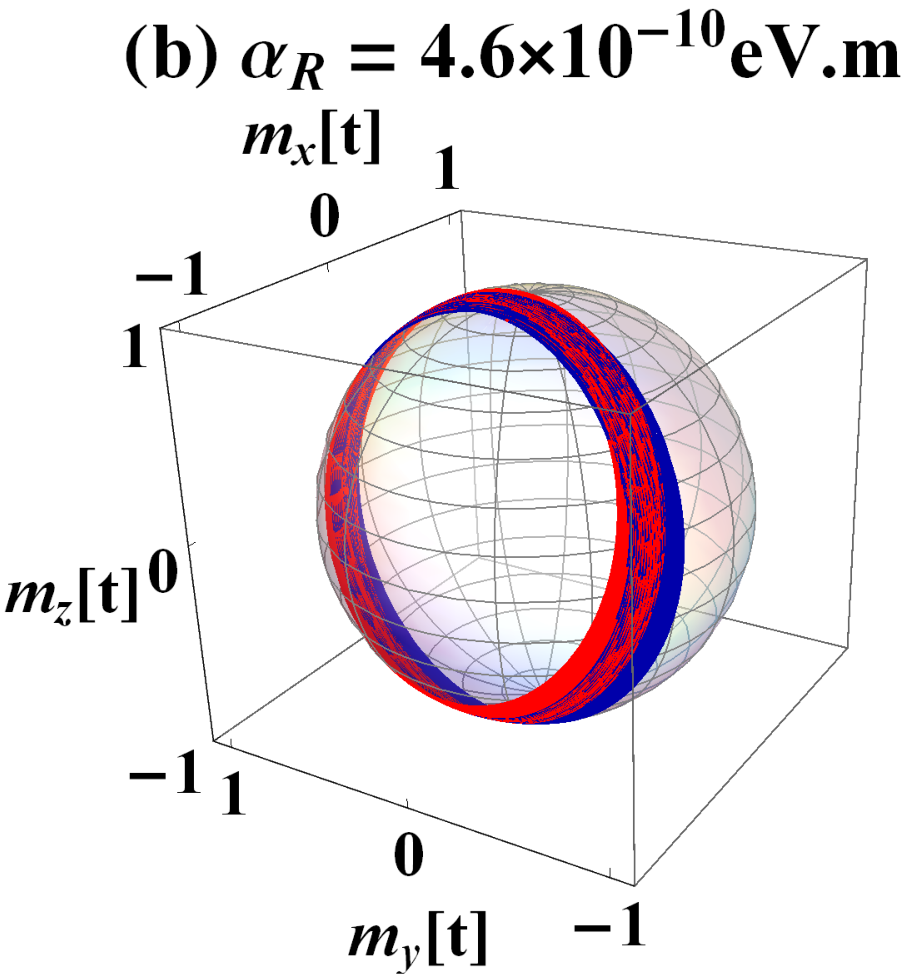}
\hspace{-0.09cm}
\vspace{0.15cm}
\includegraphics[scale = 0.30]{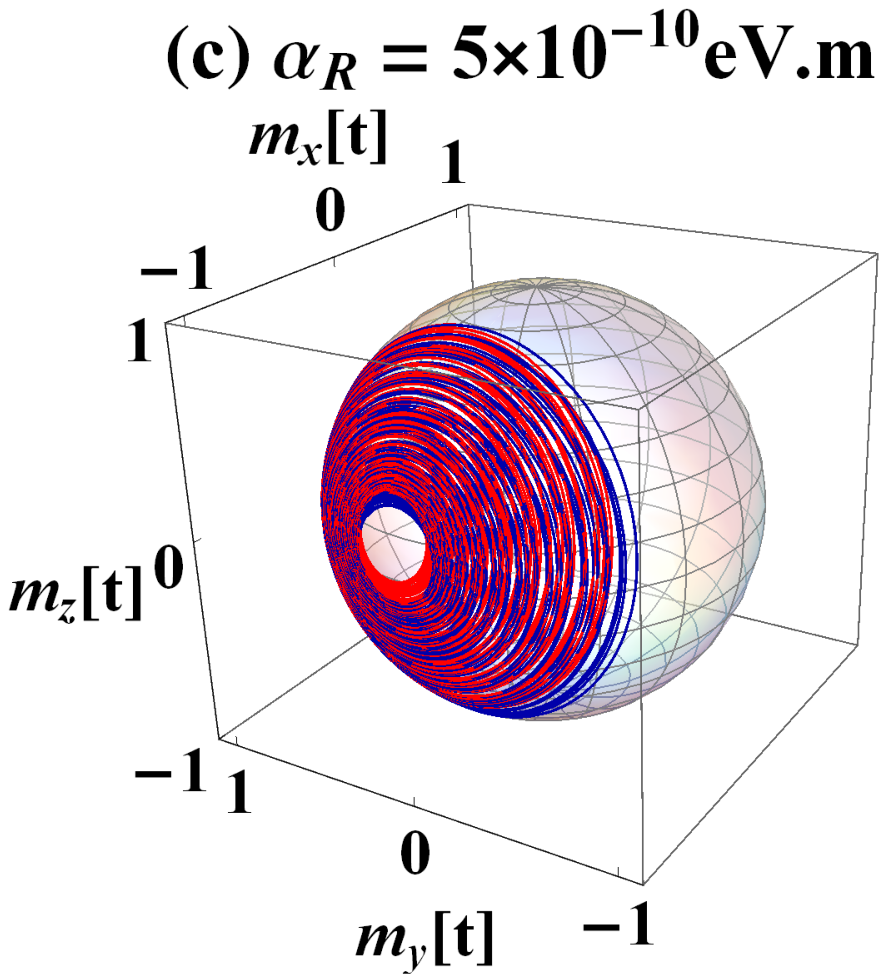}
\hspace{-0.09cm}
\vspace{0.15cm}
\includegraphics[scale = 0.275]{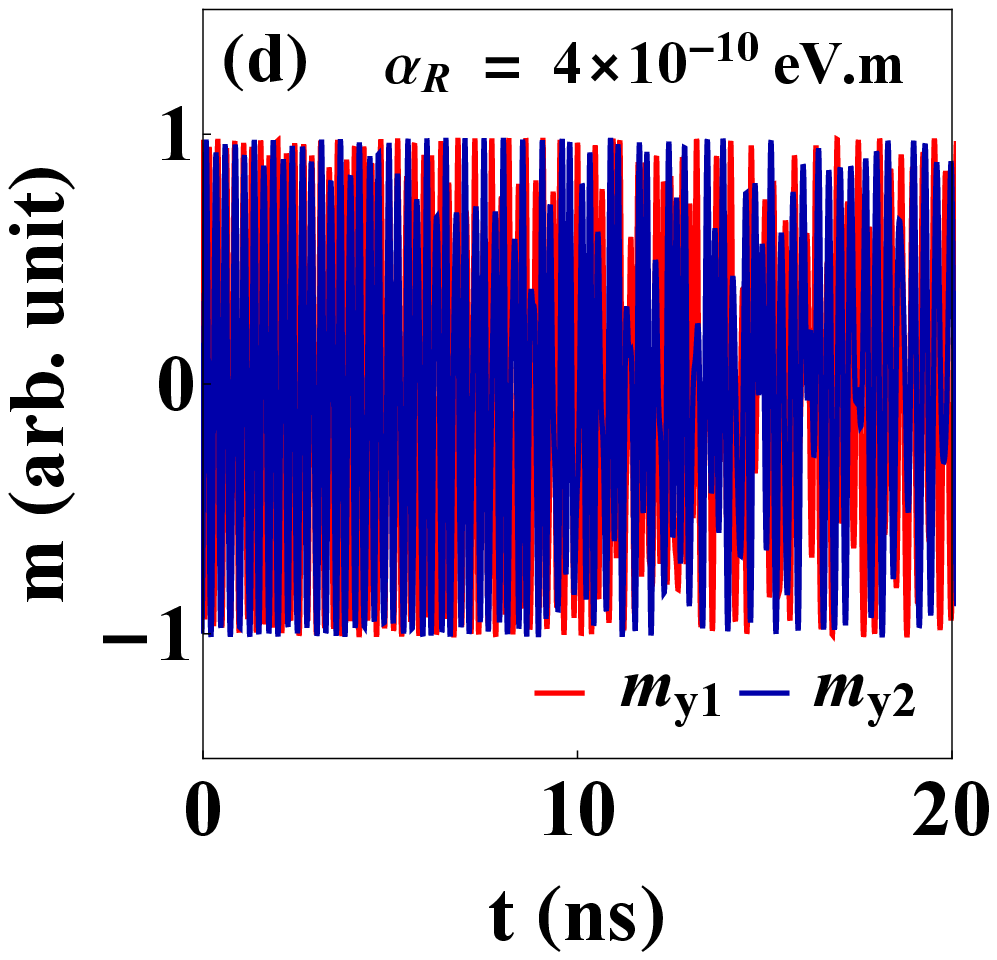}
\hspace{-0.09cm}
\includegraphics[scale = 0.275]{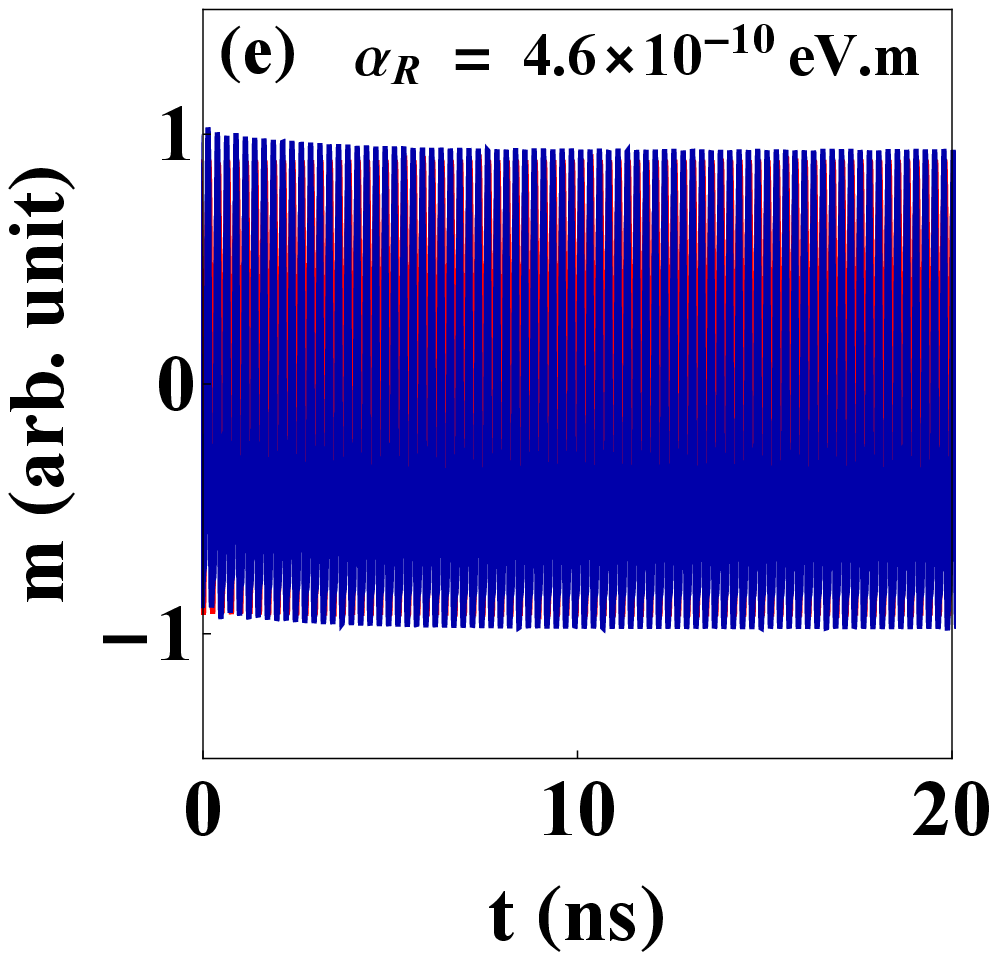}
\hspace{-0.09cm}
\includegraphics[scale = 0.275]{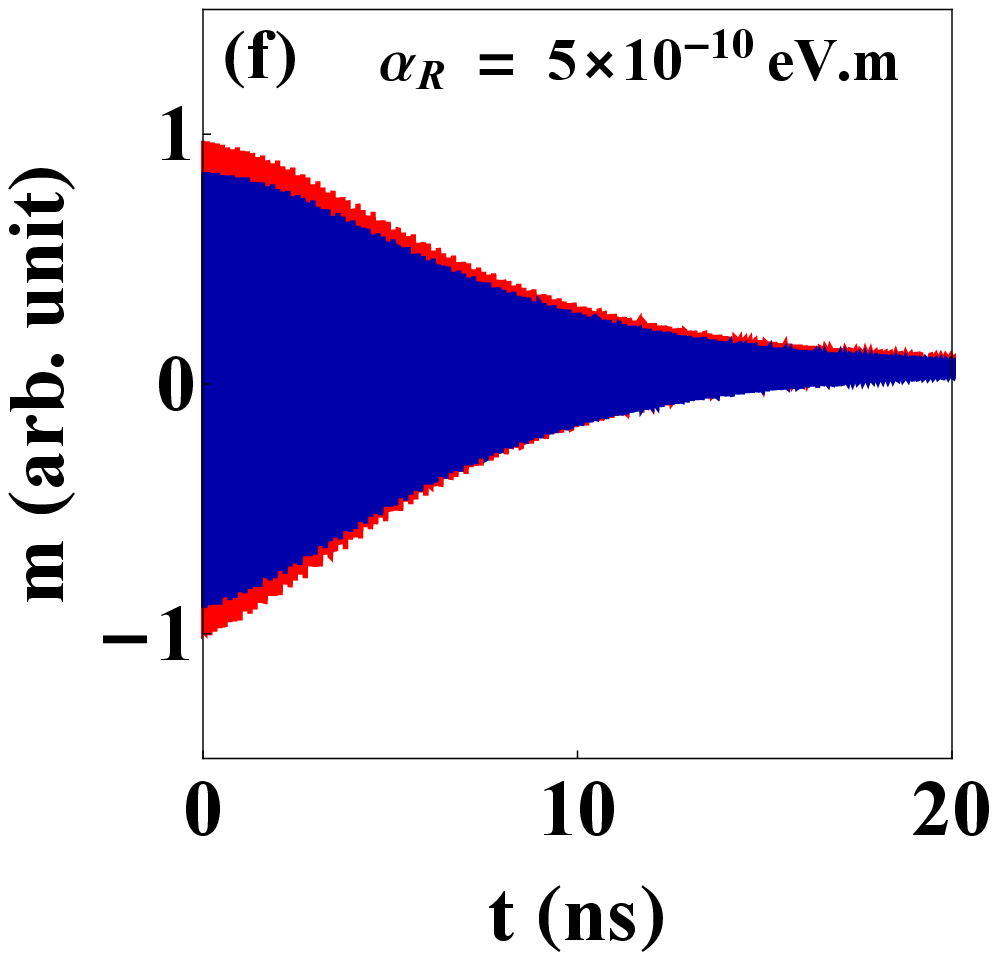}
\hspace{-0.09cm}
\includegraphics[scale = 0.275]{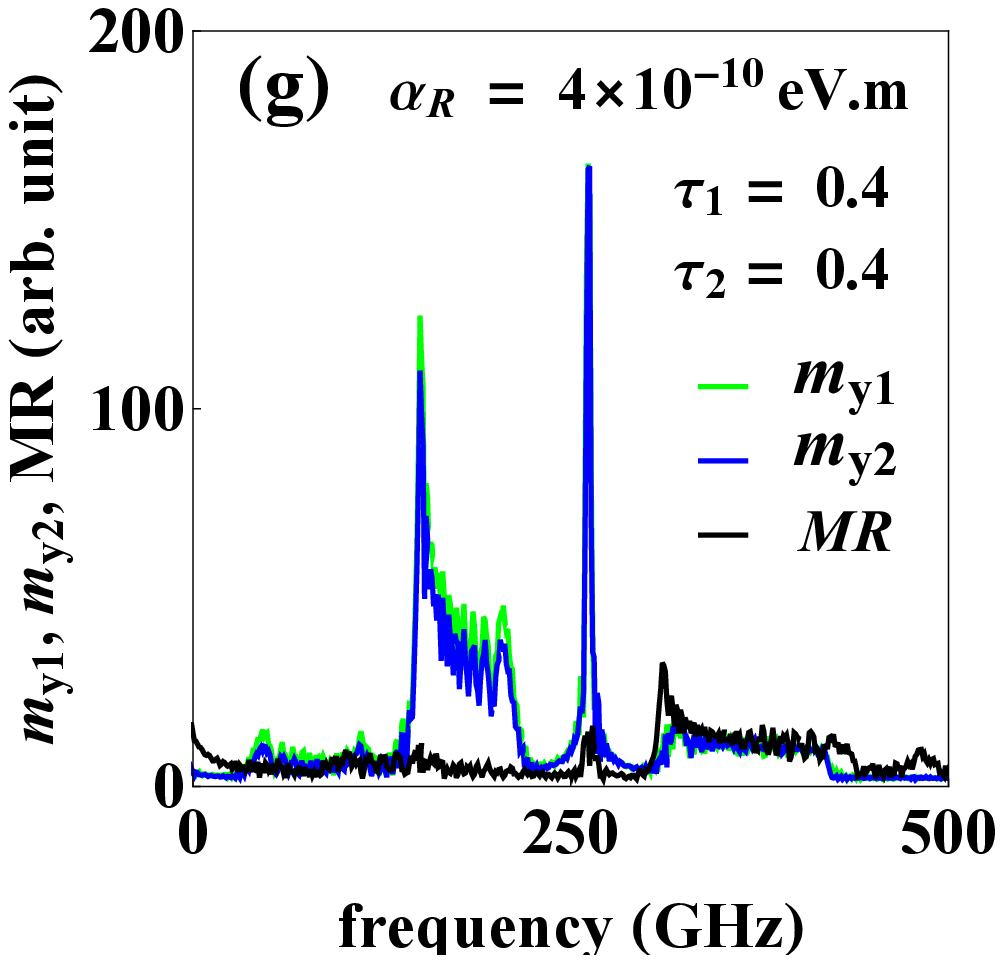}
\hspace{-0.09cm}
\includegraphics[scale = 0.275]{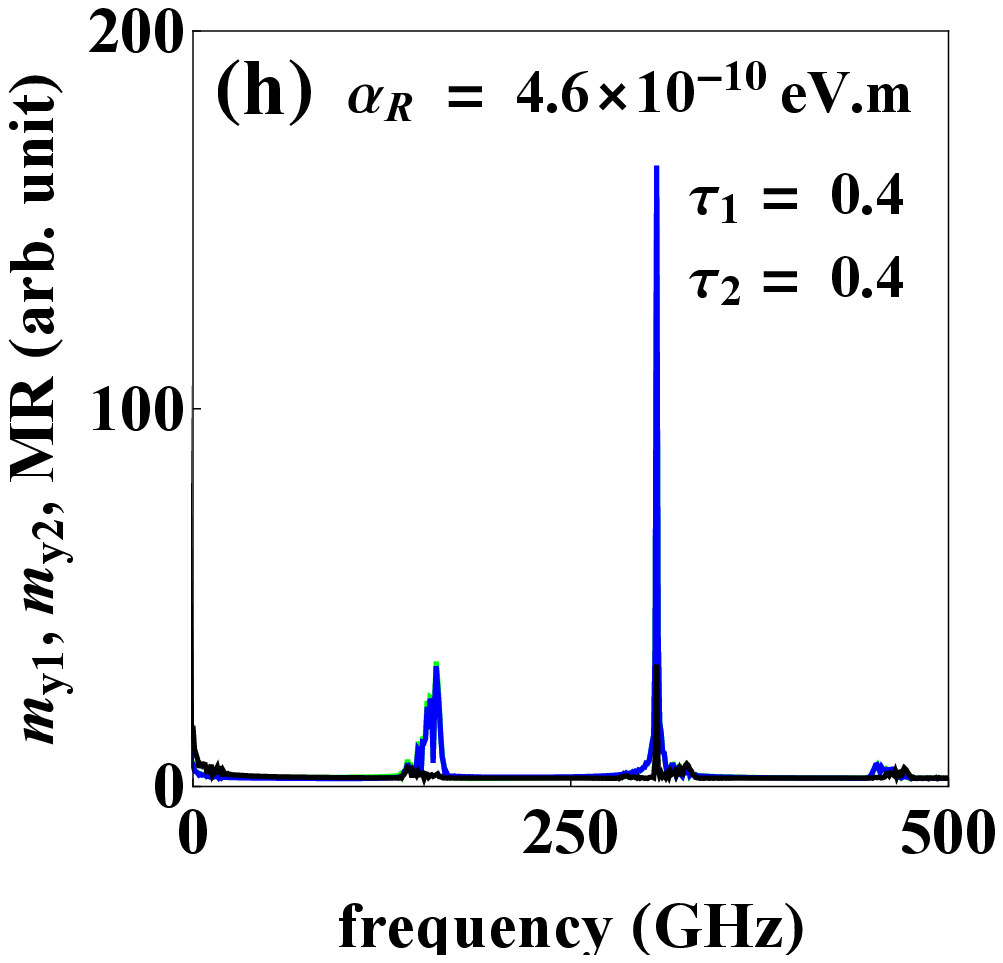}
\hspace{-0.09cm}
\includegraphics[scale = 0.275]{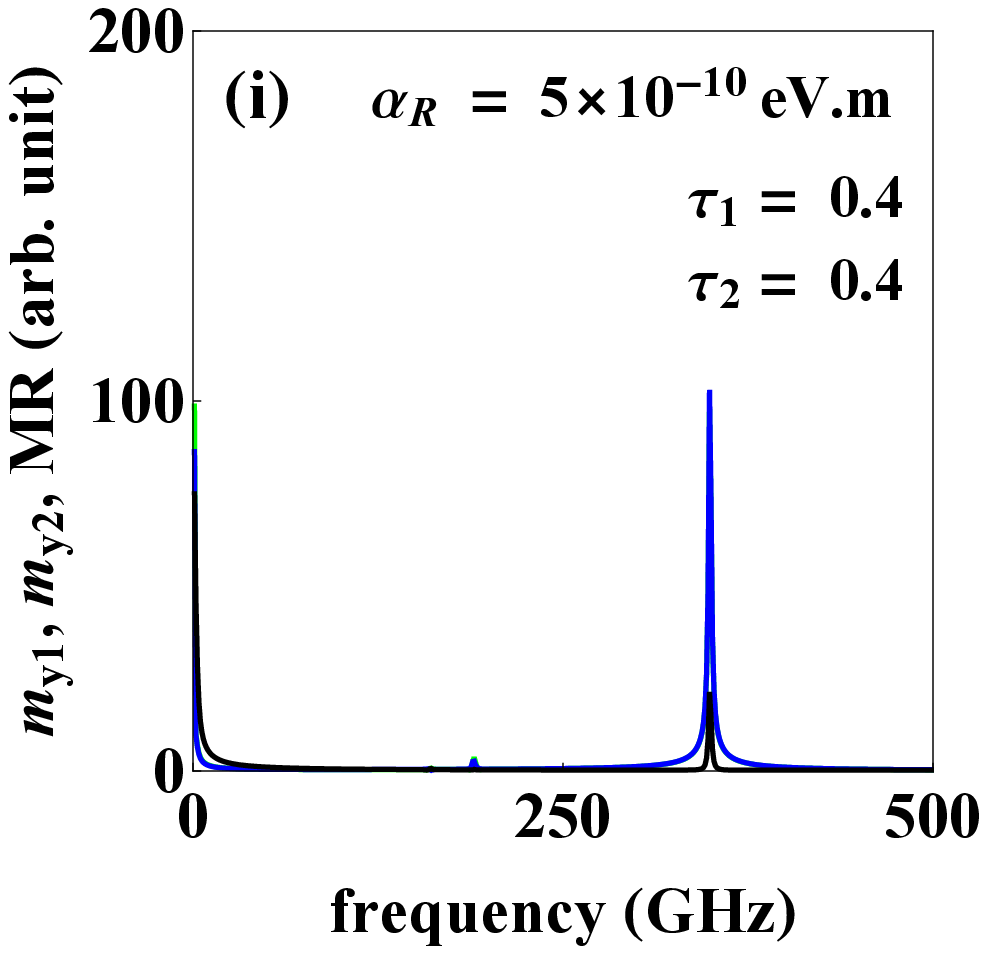}
\hspace{-0.09cm}
}
\caption{Oscillation trajectories of $\textbf{m}_1$ (blue) and $\textbf{m}_2$ (red) for (a) $\alpha_\text{R} = 4 \times 10^{-10}$ eV.m, (b) $\alpha_\text{R} = 4.6 \times 10^{-10}$ eV.m and (c) $\alpha_\text{R} = 5 \times 10^{-10}$ eV.m  considering $\tau_1 = \tau_2 = 0.4$ and  K$_\text{R}$ = $10$ K$_\text{DM}$. Plots (d) - (f) are the time evolution of the magnetization component $\text{m}_{y1}$ (red) and $\text{m}_{y2}$ (blue) while plots (g) - (i) are the Fourier transform of the magnetization.}
\label{fig5}
\end{figure}
\begin{figure}[hbt]
\centerline
\centerline{
\includegraphics[scale = 0.30]{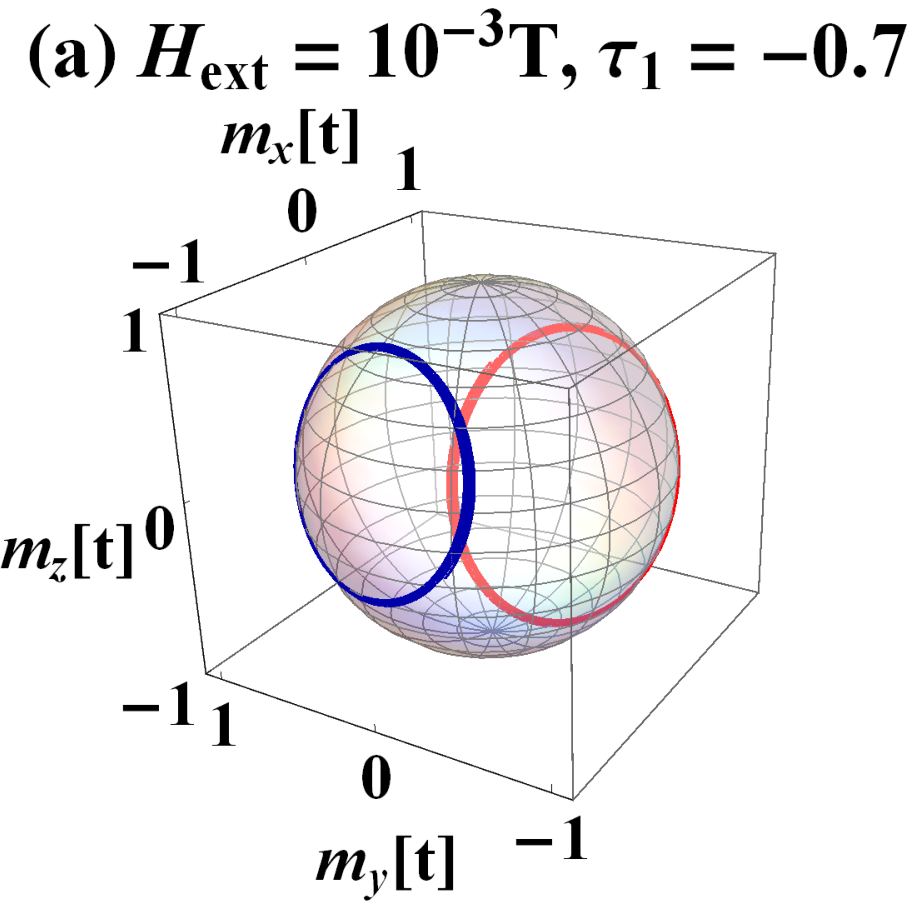}
\hspace{-0.09cm}
\vspace{0.15cm}
\includegraphics[scale = 0.30]{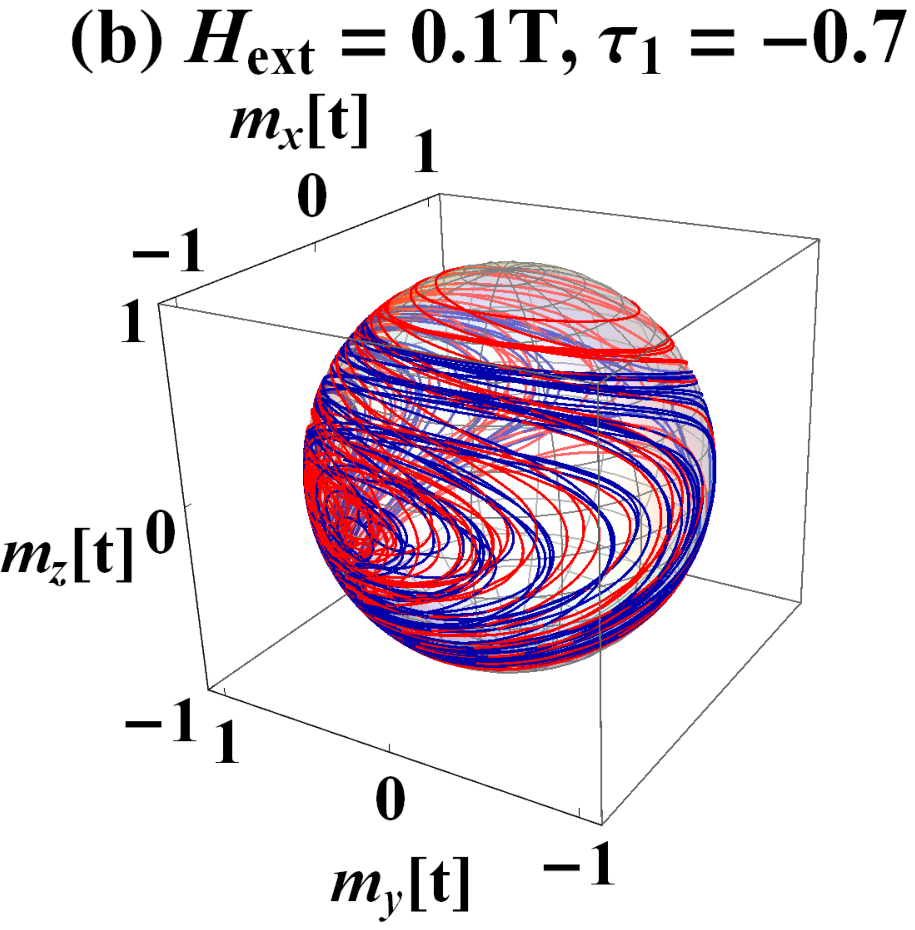}
\hspace{-0.09cm}
\vspace{0.15cm}
\includegraphics[scale = 0.30]{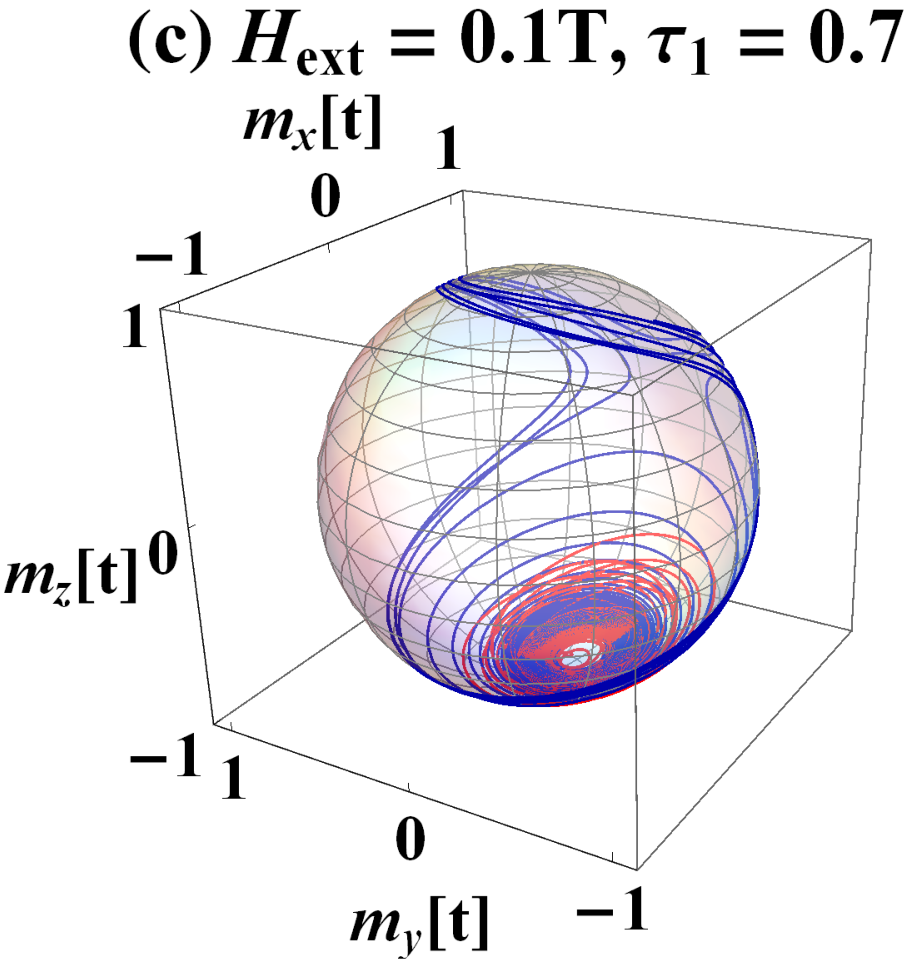}
\hspace{-0.09cm}
\vspace{0.15cm}
\includegraphics[scale = 0.275]{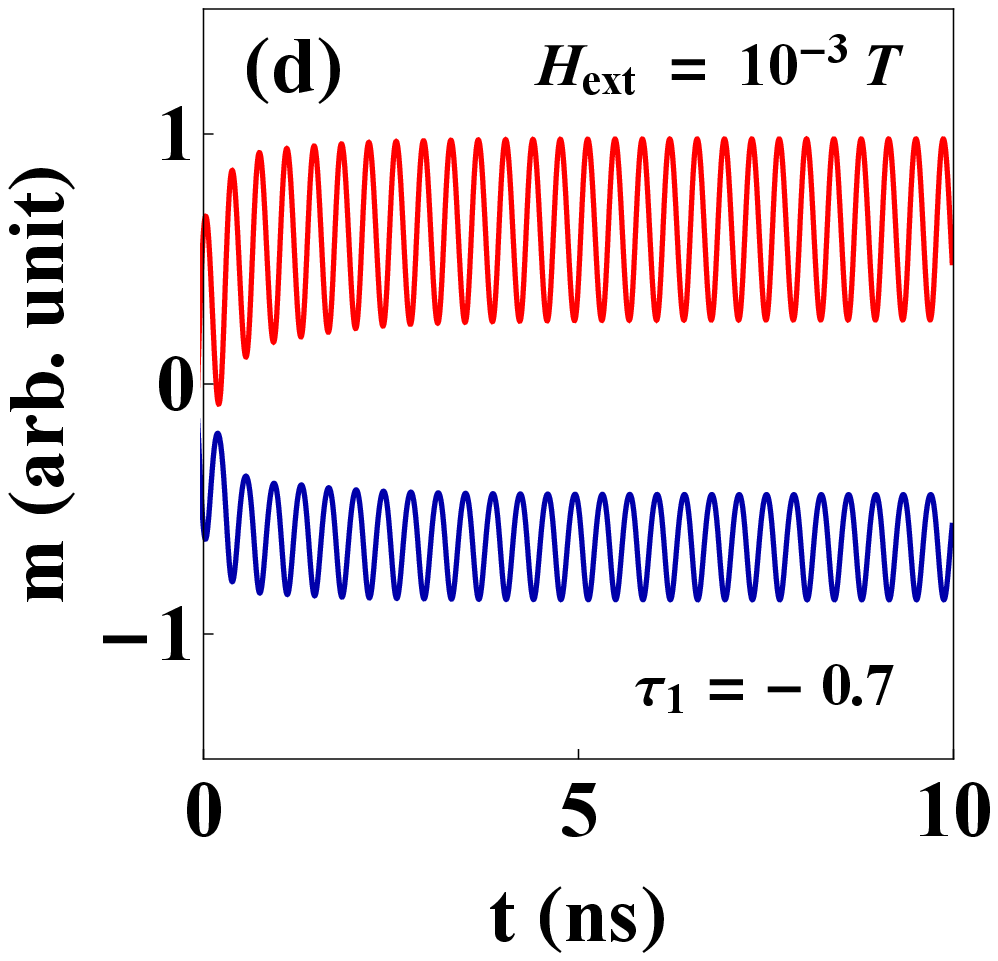}
\hspace{-0.09cm}
\includegraphics[scale = 0.275]{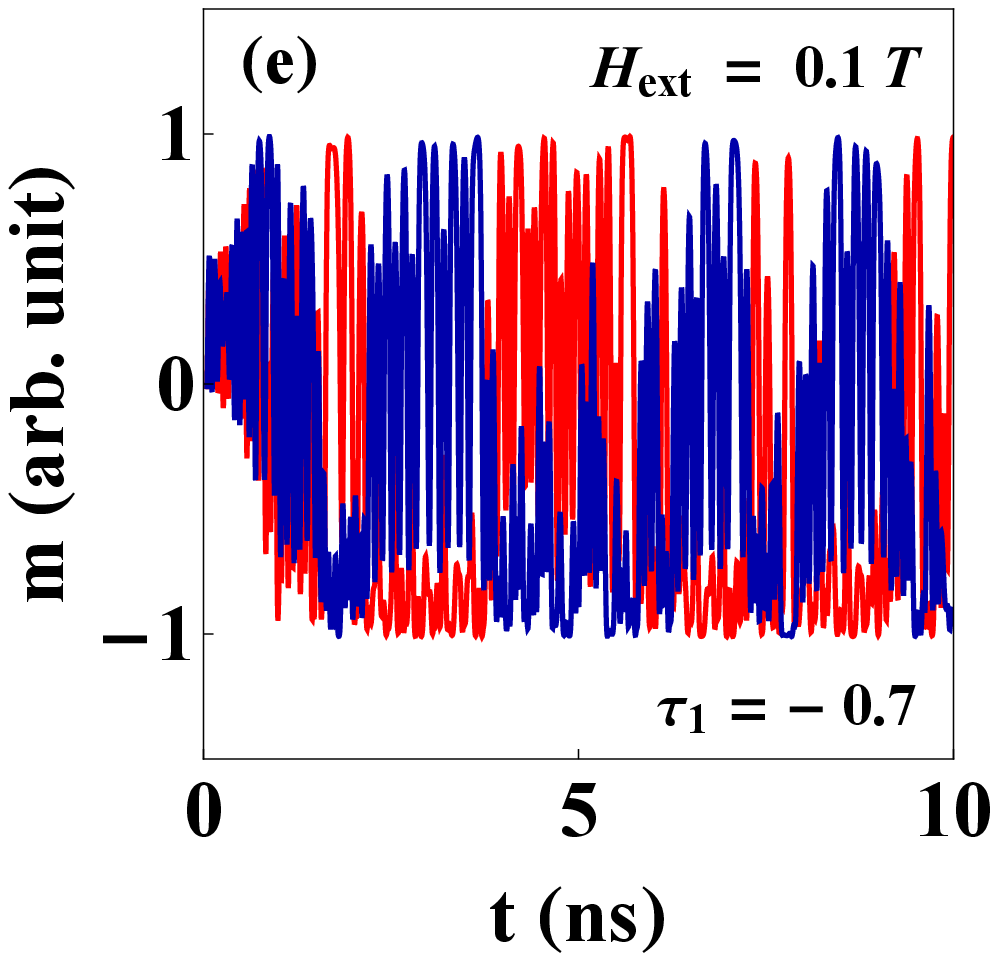}
\hspace{-0.09cm}
\includegraphics[scale = 0.275]{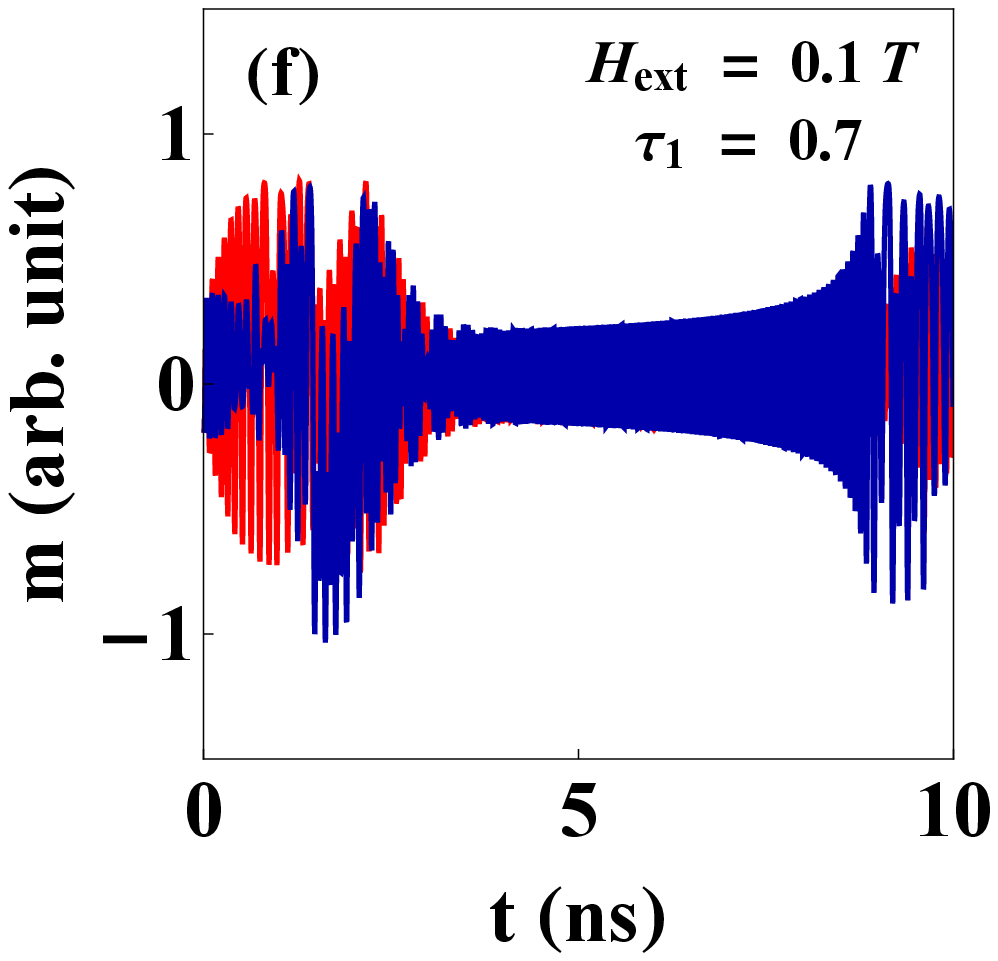}
\hspace{-0.09cm}
\includegraphics[scale = 0.275]{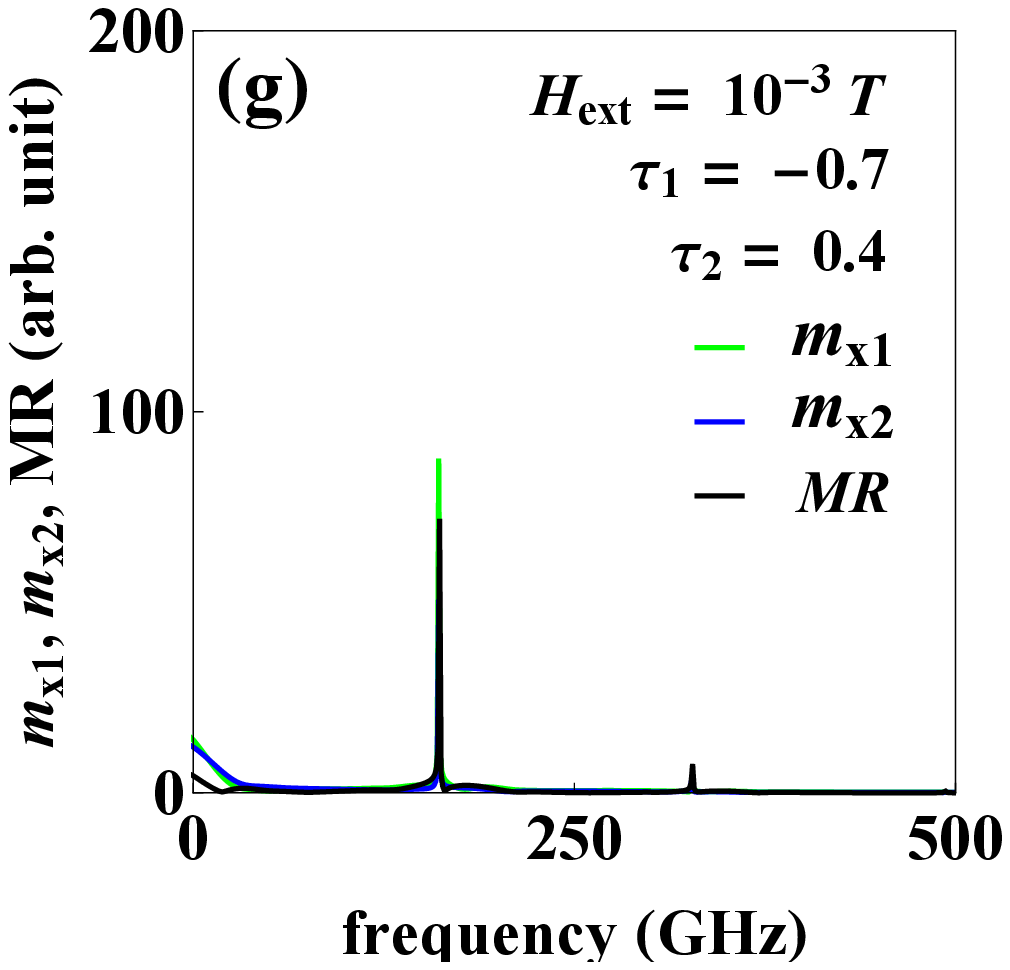}
\hspace{-0.09cm}
\includegraphics[scale = 0.275]{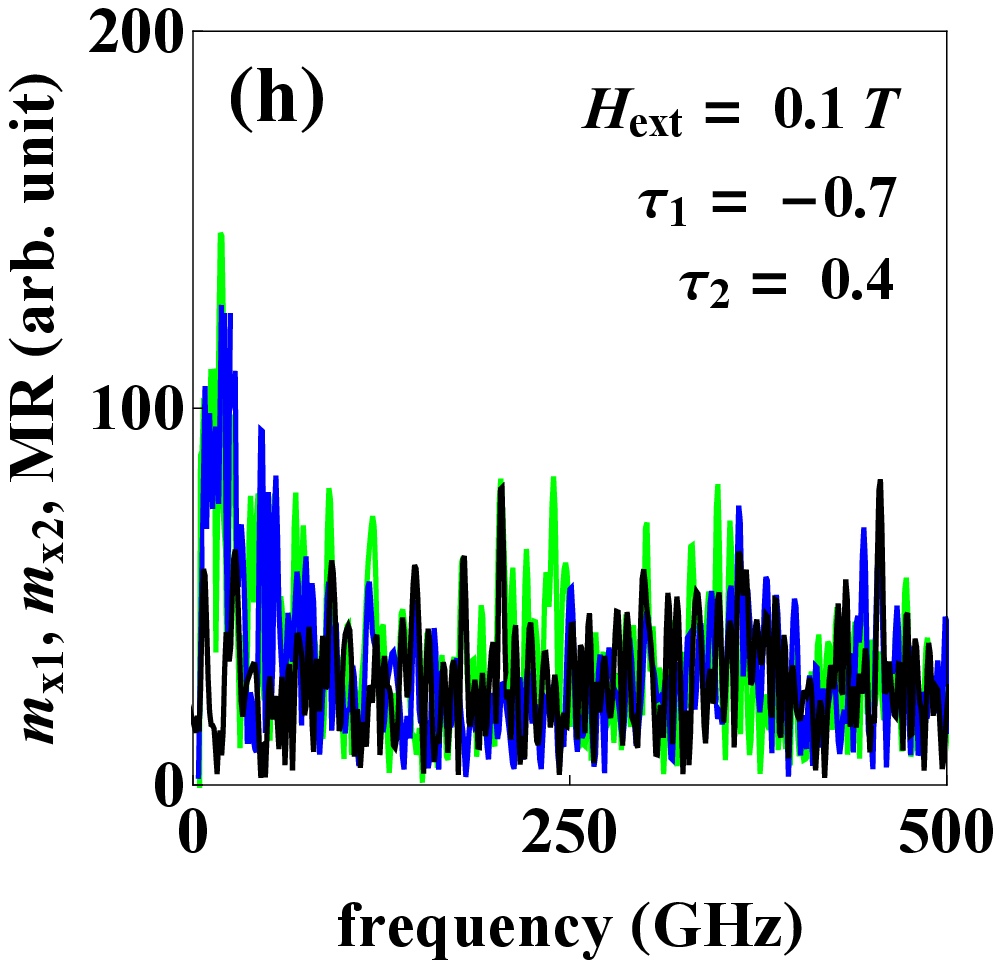}
\hspace{-0.09cm}
\includegraphics[scale = 0.275]{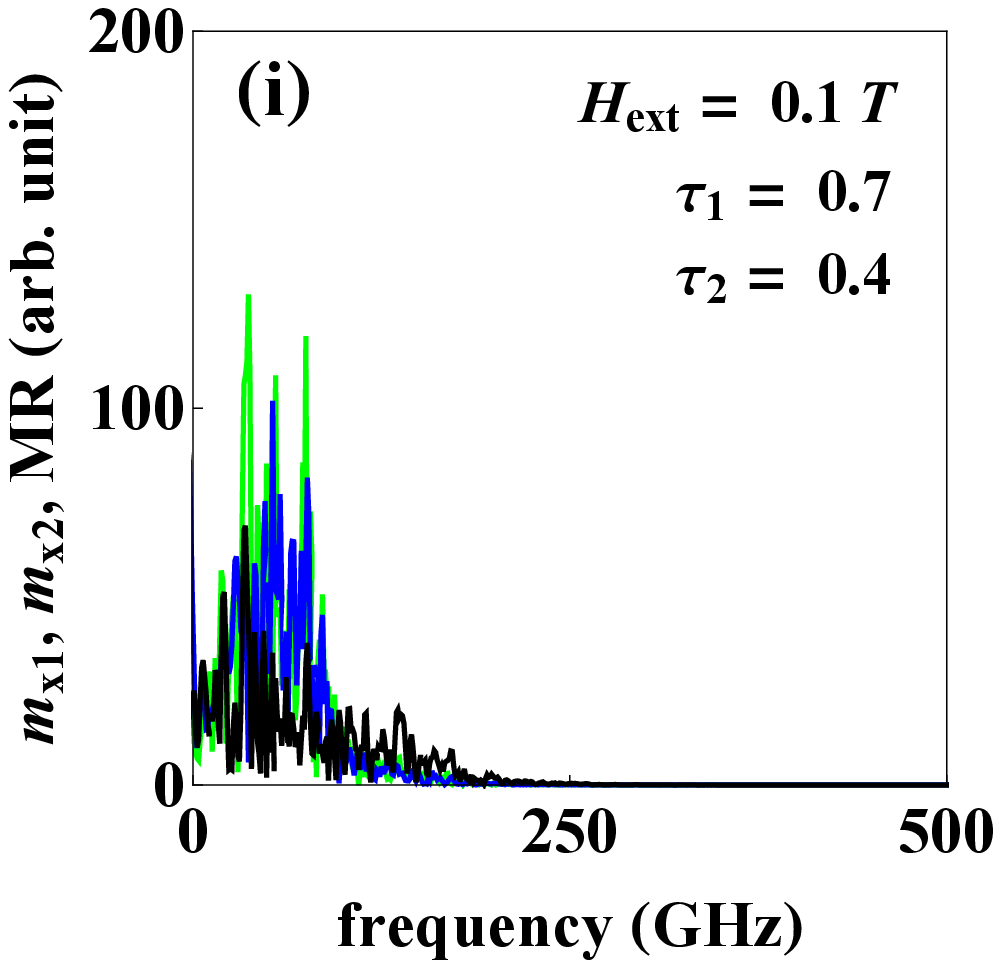}
\hspace{-0.09cm}
}
\caption{Oscillation trajectories of $\textbf{m}_1$ (blue) and $\textbf{m}_2$ (red) for (a) $H_\text{ext} = 10^{-3}$T, $\tau_1 = 0.7$ (b) $H_\text{ext} = 0.1$T, $\tau_1 = 0.7$ and (c) $H_\text{ext} = 0.1$T, $\tau_1 = -0.7$ considering $\tau_2 = 0.4$ and K$_\text{R}$ = $10$ K$_\text{DM}$. Plots (d) - (f) are the time evolution of the magnetization component $\text{m}_{x1}$ (red) and $\text{m}_{x2}$ (blue) while plots (g) - (i) are the Fourier transform of the magnetization.}
\label{fig6}
\end{figure}

\begin{figure*}[hbt]
\centerline
\centerline{
\includegraphics[scale = 0.55]{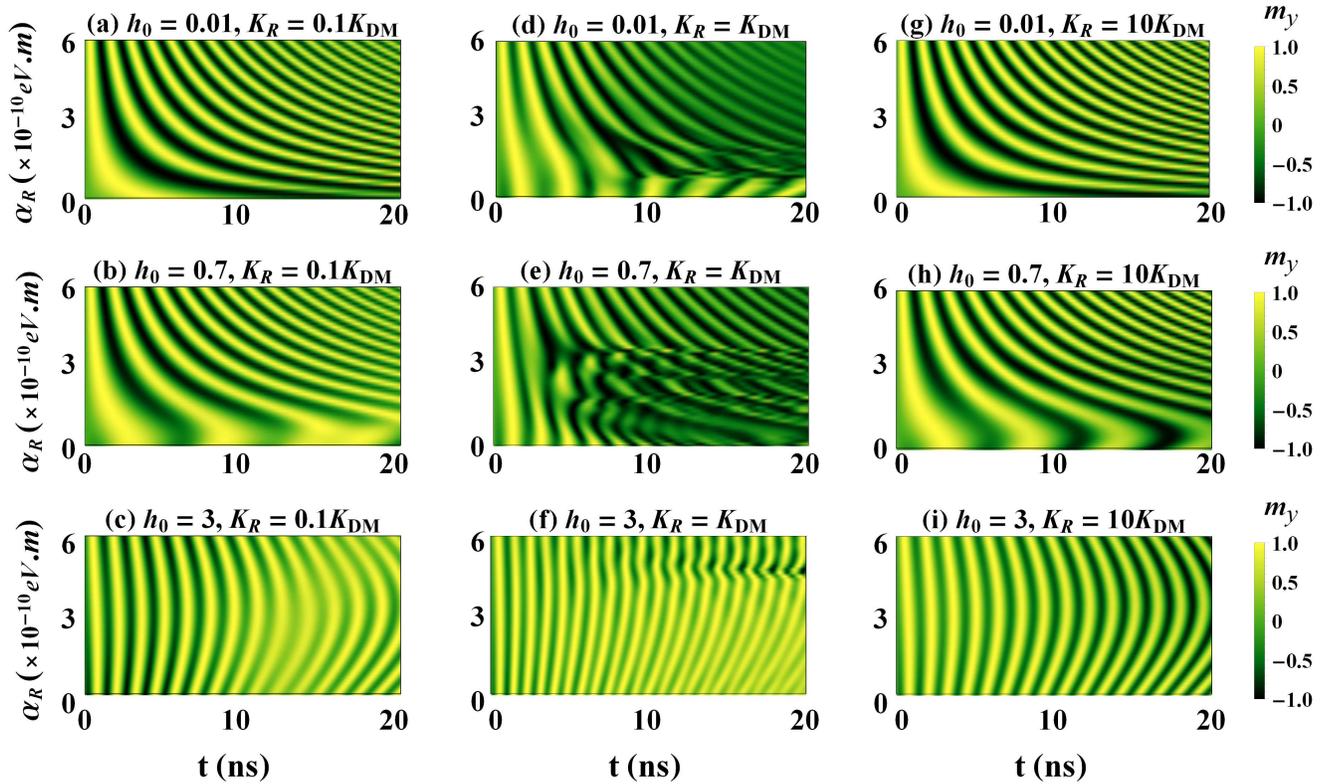}
\hspace{-0.1cm}
}
\caption{Density plot of $m_y$ with $\alpha_\textbf{R}$ and $t$ for $h_0 = 0.01$ (top), $h_0 = 0.7$ (middle) and $h_0 = 0.01$ (bottom) with different choices of K$_\text{R}$ and K$_\text{DM}$. The plots (a)-(c) in the left panel are for K$_\text{R} = 0.1K_\text{DM}$, (d)-(f) in the middle panel are for K$_\text{R} =$ K$_\text{DM}$ while (g)-(i) in the right panel are for K$_\text{R} = 10$K$_\text{DM}$.}
\label{fig7}
\end{figure*}
\subsection{Effect of RSOC}
Even though magnetization dynamics in conventional ferromagnetic MTJs have been extensively explored, the impact of SOC is yet to be understood. In general, SOC is responsible for splitting the spin-up and spin-down sub-bands. However, SOC can also significantly interplay with RKKY and DM interaction and, thus, affect the magnetization oscillations and reversal. Hence, in Fig. \ref{fig5}, we investigated the oscillations for different RSOC considering $h_0 = 0.7$, $\tau_1 = \tau_2 = 0.4$ and K$_\text{R}$ = $10$K$_\text{DM}$. The system undergoes non-linear oscillations for $\alpha_\text{R} = 4 \times 10^{-10}$ eV.m as seen from Figs. \ref{fig5}(a) and (d). This can be characterized through the multiple oscillation frequencies from Fig. \ref{fig5}(g). The disappearance of non-linear oscillations for $\alpha_\text{R} = 4.6 \times 10^{-10}$ eV.m indicates stable magnetization oscillations as seen from Figs. \ref{fig5}(b) and (e). This can be confirmed from the sharp peak around $350$ GHz in Fig. \ref{fig5}(h). With the further rise in $\alpha_\text{R} = 5 \times 10^{-10}$ eV.m, both $\mathbf{m}_1$ and $\mathbf{m}_2$ precess about the z-axis and decays slowly with an increase in time as seen from Figs. \ref{fig5}(c) and (f). The synchronized nature of these oscillations indicates the dominance of RKKY interaction in this system. So, RSOC can significantly impact RKKY and DM interactions, which indeed control the oscillations. 

\subsection{Effect of external magnetic field}
To understand the interplay of an external magnetic field on RZ material, we study the magnetization dynamics and oscillations under low and moderate external magnetic field in Fig. \ref{fig6}, considering $\tau_2 = 0.4$, $\alpha_\text{R} = 3 \times 10^{-10}$ eV.m, $h_0 = 0.7$ and K$_\text{R}$ = 10K$_\text{DM}$. The system executes stable magnetization oscillations around frequency $180$ GHz for $H_\text{ext} = 10^{-3}$T as seen from Figs. \ref{fig6}(a), (d) and (g). This is because the impact of DM interaction and the external magnetic field is minimized for strong K$_\text{R}$. However, as the magnetic field is increased to $0.1$T, the synchronization of $m_{x1}$ and $m_{x2}$ due to K$_\text{R}$ is broken. Thus, the system executes nonlinear oscillations as seen from Figs. \ref{fig6}(b) and (e). These nonlinear oscillations can be understood through the Fourier transform of the magnetization component in Fig. \ref{fig6}(h). Though the non-linear characteristics of the magnetization oscillations are found for $\tau_1 = -0.7$, the oscillations stabilize around the time range $\sim(3-8)$ ns for $\tau_1 = 0.7$ as seen from Fig. \ref{fig6}(f). As indicated by the Fourier spectra in Fig. \ref{fig6}(i), the magnetization display nonlinear oscillations under low low-frequency condition. This indicates that the oscillations not only depend upon the biasing current but can also be controlled through a suitable external magnetic field. 

\subsection{Interplay of RKKY, DM interaction with RSOC}
Though the interplay of RKKY and DM interactions have been investigated for STT switching and oscillation in various previous works \cite{li,emori,pizzini,ryu,baumgartner,cao,boulle,sampaio,jang} but the impact of RZ effect is yet to be understood. Fig. \ref{fig7} depicts the magnetization dynamics in presence of STT for double-barrier RZ-MTJ under different $h_0$ and K$_\text{R}$ and K$_\text{DM}$. It is observed that the magnetization oscillation under K$_\text{R}$ = 0.1 K$_\text{DM}$  is quite similar with that of K$_\text{R}$ = 10 K$_\text{DM}$ for corresponding choices of $h_0$ as seen from Figs. \ref{fig7}(a)-(c) and Figs. \ref{fig7}(g)-(i). We observe that with the increase in RSOC the magnetization oscillation also increases for low ($h_0 = 0.01$) and moderate ($h_0 = 0.7$) Zeeman strength. It is to be noted that the magnetization oscillation decays too rapidly as $\alpha_\text{R} \rightarrow 0$ for low values of $h_0$ while an oscillatory decay is observed at moderate $h_0$. This is  due to the reason that under low RZ energy, the Gilbert damping dominates the oscillations. However, the oscillations are still present in the system for suitable STT. The oscillations are too prominent under strong Zeeman condition even at low RSOC as seen from Fig. \ref{fig7}(c) as the required energy can be achieved from the exchange field. The oscillations slow down with time under moderate RSOC and high Zeeman energy. A non-linear characteristic of the magnetization orientation is observed as K$_\text{R}$ =  K$_\text{DM}$ as seen from the Figs. \ref{fig7}(d)-(f). This non-linear characteristics of $m_y$ is too prominent for $h_0$ = $0.7$ under moderate RSOC. This is due to the inconsistency of anisotropy of the corresponding magnetic layers at K$_\text{R}$ =  K$_\text{DM}$ \cite{li}. However, this inconsistency can be eliminated by using Zeeman field strength $h_0$ = $3$ as seen from Fig. \ref{fig7}(f) which can be used as a good oscillator. 

\section{Conclusions}
In summary, in this work the magnetization oscillations and switching have been investigated in the double-barrier RZ$|$HM$|$RZ MTJ in presence of  Rashba-Zeeman, RKKY and DM interactions for different STTs. We observed that the system behaves as a good oscillator or a switcher when there exist a significant difference in the strength of RKKY and DM interaction. A nonlinear characteristics of the magnetization oscillation is observed as the order of RKKY interaction matches with the DM interaction.  This double interface MTJ can also behave like a switcher for low and high STT when RKKY interaction is weaker than DM interaction. A similar characteristics can be seen for systems having strong RKKY interaction which signifies the interplay of RSOC on RKKY and DM interaction in RZ$|$HM$|$RZ MTJ. Though the magnetization oscillations are mostly nonlinear for systems having RKKY $\sim$ DMI, the magnetization switching can still be tuned using suitable combinations of STT and RSOC. The oscillations are found to be sensitive to the coupling between external magnetic field and RZ interaction. The nonlinearity of oscillations can be significantly reduced by using suitable combination of external magnetic field and RZ interaction. It is noted that there exist a dependence of switching time on layer thickness. The switching speed increases with thickness of the layers for systems having RKKY $\geq$ DM interactions while an opposite characteristics is seen for systems having RKKY $<$ DM interaction. Thus by choosing materials with suitable RZ, RKKY and DM interactions, a double-barrier RZ$|$HM$|$RZ MTJ can be realized as a switcher or an oscillator using suitable STT.

\end{document}